\pdfoutput=1
\documentclass[useAMS, usenatbib, a4paper, fleqn]{mn2e}
\usepackage{times, aas_macros, graphicx, amsmath, url, natbib, acronym, threeparttable, rotating, flushend}
\usepackage{hyperref}
\usepackage[noabbrev]{cleveref}
\usepackage[T1]{fontenc}

\acrodef{aals}[AALs]{Associated absorption lines}
\acrodef{bals}[BALs]{broad absorption lines}
\acrodef{qso}[QSO]{quasar}
\acrodef{agn}[AGN]{active galactic nuclei}
\acrodef{smbh}[SMBH]{super massive black hole}
\acrodef{igm}[IGM]{intergalactic medium}
\acrodef{cgm}[CGM]{circumgalactic medium}
\acrodef{cos}[COS]{Cosmic Origins Spectrograph}
\acrodef{fwhm}[FWHM]{full width at half maximum}
\acrodef{hst}[\emph{HST}]{Hubble Space Telescope}
\acrodef{stsci}[STScI]{Space Telescope Science Institute}
\acrodef{lsf}[LSF]{line-spread function}
\acrodef{snr}[SNR]{signal-to-noise ratio}
\acrodef{uv}[UV]{ultraviolet}
\acrodef{nuv}[NUV]{near ultraviolet}
\acrodef{fuv}[FUV]{far ultraviolet}
\acrodef{blr}[BLR]{broad-line region}
\acrodef{sed}[SED]{spectral energy distribution}
\acrodef{cie}[CIE]{collisional ionization equilibrium}
\acrodef{ism}[ISM]{interstellar medium}
\newcommand{\calcos}{\textsc{calcos}}
\newcommand{\vpfit}{\textsc{vpfit}}
\newcommand{\cloudy}{\textsc{cloudy}}
\newcommand{\xspec}{\textsc{xspec}}
\newcommand{\optxagnf}{\textsc{optxagnf}}

\crefname{figure}{fig.}{figs.}
\Crefname{figure}{Fig.}{Figs.}

\title[A kpc-scale outflow associated with a QSO at $z \sim 1$]{A compact, metal-rich, kpc-scale outflow in FBQS J0209-0438: Detailed diagnostics from \emph{HST}/COS extreme UV observations}
\author[Charles W. Finn et al.]{Charles W. Finn$^{1,2}$\thanks{E-mail: c.w.finn@durham.ac.uk}, Simon L. Morris$^{1}$, Neil H. M. Crighton$^{3,4}$, Fred Hamann$^{5}$, \newauthor Chris Done$^{1}$, Tom Theuns$^{2,6}$, Michele Fumagalli$^{7,8}$\thanks{Hubble Fellow}, Nicolas Tejos$^{1,9}$, \newauthor and Gabor Worseck$^{3}$ \\
$^{1}$Department of Physics, Durham University, South Road, Durham, DH1 3LE, UK \\
$^{2}$Institute for Computational Cosmology, Department of Physics, Durham University, South Road, Durham, DH1 3LE, UK \\
$^{3}$Max Planck Institute for Astronomy, K\"{o}nigstuhl 17, D-69117 Heidelberg, Germany \\
$^{4}$Centre for Astrophysics and Supercomputing, Swinburne University of Technology, PO Box 218, Victoria 3122, Australia \\
$^{5}$Department of Astronomy, University of Florida, 211 Bryant Space Sciences Center, Gainesville, FL 32611-2055 \\
$^{6}$Department of Physics, University of Antwerp, Campus Groenenborger, Groenenborgerlaan 171, B-2020 Antwerp, Belgium \\
$^{7}$Carnegie Observatories, 813 Santa Barbara Street, Pasadena, CA 91101, USA \\
$^{8}$Department of Astrophysics, Princeton University, Princeton, NJ 08544-1001, USA \\
$^{9}$University of California Observatories-Lick Observatory, University of California, Santa Cruz, CA 95064, USA
}

\begin{document}
\date{Accepted 2014 March 13. Received 2014 February 17; in original form 2013 October 29}
\pagerange{\pageref{firstpage}--\pageref{lastpage}} \pubyear{2013}
\maketitle
\label{firstpage}

\begin{abstract}
We present \emph{HST}/COS observations of highly ionized absorption lines associated with a radio-loud QSO at $z = 1.1319$. The absorption system has multiple velocity components, with an overall width of $\approx 600$~\kms, tracing gas that is largely outflowing from the QSO at velocities of a few 100~\kms. There is an unprecedented range in ionization, with detections of \ion{H}{1}, \ion{N}{3}, \ion{N}{4}, \ion{N}{5}, \ion{O}{4}, \ion{O}{4}*, \ion{O}{5}, \ion{O}{6}, \ion{Ne}{8}, \ion{Mg}{10}, \ion{S}{5} and \ion{Ar}{8}. We estimate the total hydrogen number density from the column density ratio $N(\textrm{O}\;\textsc{iv*}) / N(\textrm{O}\;\textsc{iv})$ to be $\log (n_{\textrm{H}} / \textrm{cm}^{-3}) \sim 3$. Combined with constraints on the ionization parameter in the \ion{O}{4} bearing gas from photoionization equilibrium models, we derive a distance to the absorbing complex of $2.3 \lesssim R \lesssim 6.0~\textrm{kpc}$ from the centre of the QSO. A range in ionization parameter, covering $\sim 2$ orders of magnitude, suggest absorption path lengths in the range $10^{-4.5} \lesssim l_{\textrm{abs}} \lesssim 1~\textrm{pc}$. In addition, the absorbing gas only partially covers the background emission from the QSO continuum, which suggests clouds with transverse sizes $l_{\textrm{trans}} \lesssim 10^{-2.5}~\textrm{pc}$. Widely differing absorption path lengths, combined with covering fractions less than unity across all ions pose a challenge to models involving simple cloud geometries in associated absorption systems. These issues may be mitigated by the presence of non-equilibrium effects, which can be important in small, dynamically unstable clouds, together with the possibility of multiple gas temperatures. The dynamics and expected lifetimes of the gas clouds suggest that they do not originate from close to the AGN, but are instead formed close to their observed location. Their inferred distance, outflow velocities and gas densities are broadly consistent with scenarios involving gas entrainment or condensations in winds driven by either supernovae, or the supermassive black hole accretion disc. In the case of the latter, the present data most likely does not trace the bulk of the outflow by mass, which could instead manifest itself as an accompanying warm absorber, detectable in X-rays.
\end{abstract}


\section{Introduction}
\label{introduction}
\ac{aals} seen in \ac{qso} spectra offer a unique physical perspective on the gaseous environments in the vicinity of \ac{qso}s. Many \ac{aals} are thought to arise in material that has been ejected from a region close to the \ac{smbh} \cite[within a few pc, e.g.][]{Nestor:2008gb,Wild:2008hn,Muzahid:2012kl}. The resulting outflows might play a major role in the quenching of star formation and in regulating the growth of \ac{smbh}s \citep{Silk:1998up,King:2003gt,Scannapieco:2004es,DiMatteo:2005hl,Ostriker:2010ik,Hopkins:2010cf}. Some may arise from material ejected in supernova explosions \cite[e.g.][]{Veilleux:2005ec}. In addition, some \ac{aals} appear to probe gas that is part of the host galaxy halo \cite[e.g.][]{Williams:1975ge,Sargent:1982kl,Morris:1986fa,Tripp:1996ge,Hamann:2001eg,DOdorico:2004jd}. In some cases, this gas may eventually condense in the disc to form new generations of stars via Galactic fountain processes \citep{Bregman:1980gn,Fraternali:2013dd}. The balance of gas accretion and outflow shapes the galaxy luminosity function and drives the evolution of galaxies \cite[e.g.][]{Benson:2003ch,Bower:2006fj}. Observations of \ac{aals} therefore provide a detailed snapshot of these forces at work. Constraints on the metallicity of these absorbers also provides a direct measure of the star formation and chemical enrichment histories in the centres of active galaxies \citep{Hamann:1993jb,Hamann:1999ky}.

\ac{aals} are loosely defined as having absorption redshifts within a few thousand \kms\ of the \ac{qso} emission redshift, and velocity widths of less than a few hundred \kms. These are narrow when compared to the so-called \ac{bals}, which have velocity widths and displacements from the \ac{qso} redshift that often exceed $10^4$~\kms \citep{Weymann:1979bt,Foltz:1986fg,Weymann:1991cn,Trump:2006ht}. The origin of the \ac{bals} is presumably in a wind, driven by accretion processes close to a \ac{smbh}. However, the exact origin of the \ac{aals} is far less clear. In addition, not all \ac{aals} are necessarily intrinsic to the \ac{qso} \cite[e.g.][]{Tripp:1996ge,Ganguly:2013hd}. Those that may be intrinsic show: (i) Absorption strength that is seen to vary on time-scales of around a year \cite[e.g.][]{Hamann:1995ff,Srianand:2001hh,Hall:2011ej,Vivek:2012hh}. (ii) Metallicities $\gtrsim Z_{\odot}$ \cite[e.g.][]{Petitjean:1994ti,Hamann:1997iu,Muzahid:2013dm}. (iii) Partial coverage of the \ac{qso} accretion disc continuum and/or \ac{blr} \cite[e.g.][]{Barlow:1997eb,Srianand:1999hj,Gabel:2006fw,Arav:2008fa}. (iv) The presence of excited fine structure lines \cite[e.g.][]{Morris:1986fa,Srianand:2000tq,Hamann:2001eg,Edmonds:2011fz}. These properties are rarely seen in intervening absorption-line systems \cite[see][for an exceptional case]{Balashev:2011cc}, and so \ac{aals} with these properties are believed to trace gas that originates near the \ac{qso}, or in the halo of the host galaxy.

\ac{aals} have been observed in optical, \ac{uv} and X-ray spectra of local \ac{agn} and \ac{qso}s, with the X-ray observations often revealing a plethora of absorption lines and K-shell absorption edges from species with ionization potentials of a few hundred eV (e.g. \ion{O}{7} and \ion{O}{8}). Collectively, these are usually referred to as `warm absorbers' \cite[in $\sim 50$\% of Seyfert galaxies;][]{Crenshaw:2003hz}. Many authors have suggested that the presence of warm absorbers is correlated with the detection of \ac{aals} and \ac{bals} in optical and \ac{uv} spectra, usually through species like \ion{C}{3}, \ion{C}{4} and \ion{N}{5}, with ionization potentials $\lesssim 100~\textrm{eV}$ \cite[e.g.][]{Mathur:1994ds,Mathur:1998fo,Brandt:2000de,Kaspi:2002cr,Arav:2007dx,DiGesu:2013cf}. However, at present, it is not clear whether these correlations imply a physical connection between the gas clouds traced by these ions \cite[see, for example,][]{Srianand:2000jd,Hamann:2000bi,Hamann:2013cq}.

To better understand the nature of associated absorbing clouds, more observations of the most highly ionized \ac{uv} species (ionization potentials $> 100~\textrm{eV}$) are required, so that the ionization structure of the absorbing gas can be more extensively characterised. At low redshift, observations of many \ac{uv} ions are impossible due to the presence of Galactic Lyman limit absorption (the relevant transitions have rest-frame wavelengths $< 912$~\AA). Observations in the optical, which are limited to high redshifts, are complicated by contamination from the Lyman alpha forest, together with a higher incidence rate of Lyman limit systems \citep{Fumagalli:2013ee}. At intermediate redshifts $0.5 \lesssim z \lesssim 1.5$, the problem of Galactic absorption is virtually eliminated, and the Lyman alpha forest contamination is less severe, making this a profitable redshift range to study highly ionized \ac{aals}. Observations must be conducted in the \ac{fuv}, and with the advent of the \ac{cos} on-board the \ac{hst}, hundreds of \ac{qso}s are now observable in this wavelength regime, thanks largely to a sensitivity more than ten times that of the previous generation medium resolution \ac{uv} spectrograph \citep{Green:2012dj}. Together with the \ac{nuv} modes of \ac{cos}, \ac{aals} with ionization parameters of a few, to a few hundred eV are accessible. Detailed diagnostics on the ionization structure of associated gas clouds are thus available in a large number of \ac{qso}s for the first time. In addition, coverage of strong transitions due to fine-structure excited states in ions such as \ion{O}{4} and \ion{O}{5} (see fig. 1 in \citet{Arav:2013be} for a full summary) provide powerful density diagnostics in highly ionized associated gas clouds, which provide crucial constraints on the physical conditions in and around the absorbing regions.

In this paper, we present observations of the radio-loud \ac{qso} FBQS J0209-0438 obtained with \ac{cos}. This \ac{qso} was targeted as part of a larger programme of observations to study two-point correlation statistics between \ac{igm} absorbers and galaxies at $z \lesssim 1$ \cite[PID 12264, PI: S.L. Morris;][]{2014MNRAS.437.2017T}. A highly ionized system of \ac{aals} is present, with complex velocity structure and an overall velocity width $\approx 600$~\kms. We also report the detection of absorption due to the fine-structure \ion{O}{4}* transition. A summary of the observations and data reduction, together with a characterisation of the rest-frame \ac{qso} \ac{sed} is presented in \Cref{observations}. A complete analysis of the \ac{aals}; their covering fractions, column densities and line widths is presented in \Cref{analysis}. In \Cref{properties} we present the results of photoionization and collisional ionization models in an attempt to characterise the physical properties of the gas. In particular we examine the ionization state, metallicity and density of the gas, and use these properties to put constraints on the absorbing geometry and distance from the \ac{qso}. In \Cref{discussion} we present a discussion of these results and draw conclusions.
 
\section{Observations of FBQS J0209-0438}
\label{observations}
The \ac{qso} FBQS J0209-0438 (hereafter Q0209) was observed with \ac{hst}/\ac{cos} in December 2010. The observations made use of both the medium-resolution ($R \sim 18000$) \ac{fuv} and low-resolution ($R \sim 3000$) \ac{nuv} modes of COS, giving wavelength coverage free from second-order light in the range 1240 -- 3200~\AA. Four central wavelength settings were used in the \ac{fuv}, and three in the \ac{nuv}, to ensure that the resulting spectrum had no gaps. For each central wavelength setting, multiple exposures were obtained at a number of positions offset along the dispersion direction from the nominal one (FP-POS=3; see Table~\ref{cos_config}). Each offset position is separated by $\sim 250$ pixels in the \ac{fuv} channels and $\sim 52$ pixels in the \ac{nuv} channels, with FP-POS=1, 2 offset to lower wavelengths from FP-POS=3 and FP-POS=4 offset to higher wavelengths. Merging these offsets minimises the effects of fixed-pattern noise in the \ac{cos} \ac{fuv} and \ac{nuv} detectors by effectively dithering around these features, allowing them to be subtracted in the final co-added spectra. They are particularly crucial for the \ac{fuv} modes, which suffer from additional fixed-pattern noise attributable to grid wires that produce shadows on the face of the detector. For more details, including a complete description of the design and in-flight performance of \ac{cos}, see \citet{Osterman:2011gj} and \citet{Green:2012dj}.

On the 7th October 2013, under clear skies and excellent seeing conditions ($\sim 0.55''$), we obtained a near-infrared spectrum of Q0209 with FIRE \citep{Simcoe:2013kh}, mounted on the Magellan Walter Baade 6.5m telescope at Las Campanas Observatory. In echelle mode, and for the adopted $0.6''$ slit, FIRE delivers a continuous spectrum across the wavelength range $0.82-2.51\mu\textrm{m}$ at a spectral resolution of $\sim 50$~\kms. Data were collected in $2 \times 729~\textrm{s}$ exposures while Q0209 was at an airmass of 1.1. To correct for spectral features arising from the Earth's atmosphere, we also acquired 2 spectra of the A0V star HD25266 at similar airmass, with exposure times of $729~\textrm{s}$ each.

\subsection{COS data reduction}
Individual exposures from \ac{cos} were downloaded from the \ac{stsci} archive and reduced using \calcos\ v2.18.5. The boxcar extraction implemented in \calcos\ was optimised by narrowing the source extraction box to match the apparent size of the source in the cross-dispersion direction upon inspection of each flat-fielded image. This amounted to 25 pixels for all G130M exposures, and 20 pixels for all G160M and G230L exposures. The background extraction boxes were also enlarged to encompass as much of the background signal as possible, whilst avoiding regions close to the detector edges. The \calcos\ reduction procedure performs a boxcar smoothing on the background counts at each pixel along the dispersion axis to provide a robust measure of the background. This background smoothing is applied everywhere, including areas affected by scattered light from strong geocoronal emission lines, which leads to an overestimation of the background level in these regions and their immediate vicinity due to `smearing' of the light. To avoid this, we set the background smoothing length in \calcos\ to 1 pixel and perform our own background smoothing procedure in post-processing. This procedure masks out affected portions of the spectrum, namely the \lya~and \ion{O}{1} $\lambda\lambda 1302 + 1306$ geocoronal emission lines, then interpolates across the gap to estimate the actual background level.

\begin{table}
\caption{Summary of the \ac{hst}/\ac{cos} observations.}
\label{cos_config}
\begin{threeparttable}
\begin{tabular}{c c c c c}
	\hline
	grating & $\lambda_{\textrm{centre}}$ (\AA)\tnote{a} & FP-POS & $t_{\textrm{exp}}$ (s)\tnote{b} & x1d rootname \\
	\hline
	G130M & 1291 & 2 & 2321 & \texttt{lbj011ucq} \\
	G130M & 1291 & 3 & 2948 & \texttt{lbj011v4q} \\
	G130M & 1291 & 4 & 2948 & \texttt{lbj011vcq} \\
	G130M & 1318 & 3 & 2948 & \texttt{lbj011vjq} \\
	G130M & 1318 & 4 & 2948 & \texttt{lbj011vsq} \\
	G160M & 1600 & 1 & 2276 & \texttt{lbj012vpq} \\ 
	G160M & 1600 & 2 & 2948 & \texttt{lbj012vvq} \\
	G160M & 1600 & 3 & 2948 & \texttt{lbj012w2q} \\
	G160M & 1600 & 3 & 2948 & \texttt{lbj012waq} \\
	G160M & 1600 & 4 & 2948 & \texttt{lbj012wiq} \\
	G160M & 1623 & 1 & 2276 & \texttt{lbj013k8q} \\
	G160M & 1623 & 2 & 2948 & \texttt{lbj013keq} \\
	G160M & 1623 & 3 & 2948 & \texttt{lbj013kmq} \\ 
	G160M & 1623 & 3 & 2948 & \texttt{lbj013ktq} \\
	G160M & 1623 & 4 & 2948 & \texttt{lbj013l0q} \\
	G230L & 2950 & 3 & 2373 & \texttt{lbj014vkq} \\
	G230L & 2950 & 4 & 2985 & \texttt{lbj014vuq} \\
	G230L & 2635 & 3 & 2985 & \texttt{lbj014w3q} \\
	G230L & 3360 & 3 & 2985 & \texttt{lbj014wcq} \\
	G230L & 3360 & 4 & 2985 & \texttt{lbj014wlq} \\
	\hline
\end{tabular}
\begin{tablenotes}
    \item[a] Central wavelength setting.
    \item[b] Exposure time.
\end{tablenotes}
\end{threeparttable}
\end{table}

Alignment and co-addition procedures were performed using custom built Python routines\footnote{\url{https://github.com/cwfinn/COS/}}. These are loosely based upon IDL routines developed by the \ac{cos} GTO team\footnote{\url{http://casa.colorado.edu/~danforth/science/cos/costools.html}}. They work as follows:

Individual \verb|x1d| files (\verb|*_x1d.fits|) produced in the \calcos\ reduction are collated, along with their header information (files are listed in \Cref{cos_config}). Each of these files corresponds to a one-dimensional extracted spectrum from a single central wavelength setting and FP-POS position, and contains two data extensions: an `A' stripe and a `B' stripe. Data quality flags\footnote{\url{http://www.stsci.edu/hst/cos/pipeline/cos_dq_flags}} for all pixels are assigned new flags to mean one of three options: (i) retain pixel for co-addition with a weight equal to 1, (ii) retain pixel for co-addition with a weight equal to 0.5, or (iii) discard pixel from the co-addition process. Flags assigned with option (i) are those where no anomalous condition is noted, or where unusual features have been identified in long background exposures. The latter is not expected to have any effect on the final data products except perhaps where the count rate is very low or the background higher than normal. Flags assigned with option (ii) are those in regions where the background count rate is apparently higher than the surrounding region and/or is unstable, and in regions on the \ac{fuv} detector where the gain is low enough so as to affect the calibration. For the \ac{nuv} channel only, regions affected by detector shadows (vignetting in this case) are assigned option (ii). These vignetted areas affect significant portions of the \ac{nuv} spectrum and so are retained to avoid large gaps in the final co-added spectrum. A weight of 0.5 ensures that the data in these regions contributes significantly towards increasing the \ac{snr}, whilst minimising any additional error in the flux calibration. All other flags are assigned option (iii). Data quality flags that are assigned either of the first two options are referred to as `good' and those assigned the latter option are referred to as `bad.'

Next, the background counts are re-estimated, boxcar smoothing only across pixels not affected by scattered geocoronal light and with good data quality flags. The error array from \calcos\ is then re-calculated with the new background smoothing lengths. These smoothing lengths are set to 1000 pixels for the \ac{fuv}A stripes, 500 pixels for the \ac{fuv}B stripes and 100 pixels for all \ac{nuv} stripes. The numbers are arbitrary, but chosen to be large enough to ensure a robust estimation of the background, and small enough not to destroy large-scale features in the background light profile. Each spectrum is then flux calibrated using the time dependent sensitivity curves provided by \ac{stsci}, corrected to the epoch of observation.

Exposures are now co-aligned by cross-correlating regions centred on strong Galactic absorption features. Specifically, these are \ion{C}{2} $\lambda 1334$, \ion{Al}{2} $\lambda 1670$, \ion{Si}{2} $\lambda 1260$, \ion{Si}{2} $\lambda 1526$ and \ion{Mg}{2} $\lambda\lambda 2796, 2803$. Using these features allows for co-alignment between all settings in the \ac{fuv} gratings (assuming the \ac{fuv} wavelength scale from \calcos\ is relatively correct) and between the 2635~\AA\ and 2950~\AA\ central wavelength settings of the G230L grating `B' stripes. For each grating we pick the central wavelength setting and FP-POS position with the most accurately determined wavelength solutions from \ac{stsci} as a reference. This is FP-POS=3 for all gratings and central wavelength settings of 1309~\AA, 1600~\AA\ and 2950~\AA\ (just the `B' stripe) for the G130M, G160M and G230L gratings respectively. We assume that \calcos\ correctly shifts these configurations into a heliocentric reference frame. All other settings for each grating are then cross-correlated with these ones if the reference and comparison settings both contain one of the absorption features specified. Wavelength offsets are then applied to the comparison settings to match the reference ones. These offsets typically amount to a resolution element or less. For those settings that could not be aligned on any of the Galactic features specified, we searched for other strong absorption lines on which to perform the cross-correlation and found at least one absorption line for each setting. Once each exposure has been aligned, they are then scaled so that their median flux values match in overlap regions.

Before performing the final co-addition of the data, flux and error values assigned to pixels with bad data quality flags are set to zero, and pixels flagged for de-weighting have their exposure time reduced by a factor of two. Flux and error values are then placed on a linear wavelength scale using a nearest-neighbour interpolation. The wavelength spacing is set to the dispersion of the G130M grating at $\lambda \leq 1460$~\AA, the dispersion of the G160M grating at 1460~\AA\ $< \lambda \leq 1795$~\AA, and the dispersion of the G230L grating at $\lambda > 1795$~\AA. Performing the interpolation in this way prior to co-addition has been shown to minimise the effects of non-Poissonian noise in the co-added data \citep{Keeney:2012kg}. The co-addition is then performed using modified exposure-time weighting, i.e. flux values are co-added according to the following rule:
\begin{equation}
	F_i = \frac{\sum_j{F_j \times t_{\textrm{exp;~}j}}}{\sum_j{t_{\textrm{exp;~}j}}},
\end{equation}
where $i$ represents the $i$th pixel in the final, co-added spectrum, and $j$ represents the $j$th pixel that is co-added to make pixel $i$. Similarly, error values are co-added as follows:
\begin{equation}
	\delta F_i = \frac{\sqrt{\sum_j{(\delta F_j \times t_{\textrm{exp;~}j}}) ^ 2}}{\sum_j{t_{\textrm{exp;~}j}}}	
\end{equation}
Finally, the combined \ac{fuv} and \ac{nuv} spectra were binned to ensure Nyquist sampling, i.e. two pixels per resolution element. This corresponds to 0.0395~\AA~pixel$^{-1}$ for the \ac{fuv}, and a 0.436~\AA~pixel$^{-1}$ for the \ac{nuv}.

\subsection{FIRE data reduction}
Data were reduced with the FIREHOSE pipeline, which optimally extracts 1D spectra and associated errors in each order from flat-fielded 2D spectral images. The pipeline also computes the wavelength calibration using OH sky lines and ThAr arc lamp spectra obtained after each science exposure. Slit tilts in each order are measured and accounted for in the final wavelength solution, which is in vacuum and includes the heliocentric correction of  $6.8~\rm km~s^{-1}$. Telluric lines are corrected for in the final spectra, and each order is flux calibrated using the Spextool software package \citep{Cushing:2004bq}. Finally, the 1D spectra are optimally combined, and each order is merged in a single spectrum. The resulting signal-to-noise (order dependent) ranges between 18 and 36 per spectral pixel.

\begin{figure}
    \centering
    \includegraphics[width=8.4cm, natwidth=20.32cm, natheight=16.27cm]{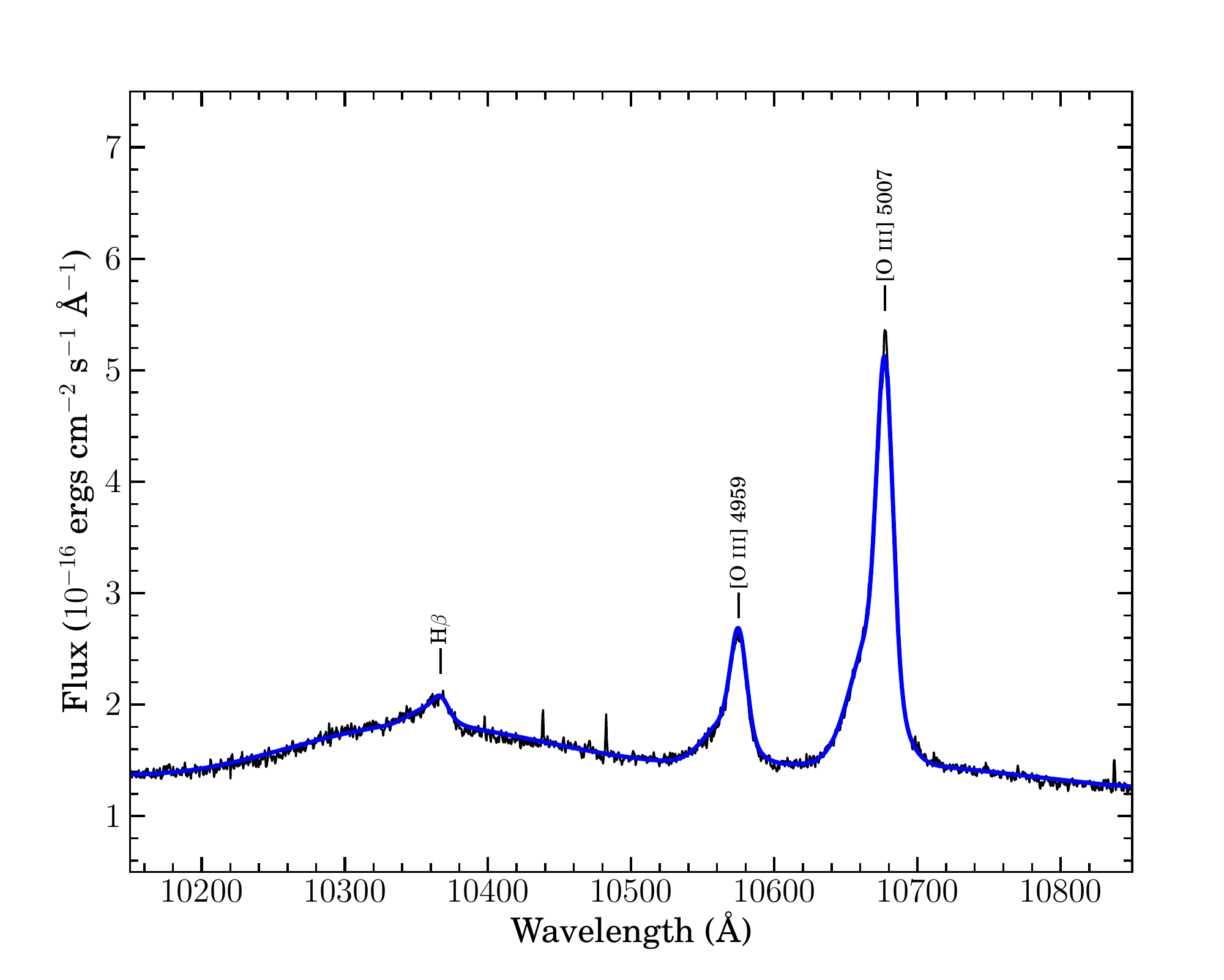}
    \caption{A region of the Magellan/FIRE spectrum of Q0209, centred on the [O$\;$\textsc{iii}] $\lambda\lambda$5008,4960 and \hbeta\ emission lines. The blue curve shows the total fitted emission profile.}
\label{q0209_OIII}
\end{figure}

\subsection{Redshift measurement and black hole mass}
\label{redshift}
We measure the redshift of Q0209 using the [\ion{O}{3}] $\lambda\lambda$5008,4960 doublet emission lines, seen in the FIRE spectrum. The region around the [\ion{O}{3}] and \hbeta\ emission is shown in \Cref{q0209_OIII}. An asymmetric profile in the [\ion{O}{3}] lines is apparent, with an extended blue wing, indicative of outflowing gas. We decompose the spectrum in this region into the contribution from a power-law continuum, blended \ion{Fe}{2} emission lines (Gaussian smoothing over the template of \citet{VeronCetty:2004ix}) and a multiple Gaussian fit to the [\ion{O}{3}] and \hbeta\ emission, using a similar method to that described in \citet{Jin:2012eu}. The blue curve shows the resulting best-fit emission model. A two component Gaussian fit to each of the [\ion{O}{3}] lines effectively removes the outflowing component. From the line centre in the stronger Gaussian component, we then derive a systemic redshift measurement of $z_{\textrm{QSO}} = 1.13194 \pm 0.00001$, corresponding to a statistical velocity uncertainty of $\Delta v \approx 3$~\kms. The black hole mass is estimated from the \ac{fwhm} of the broad \hbeta\ component together with the rest-frame 5100~\AA\ flux, using equation (3) in \cite{Woo:2002kw}. We find a value $M_{\textrm{BH}} \approx 1.9 \times 10^9 M_{\odot}$. This value compares favourably with an earlier estimation of $M_{\textrm{BH}} \approx 1.4 \times 10^9 M_{\odot}$ measured from a Keck/HIRES spectrum using the broad \ion{Mg}{2} emission line, following the method described in \citet{Matsuoka:2013ey}. The latter black hole mass estimate is used as a constraint on the accretion disc models presented in \Cref{sed_section}, which define the \ac{sed} shape of Q0209 used in subsequent photoionization modelling. More details on the black hole mass estimation can be found in Done et al. (in prep).

\begin{figure*}
\begin{minipage}{18cm}
	\centering
	\includegraphics[width=18cm, natwidth=30.48cm, natheight=18.31cm]{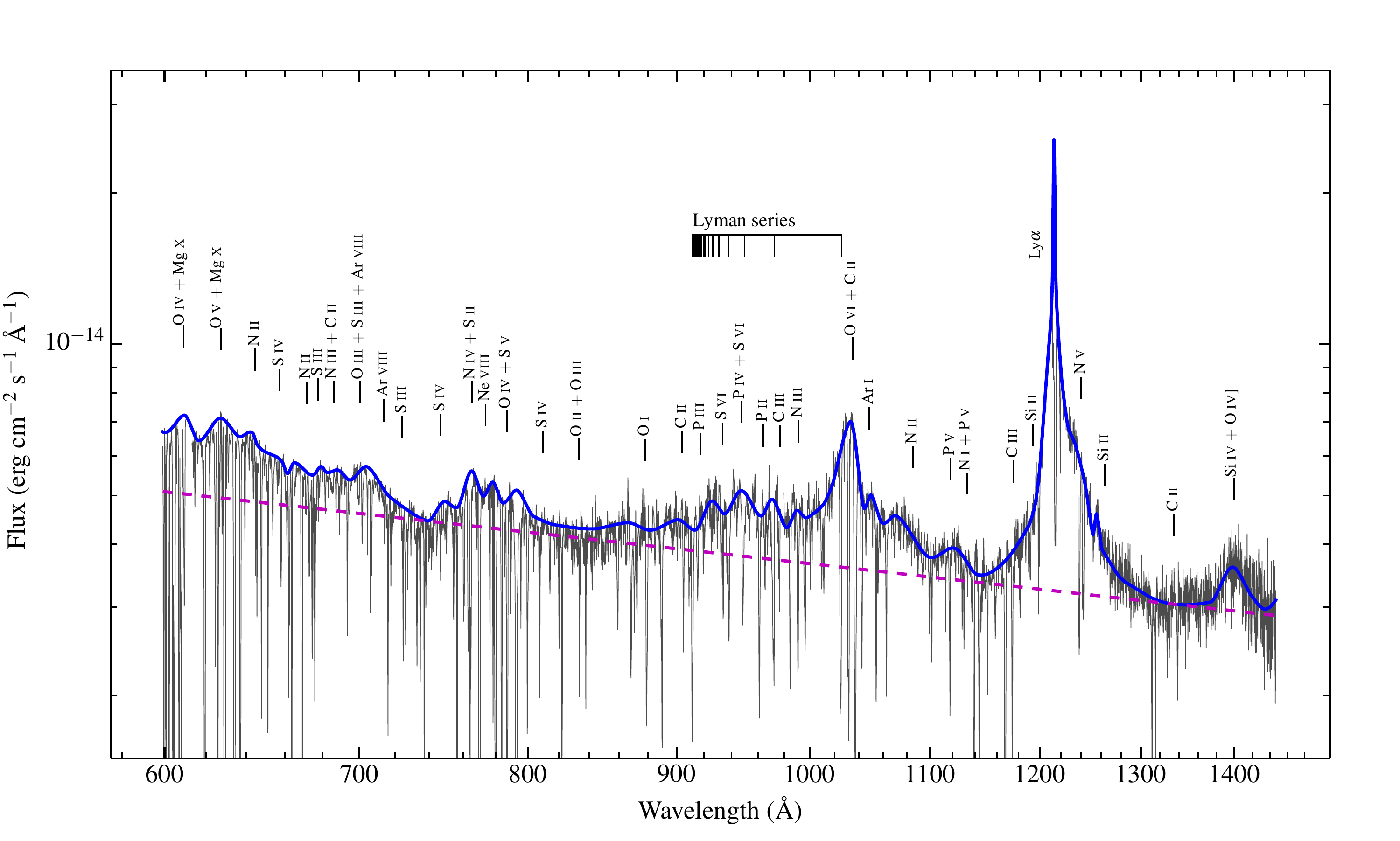}
	\caption{Rest-frame \ac{hst}/\ac{cos} spectrum of Q0209, corrected for Galactic extinction and plotted in 0.5~\AA\ bins. The blue curve shows a spline fit to the total emission profile, as described in \Cref{continuum_fitting}. The dashed magenta curve is a power law fit to the continuum profile. A large number of broad emission lines are present. The expected positions of prominent broad emission lines are labelled.}
\label{q0209}
\end{minipage}
\end{figure*}

\subsection{Continuum fitting}
\label{continuum_fitting}
In order to perform a proper absorption line analysis, we require an estimate of the unabsorbed \ac{qso} continuum (including emission lines). The spectrum is then normalised using this continuum. To fit the continuum we adopted a method based on that described in \citet{Young:1979if}, \citet{Carswell:1982uz} and \citet{Aguirre:2002is}. The technique is as follows. First, the spectrum is split into an arbitrary number of wavelength intervals, with the flux median and standard deviation calculated in each. These intervals are chosen based on trial and error, and are on average around 12 \AA\ wide shortward of the \lya\ emission line and generally smaller across emission lines, where flux gradients vary most markedly. Larger intervals are chosen in regions free of emission lines longward of \lya. A first-order spline is then fitted through a set of points defined by the central wavelength and median flux value in each interval. Pixels falling more than three standard deviations below the spline are rejected, then the median and standard deviation are recalculated using the remaining pixels. This process is iterated over until the remaining pixel fluxes above the spline have an approximately Gaussian distribution, with standard deviation equal to the expected $1 \sigma$ flux errors. A cubic spline is then fitted to the entire spectrum to a give a smoothly varying result. Finally, the continuum has to be manually adjusted by hand in regions where the fit still appears poor, typically over strong absorption features and emission lines. The \lya\ emission line strength is highly uncertain, due to strong absorption disguising the peak in the line. However, the ratio of \lyb\ to \lya\ line strengths in the fitted spline compares favourably with the same ratio seen in the \citet{Shull:2012ki} \ac{hst}/\ac{cos} composite spectrum of active Galactic nuclei, and the results of this paper are not sensitive to the exact placement of this peak.

The spline continuum is shown as a blue line on top of the rest frame spectral data in \Cref{q0209}, corrected for Galactic extinction using the empirical mean extinction curve of \citet{Cardelli:1989dp}. We calculate the extinction as a function of wavelength using a Galactic \ion{H}{1} column density of $2 \times 10^{20}~\textrm{cm}^{-2}$, which sits between the measured values of $1.85 \times 10^{20}~\textrm{cm}^{-2}$ \citep{Kalberla:2005de} and $2.44 \times 10^{20}~\textrm{cm}^{-2}$ \citep{Dickey:1990df} in this direction. We assume an $E(B - V)$ to $N_\textrm{H}$ ratio of 1.5 \citep{Gorenstein:1975jp}, which gives $E(B - V) = 0.028$. The spectrum is decomposed into the contribution from emission lines, plus that from the accretion disc continuum. We do this by choosing regions of emission-line free continuum, taking the minimum value inferred from the fitted spline in each of these regions, and fitting a power law through the resulting data points, giving a spectral index of $\alpha_{\lambda} = -0.64    $. The flux shortward of $\sim 600$~\AA\ (1280~\AA\ observed frame) falls to zero due to a Lyman-limit system at $z \simeq 0.39$. A large number of broad emission lines are present, including most of the lines seen in the composite spectrum of \citet{Shull:2012ki} over the same wavelength range. The expected locations of many prominent broad emission lines are labelled.

\begin{figure}
    \centering
    \includegraphics[width=8.4cm, natwidth=20.32cm, natheight=20.32cm]{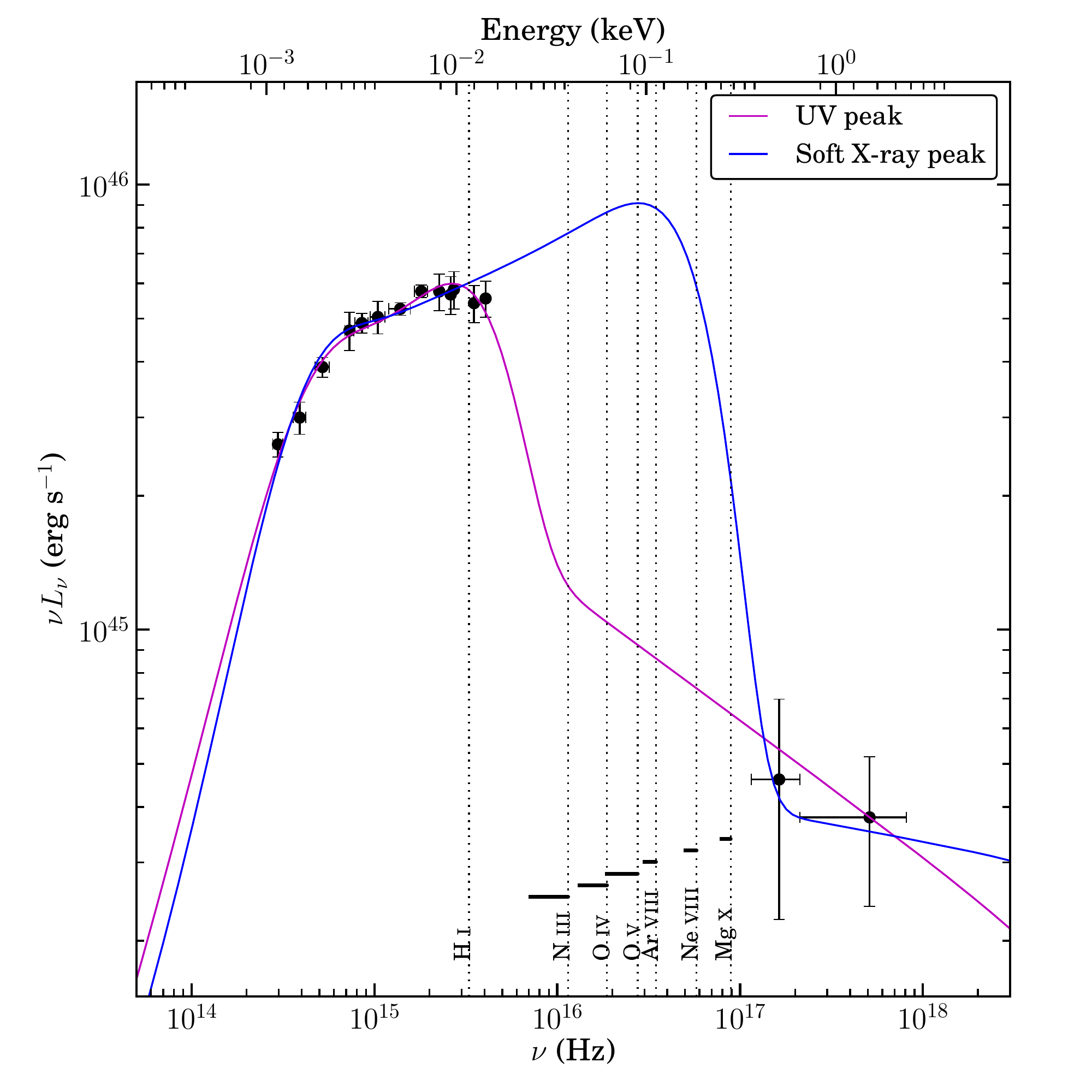}
    \caption{Rest-frame model \ac{sed}s from \optxagnf, fitted to extinction-corrected photometric data and line-free regions of the \ac{hst}/\ac{cos} spectrum using \xspec. The magenta curve is the best fitting model, which peaks in the \ac{uv}. The blue curve is also a good fit to the data, but peaks in the soft X-ray bandpass. Together, these models represent the uncertainty in the \ac{sed} shape over the soft X-ray region, where there is a large correction for Galactic extinction. Dotted lines represent the ionization destruction potentials for a range of ions that are used as constraints in photoionization modelling. Solid black lines represent the range of energies over which these ions are formed. Crucially, these ions are produced by photons with energies in a region where the \ac{sed}s are most markedly different, both in terms of luminosity and spectral slope.}
\label{sed}
\end{figure}

\subsection{Spectral energy distribution}
\label{sed_section}
For the purposes of photoionization modelling, we construct an \ac{sed} that extends from the far infra-red ($\sim 10^{-5}~\textrm{keV}$), through to the hard X-ray bandpass ($\sim 100~\textrm{keV}$), which represents the range in photon energy over which most of the emission is generated by gas accretion. This emission forms the dominant contribution to the ionising photon flux. Data points in the \ac{uv} are taken from line free regions of the \ac{cos} spectrum as described in \Cref{continuum_fitting}. Optical data in the $u$, $g$, $r$, $i$ and $z$ photometric bands come from the SDSS \citep{Ahn:2013wx}, CFHTLS \citep{Cuillandre:2012jv} and PanSTARRS\footnote{\url{http://www.ps1sc.org}} surveys. Near-infrared data in the $J$, $H$, and $K_{\textrm{s}}$ bands comes from 2MASS \citep{Skrutskie:2006hl}. X-ray points are simulated from the ROSAT all-sky survey \citep{Voges:1999ws} flux and spectral index using the ROSAT detector response matrix\footnote{available from \url{http://heasarc.gsfc.nasa.gov/docs/rosat/pspc_matrices.html}}. The data points are corrected for Galactic dust extinction using the method described in \Cref{continuum_fitting}, and for the Galactic absorption cross-section due to gas, grains and molecules using the model presented in \citet{Wilms:2000en}. We then fit the data with \optxagnf\ -- an energetically self-consistent accretion disc model described fully in \citet{Done:2012eq} -- using the spectral fitting package \xspec \footnote{\url{http://heasarc.nasa.gov/xanadu/xspec/}}. Briefly, the model consists of three main components: (i) a colour-temperature corrected blackbody spectrum powered by the outer regions of the black hole accretion disc, (ii) a soft X-ray excess, attributable to Compton up-scattering of seed photons in the hotter, optically thick inner region of the accretion disc, and (iii) an additional X-ray component formed through Compton up-scattering in a hot, optically thin corona above the disc, creating a power law tail that extends through the hard X-ray bandpass. The model assumes that all the energy used to power these three components is produced through mass accretion. Therefore, the soft and hard X-ray components are physically constrained even though their origin is poorly understood. In modelling the spectrum, we assume a black hole mass of $1.4 \times 10^9 M_{\odot}$ (see \Cref{redshift}). The resulting rest-frame \ac{sed} is shown in \Cref{sed}. Two models in blue and magenta are shown that fit the data well: one peaking in the soft X-rays, the other in the UV, with normalised $\chi^2$ values of 1.218 and 0.526 respectively. The models differ quite dramatically over the soft X-ray bandpass, representing our ignorance of the true SED shape in this region due to modelling uncertainties on the Galactic extinction across the extreme \ac{uv} bandpass, and a lack of high quality X-ray observations. Crucially, it is over this energy range where the ions considered in this paper are created (and destroyed). We therefore consider both models in later analysis. The models represent the extremes allowed by the data, and so it is useful to bear in mind that the true \ac{sed} may be lie somewhere between these two possibilities. Dotted vertical lines represent the ionization destruction potentials for a range of ions later considered in photoionization modelling. Solid black horizontal lines represent the range in energy where these ions are present (extending down to their ionization creation potentials).

\begin{table}
\caption{\ac{aals} detected at a $> 3 \sigma$ significance level, listed first in order of decreasing solar abundance relative to hydrogen, second in order of increasing ionization potential, and third in order of decreasing oscillator strength.$^1$}
\begin{threeparttable}
\begin{tabular}{l c c c c}
	\hline
	ion & $\lambda_{\textrm{rest}}$ (\AA)\tnote{a} & IPc (eV)\tnote{b} & IPd (eV)\tnote{c} & $W_{\lambda}$ (\AA)\tnote{d} \\
	\hline
	H$\;$\textsc{i} & 1215.7 & ... & 13.6 & $2.365 \pm 0.031$\tnote{*} \\
	& 1025.7 &  &  & $1.030 \pm 0.032$ \\
	& 972.5 &  &  & $0.967 \pm 0.074$\tnote{*} \\
	& 949.7 &  &  & $0.691 \pm 0.045$\tnote{*} \\
	& 937.8 &  &  & $0.757 \pm 0.059$\tnote{*} \\
	O$\;$\textsc{iv} & 787.7 &  &  & $0.937 \pm 0.015$\tnote{*} \\
	& 608.4 & 54.9 & 77.4 & $0.633 \pm 0.008$\tnote{*} \\
	O$\;$\textsc{iv}* & 609.8 & ... & ... & $0.362 \pm 0.009$\tnote{*} \\
	& 790.2 &  &  & $0.288 \pm 0.020$\tnote{**} \\
	& 790.1 &  &  & $0.288 \pm 0.020$\tnote{**} \\
	O$\;$\textsc{v} & 629.7 & 77.4 & 113.9 & $0.914 \pm 0.008$\tnote{*} \\
	O$\;$\textsc{vi} & 1031.9 & 113.9 & 138.1 & $1.185 \pm 0.040$ \\
	& 1037.6 &  &  & $1.690 \pm 0.039$\tnote{*} \\
	Ne$\;$\textsc{viii} & 770.4 & 154.2 & 207.3 & $1.072 \pm 0.011$\tnote{*} \\
	& 780.3 &  &  & $0.981 \pm 0.011$\tnote{*} \\
	N$\;$\textsc{iii} & 685.0 & 29.6 & 47.4 & $0.126 \pm 0.011$\tnote{**} \\
	& 685.5 &  &  & $0.092 \pm 0.011$\tnote{**} \\
	N$\;$\textsc{iv} & 765.1 & 47.4 & 77.5 & $0.685 \pm 0.014$ \\
	N$\;$\textsc{v} & 1238.8 & 77.5 & 97.9 & $1.076 \pm 0.051$ \\
	& 1242.8 &  &  & $0.768 \pm 0.054$ \\
	Mg$\;$\textsc{x} & 609.8 & 328.0 & 367.5 & $0.433 \pm 0.008$\tnote{*} \\
	& 625.0 &  &  & $0.362 \pm 0.009$\tnote{*} \\
	S$\;$\textsc{v} & 786.4 & 47.2 & 72.6 & $0.185 \pm 0.015$ \\
	Ar$\;$\textsc{viii} & 700.2 & 91.0 & 124.3 & $0.165 \pm 0.009$ \\
	& 713.8 &  &  & $0.014 \pm 0.011$ \\
	\hline
\end{tabular}
\begin{tablenotes}
    \item[1] Atomic data from the NIST Atomic Spectra Database (\url{http://www.nist.gov/pml/data/asd.cfm}).
    \item[a] Rest-frame transition wavelength.
    \item[b] Ionization potential for creation.
    \item[c] Ionization potential for destruction.
    \item[d] Observed equivalent width across the entire absorption trough.
    \item[*] Measured value includes a contribution from blended, unrelated absorption.
    \item[**] Measured value includes a contribution from absorption lines due to a closely separated transition of the same ion.
\end{tablenotes}
\end{threeparttable}
\label{aal_detections}
\end{table}

\begin{figure*}
\begin{minipage}{18cm}
	\centering
	\includegraphics[width=18cm, natwidth=30.48cm, natheight=36.59cm]{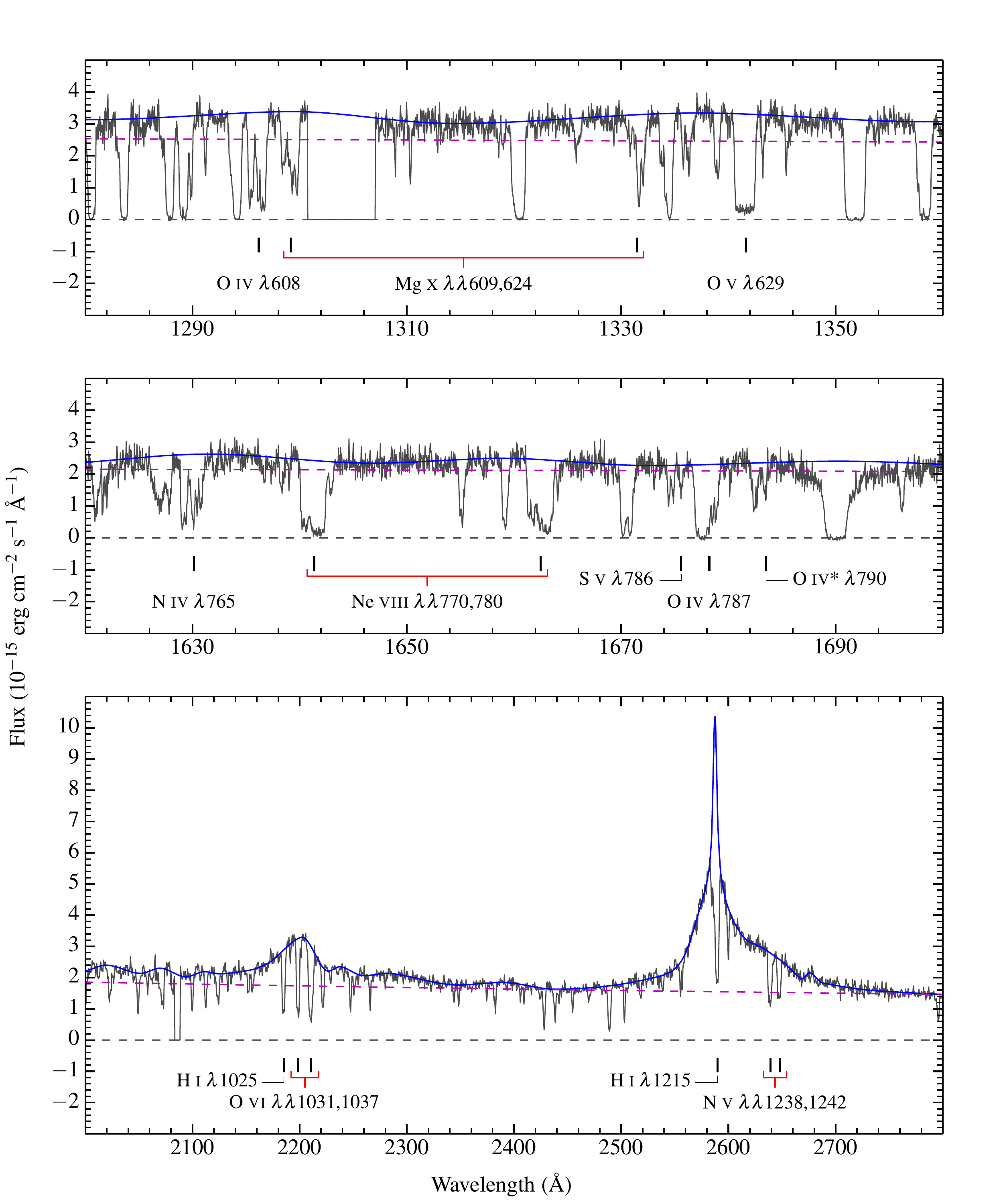}
	\caption{The \ac{hst}/\ac{cos} spectrum of Q0209 in the observed frame, shown in regions containing the most prominent \ac{aals}. The blue curve shows the unabsorbed continuum fit including emission lines. The black dashed line shows the zero-flux level. The magenta dashed line shows the power-law accretion disc continuum. Markers indicate the positions of the \ac{aals}. Labels indicate the ion (singlet or doublet) and rest-frame transition wavelengths giving rise to those absorption troughs.}
\label{q0209_AALs}
\end{minipage}
\end{figure*}

\section{Analysis of the associated absorption}
\label{analysis}
Sections of the \ac{cos} spectrum of Q0209, together with the spline continuum (blue line) and power-law accretion disc spectrum (dashed magenta line) are shown in \Cref{q0209_AALs}. The spectral resolutions up to, and above, 1750~\AA\ are $\sim 16$~\kms\ and $\sim 100$~\kms\ per resolution element respectively. The plot labels just the most prominent associated absorption troughs, but we report all of the \ac{aals} detected with $> 3 \sigma$ significance in \Cref{aal_detections}. Equivalent widths are measured by integrating over the whole absorption trough in each ion (including all discrete velocity components). Some of these troughs are blended with unrelated absorption lines at lower redshifts, making the measured equivalent widths larger than the intrinsic ones. We subtract away the effects of line blending by fitting Voigt profiles with \vpfit\footnote{\url{http://www.ast.cam.ac.uk/~rfc/vpfit.html}} (\Cref{aals}, \Cref{column_densities}). The velocity structure across all ions is tied to that of \ion{N}{4} $\lambda$765 (\Cref{vstructure}) in the fitting process, based on an empirical (by-eye) similarity between the absorption troughs. This similarity suggests that all the ions we detect in a particular velocity component are formed in regions that are co-spatial. We choose \ion{N}{4} $\lambda$765 as a reference, due to a lack of line blending and the fact that almost all components are cleanly resolved in this absorption trough. We have full coverage of the \ion{H}{1} Lyman series transitions. Ions searched for, but not detected above a $3 \sigma$ significance level in the AAL system are \ion{C}{3}, \ion{O}{3}, \ion{O}{5}*, \ion{S}{3}, \ion{S}{4}, \ion{S}{6} and \ion{Na}{9}. The expected locations of these particular lines also coincide with unrelated absorption at $z \ll z_{\textrm{QSO}}$ in some cases.

\subsection{Partial covering}
\label{partial_covering_section}
In this section we look for the presence of partial covering, i.e. indications that the absorbing clouds do not fully cover the background emission from the \ac{qso} continuum and/or \ac{blr}. The clearest evidence for partial covering comes from flat-bottomed, apparently saturated absorption troughs that do not reach zero intensity (see for example the \ion{O}{5} absorption trough in \Cref{q0209_AALs}). If the individual line components are resolved, then these profiles must be caused by optically thick absorption \emph{plus} some unabsorbed flux. Under the assumption that the absorbers are spatially homogeneous, the residual flux at an observed wavelength $\lambda$ in the normalised \ac{qso} spectrum may be written as
\begin{equation}
    R_{\lambda} = (1 - C_f) + C_f e^{-\tau_{\lambda}},
\label{residual_flux}
\end{equation}
where $\tau_{\lambda}$ is the optical depth and $C_f$ the covering fraction, defined as the ratio of occulted to total emitted photons from the background light source(s). Solving for $\tau_{\lambda}$ we have
\begin{equation}
    \tau_{\lambda} = - \ln \left(\frac{R_{\lambda} - 1 + C_f}{C_f}\right).
\label{tau}
\end{equation}
For ions with just one transition, estimating the covering fraction is only possible when the line is saturated, in which case the exponential goes to zero and $C_f = 1 - R_{\lambda}$. Otherwise, we cannot estimate $C_f$ as we do not know $\tau_{\lambda}$. However, for multiplets, where more than one transition is available, we can eliminate $\tau_{\lambda}$ by noting that
\begin{equation}
    \gamma = \frac{\tau_{\lambda 1}}{\tau_{\lambda 2}} = \frac{f_1 \lambda_1}{f_2 \lambda_2},
\label{gamma}
\end{equation}
where $f_1$ and $f_2$ are the oscillator strengths of each transition. This ratio is close to 2 in the case of doublet lines. For two transitions of the same ion, with residual flux values $R_{\lambda 1}$ and $R_{\lambda 2}$, and covering fractions $C_{f1}$ and $C_{f2}$ respectively, we may then write
\begin{equation}
    R_{\lambda 1} = 1 - C_{f1} + C_{f1} \left(\frac{R_{\lambda 2} - 1 + C_{f2}}{C_{f2}}\right)^{\gamma}
\label{partial_covering_equation}
\end{equation}
\citep{Petitjean:1999vh}. For simplicity we can assume that $C_{f1} = C_{f2}$ for each ion, although in general this may not be true \citep{Srianand:1999hj}. Complex velocity structure in the broad emission lines (e.g. the presence of both narrow and broad velocity components) can mean that absorbed photons from different parts of a broad line profile will originate from spatially distinct locations (narrow-line region versus broad-line region). Unless the background emission intensity happens to be spatially homogeneous for any given wavelength, this may imply that $C_{f1} \neq C_{f2}$, even for doublets that have a relatively small wavelength separation (see for example the \ion{O}{6} doublet in \Cref{q0209_AALs}, which spans the centre and the blue wing of the \lyb\ $+$ \ion{O}{6} emission line). In addition, over the whole wavelength range of the \ac{qso} spectrum, there are regions dominated more by the accretion disc continuum than by the \ac{blr}, and vice versa. Since the \ac{blr} is larger in size than the continuum, if the absorbing clouds have transverse sizes larger than the continuum region, this can also lead to an inequality. We expect deviations from $C_{f1} = C_{f2}$ to be small in most cases, although the relatively high \lya\ emission line flux can lead to situations where the \lya\ absorption profile has a smaller apparent optical depth than the \lyb\ profile \cite[e.g.][]{Petitjean:1999vh}. Nevertheless, we do not see strong evidence for this effect in our data, and therefore we favour a scenario where there are many clouds with transverse sizes smaller than the continuum size \cite[$\lesssim 4$ light-days;][]{JimenezVicente:2012es}. In what follows we assume $C_{f1} = C_{f2}$ within the measurement uncertainties.

\begin{figure}
    \centering
    \includegraphics[width=8.4cm, natwidth=20.32cm, natheight=10.16cm]{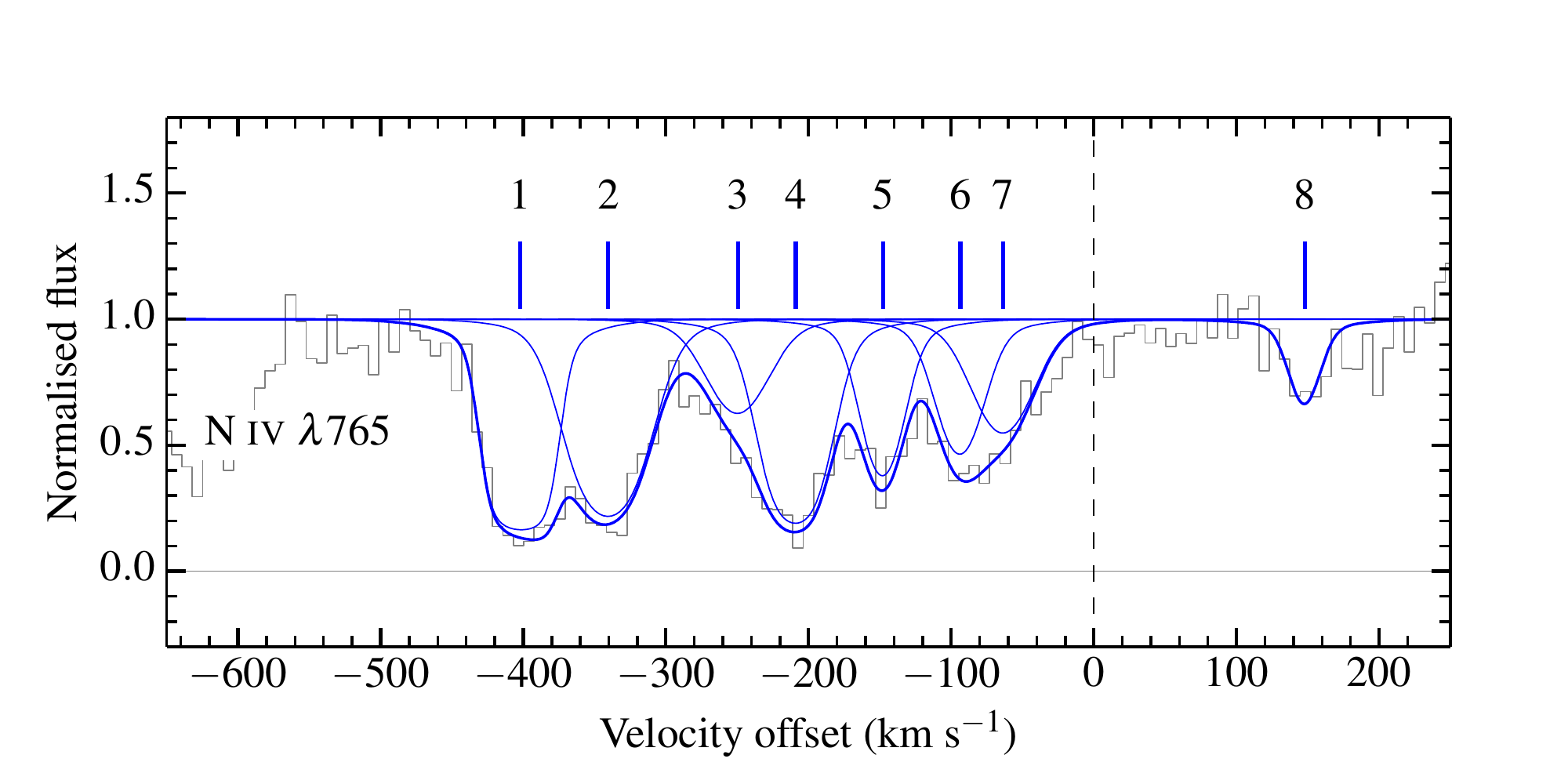}
    \caption{Velocity structure in the N$\;$\textsc{iv} absorption trough. Voigt components are at $-402$, $-340$, $-249$, $-209$, $-148$, $-94$, $-63$ and $+148$~\kms\ with respect to the \ac{qso} rest frame. The vertical dashed line marks the rest-frame velocity of the \ac{qso}. Thin blue lines are individual Voigt profile fits to the data. The thick blue line represents the overall fitted profile.}
\label{vstructure}
\end{figure}

\begin{figure}
    \centering
    \includegraphics[width=8.4cm, natwidth=20.32cm, natheight=30.48cm]{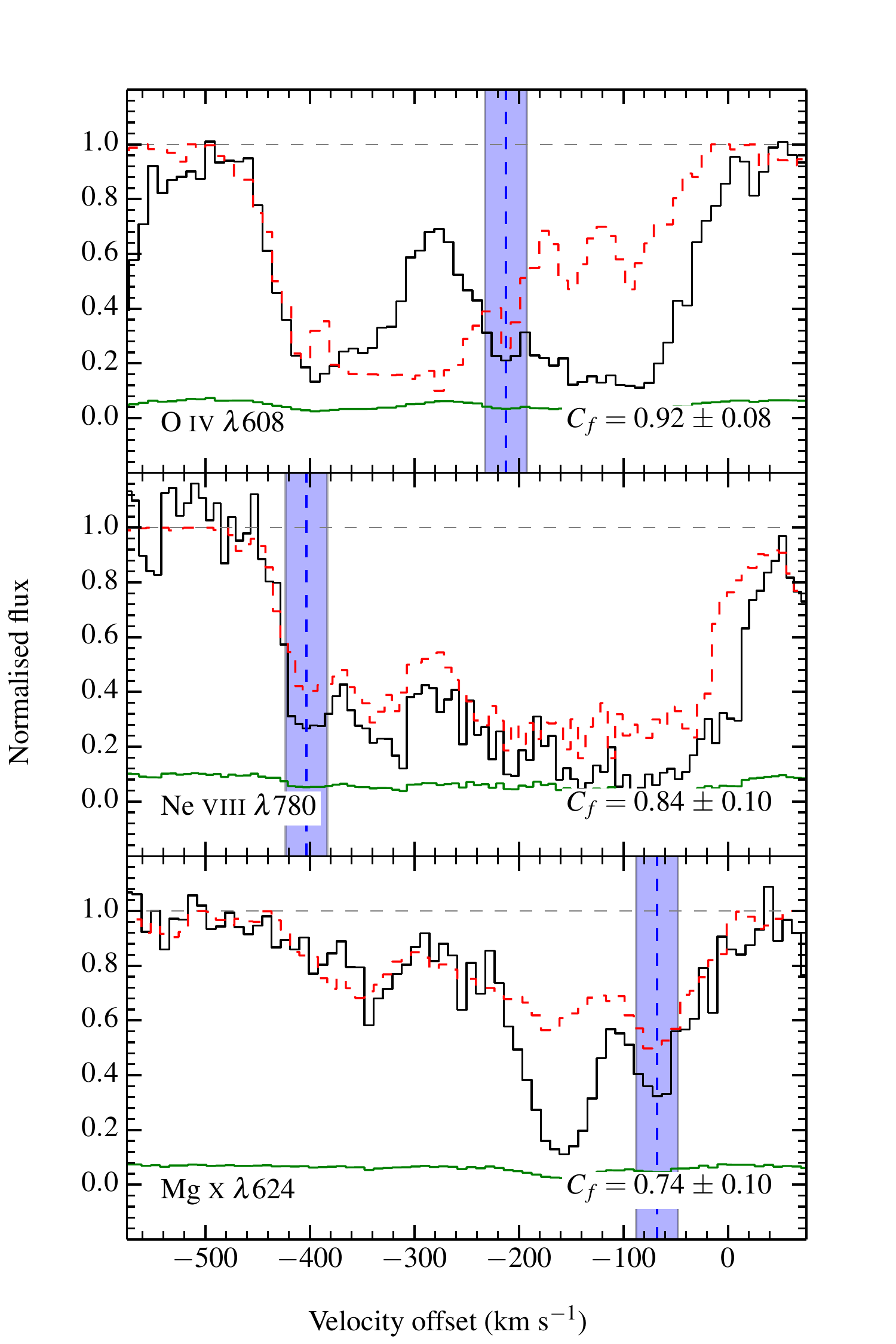}
    \caption{Observed (solid black line) compared to predicted (dashed red line) profiles of O$\;$\textsc{iv} $\lambda$608, Ne$\;$\textsc{viii} $\lambda$780 and Mg$\;$\textsc{x} $\lambda$624 based on apparent optical depth measurements of their stronger counterparts, O$\;$\textsc{iv} $\lambda$787, Ne$\;$\textsc{x} $\lambda$770 and Mg$\;$\textsc{x} $\lambda$609 respectively. The deeper observed compared to predicted profiles indicate partial covering of the background emission. This situation is occasionally inverted due to line blending. The velocity offset is with respect to the \ac{qso} rest frame. The green line is the $1\sigma$ error on the normalised flux in each pixel. Dashed blue lines represent the velocity centroids of the lines not affected (or minimally affected) by blends from unrelated absorption. Vertical shaded regions represent the range of velocities over which covering fractions are calculated (per pixel). Quoted $C_f$ values are the average of those calculated within these regions.}
\label{partial_covering}
\end{figure}

Residual flux is clearly present in the saturated absorption troughs of \ion{O}{5} $\lambda$629 and \ion{Ne}{8} $\lambda$770 (at the blue end), which implies that $C_f < 1$ for these ions (see \Cref{q0209_AALs}). Taking the average normalised residual flux across the flat portions of these profiles gives covering fractions of $0.91 \pm 0.01$ and $0.93 \pm 0.01$ respectively (with $\tau \gg 1$ in \cref{residual_flux}). In these, and all following covering fraction estimates, the quoted statistical error does not include any contribution from the error on the continuum fit. We also caution that the error bars assume Gaussian statistics, which underestimate the true flux error in absorption troughs where the number of counts in a given bin is low \cite[$\lesssim 100$;][]{Gehrels:1986cx}. Given the flat profile across the entirity of the \ion{O}{5} absorption trough, there is no strong evidence for covering fractions that change across the profiles. We can check this further by examining the apparent doublet ratio for \ion{Ne}{8} in components that are unsaturated, and unaffected by blending with unrelated absorption lines. There is one component where this measurement is possible in the case of \ion{Ne}{8}. Covering fractions can also be determined in this way for \ion{Mg}{10} and \ion{O}{4}. The former has one velocity component available that meets the aforementioned criteria, and the latter has one that is still mildly affected by line blending. We first calculate the apparent optical depth as a function of velocity across the stronger transition for each ion, then scale these optical depths by $\gamma$ (\cref{gamma}) to predict the optical depths in the weaker transitions. The difference between the observed and predicted profiles for the weaker member of each ion can be seen in \Cref{partial_covering}. The dashed blue lines show the velocity centroids of the components used to measure the covering fractions. We determine covering fractions by numerically solving \cref{partial_covering_equation} over a $\sim 40$~\kms\ region centred on each velocity component (blue shaded regions in \Cref{partial_covering}), and average the results. This procedure gives covering fractions of $0.92 \pm 0.08$ for \ion{O}{4}, $0.84 \pm 0.10$ for \ion{Ne}{8}, and $0.74 \pm 0.10$ for \ion{Mg}{10}. For \ion{O}{4}, we note that the chosen velocity component is mildly affected by blending, which may add an additional uncertainty on top of the measured one. We also note that the \ion{Ne}{8} covering fraction determined from this method is consistent with that measured from the saturated blue wing of \ion{Ne}{8} $\lambda$770 within the $1\sigma$ uncertainty, and adopt the latter result as the covering fraction for this ion.

\begin{table}
\setlength{\tabcolsep}{14pt}
\caption{Adopted covering fractions.}
\label{covering_fractions}
\begin{threeparttable}
\begin{tabular}{l l c c}
    \hline
    ion & measurement\tnote{a} & $C_f$\tnote{b} & $\Delta C_f$\tnote{c} \\
    \hline
    H$\;$\textsc{i} & inferred & 0.92 & ... \\
    N$\;$\textsc{iii} & inferred & 0.92 & ... \\
    N$\;$\textsc{iv} & inferred & 0.92 & ... \\
    N$\;$\textsc{v} & inferred & 0.91 & ... \\
    O$\;$\textsc{iv} & direct & 0.92 & 0.08 \\
    O$\;$\textsc{iv}* & inferred & 0.92 & ... \\
    O$\;$\textsc{v} & direct & 0.91 & 0.01 \\
    O$\;$\textsc{vi} & inferred & 0.91 & ... \\
    Ne$\;$\textsc{viii} & direct & 0.93 & 0.01 \\
    Mg$\;$\textsc{x} & direct & 0.74 & 0.10 \\
    S$\;$\textsc{v} & inferred & 0.91 & ... \\
    Ar$\;$\textsc{viii} & inferred & 0.91 & ... \\
    \hline
\end{tabular}
\begin{tablenotes}
    \item[a] Measurements are either direct (based on saturated lines or from comparing line ratios) or inferred (assumed to be the same as that measured directly from an ion with similar ionization potential).
    \item[b] Assuming the covering fraction of the continuum is the same as the covering fraction of the \ac{blr}. Requires many clouds smaller than the size of the continuum region.
    \item[c] $1 \sigma$ statistical uncertainty on the covering fraction (for directly measured values only, not including errors on the continuum fit).
\end{tablenotes}
\end{threeparttable}
\end{table}

For \ion{O}{4}* $\lambda\lambda$609,790 we first note that one of these transitions overlaps with that of \ion{Mg}{10} $\lambda$609. Therefore, to determine the covering fraction, we first perform a fit to the \ion{Mg}{10} $\lambda$624 absorption trough, taking into account the covering fraction already determined for this ion, and fixing the velocity structure to that from the fit to \ion{N}{4} $\lambda$765 (\Cref{vstructure}). We then re-normalise the spectrum using the calculated \ion{Mg}{10} profile, leaving just the absorption signature from \ion{O}{4}*. It is subsequently apparent that the covering fraction for \ion{O}{4}* is consistent with that of \ion{O}{4}. We do not attempt to explicitly calculate the covering fraction in this case, due to the additional uncertainty imposed by subtracting the \ion{Mg}{10} absorption. Examining multiplet ratios in \ion{N}{3} and \ion{Ar}{8} reveals that these ions are consistent with a covering fraction of unity, however their detection significance is considerably smaller than for the rest of the ions detected here. Therefore $C_f$ may still be less than 1, as found for ions with better measurements.

For the remaining \ac{aals}, covering fractions are even more difficult to determine. In the case of \ion{N}{4} and \ion{S}{5}, it is because they are singlet ions (only one transition). In all other cases it is because the lines fall in the \ac{nuv} portion of the spectrum, where the lower resolution complicates the process of determining the true residual flux in each line (since individual components are not resolved). The relative contribution of partial covering fractions and resolution effects to the line ratios are very difficult (or impossible) to disentangle. For these lines, the simplest approach is to take another ion with measured covering fraction that is assumed to trace the same gas, and adopt this covering fraction. We deem the best (available) choice of ion, for which this assumption might hold, to be that with the most similar ionization potential. We adopt this approach also in the case of \ion{N}{3} and \ion{Ar}{8}. Matching ions under this criterion gives two groups: (\ion{O}{4}, \ion{H}{1}, \ion{N}{3}, \ion{N}{4}, \ion{S}{5}) and (\ion{O}{5}, \ion{N}{5}, \ion{O}{6}, \ion{Ar}{8}), where only the first member of each group has a measured covering fraction. Covering fractions for all ions are summarised in \Cref{covering_fractions}. Despite the uncertainties in assigning covering fractions for the absorption lines in our sample, we note that all measured values are high ($C_f \sim 0.9$) with the possible exception of \ion{Mg}{10}, and produce statistically good results in Voigt profile fitting (\Cref{column_densities}). In general, we find that the results of this paper are not sensitive to the precise values of the covering fractions. Indeed, if we take the best measured covering fraction from \ion{O}{5} and apply this to all ions, the results we obtain are largely consistent with those obtained later, within the measurement uncertainties.

\subsection{Column densities and line widths}
\label{column_densities}
To measure the column densities for each ion, we first perform a simultaneous Voigt profile fit to all of the \ac{aals} in the medium resolution, \ac{fuv} part of the spectrum, using \vpfit. We do so with the assumption that all ions must share the same 8 separate velocity components, this number having been determined from an independent fit to the \ion{N}{4} $\lambda$756 absorption trough (see \Cref{partial_covering_section}). Voigt profiles are convolved with the wavelength dependent, non-Gaussian \ac{cos} \ac{lsf}. We calculate the \ac{lsf} at the observed wavelength of each transition by interpolating between the tabulated \ac{lsf}s, which are specified at a discrete set of wavelengths \citep{Ghavamian:2009tr,Kriss:2011um}.

For the remaining ions in the low resolution, \ac{nuv} part of the spectrum, individual components are not resolved, and so these data give very poor constraints on the Doppler broadening ($b$) parameters, which adds to the uncertainty on the column densities. Crucially, coverage of the \ion{H}{1} Lyman series absorption, for which we require well constrained column densities in forthcoming photoionization analysis, is limited to the \ac{nuv} spectrum. One way of reducing this uncertainty is to find pairs of ions that likely trace the same gas, then require that the $b$ values for these pairs follow some scaling relation, thus reducing the number of degrees of freedom in the $\chi^2$ minimisation. Tying $b$ values sensibly requires knowledge about both the thermal and turbulent motions in the gas. We define $b$ as
\begin{equation}
    b^2 = b_{\textrm{turb}}^2 + \frac{2kT}{m},
\end{equation}
where $b_{\textrm{turb}}$ is the turbulent contribution to the line width, $k$ is Boltzmann's constant, $T$ is the gas temperature, and $m$ is the atomic mass of the ion in question. The relationship between $b$ values in two ions, labelled $b_1$ and $b_2$, can then be expressed as
\begin{equation}
    b_1^2 = \left(\frac{m_2}{m_1}\right) b_2^2 + b_{\textrm{turb}}^2 \left(1 - \frac{m_2}{m_1}\right).
\label{b_scaling}
\end{equation}
If $b_{\textrm{turb}}$ is zero, then $b$ values can be related by a simple mass scaling. We note that if the gas is photoionized, with a nominal temperature $T \sim 10^4~\textrm{K}$, then initial fits to the \ac{fuv} data indicate that
\begin{equation}
    b_{\textrm{turb}} = \sqrt{b^2 - \frac{2kT}{m}} > 0
\end{equation}
(see \Cref{voigt_parameters}). In fact, for $T < 10^5\textrm{K}$, turbulence dominates the broadening for these lines given the measured $b$ values. Therefore, if the gas is photoionized, we cannot relate $b$ values by the masses of the ions alone.

\begin{figure}
    \centering
    \includegraphics[width=8.4cm, natwidth=20.32cm, natheight=20.32cm]{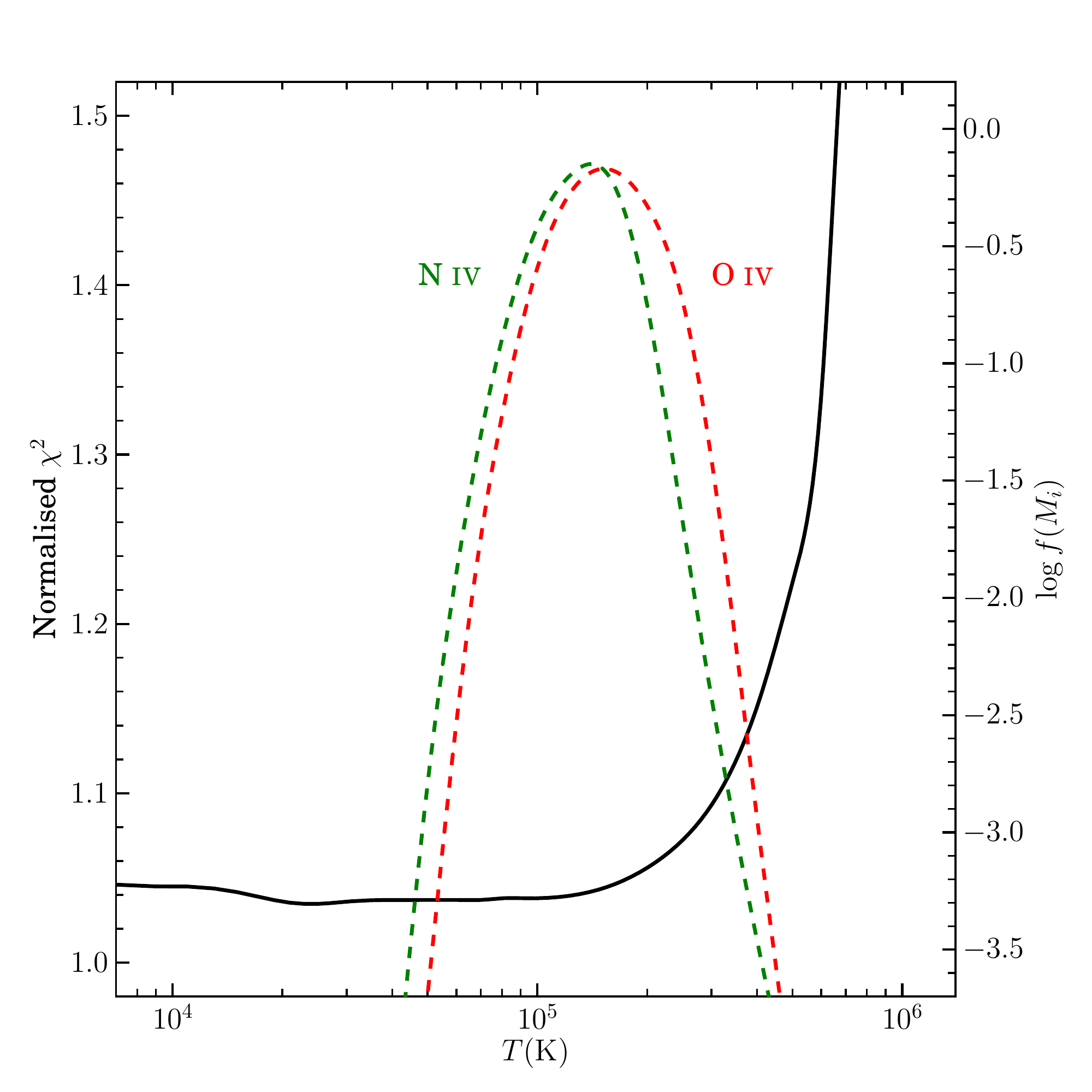}
    \caption{Normalised $\chi^2$ values for Voigt profile fits to the N$\;$\textsc{iv} and O$\;$\textsc{iv} absorption troughs as a function of temperature, assuming their $b$ values scale as in \cref{same_bturb}. The normalised $\chi^2$ value is minimised in the range $10^4 \lesssim T \lesssim 10^5~\textrm{K}$; temperatures characteristic of photoionized gas. Green and red dotted curves show, for collisional ionization equilibrium, the predicted ion fractions of N$\;$\textsc{iv} and O$\;$\textsc{iv} respectively as a function of temperature. Both peak at $\sim 10^5~\textrm{K}$, and we are therefore not able to rule out the possibility that the gas is collisionally ionized.}
\label{temp}
\end{figure}

\begin{table*}
\caption{Column density and Doppler broadening parameter measurements of the \ac{aals}.}
\label{voigt_parameters}
\begin{tabular}{| l c c c c c c c c |}
    \hline
    & \multicolumn{8}{c}{$\log \left(N / \textrm{cm}^{-2}\right)$} \\
    \\
    \cline{2-9}
    ion & $v_1$ & $v_2$ & $v_3$ & $v_4$ & $v_5$ & $v_6$ & $v_7$ & $v_8$ \\
    \hline
    H$\;$\textsc{i} & $14.64 \pm 0.32$ & $14.99 \pm 0.41$ & $14.92 \pm 0.65$ & $14.97 \pm 0.97$ & $14.94 \pm 0.97$ & $14.83 \pm 0.77$ & $14.50 \pm 0.29$ & $< 13.46 $ \\
    N$\;$\textsc{iii} & $< 12.99 $ & $< 12.99 $ & $< 12.99 $ & $13.41 \pm 0.09$ & $< 12.99 $ & $< 12.99 $ & $< 12.99 $ & $< 12.99 $ \\
    N$\;$\textsc{iv} & $> 14.11 $ & $14.03 \pm 0.05$ & $13.42 \pm 0.26$ & $14.09 \pm 0.11$ & $13.61 \pm 0.07$ & $13.50 \pm 0.11$ & $13.49 \pm 0.10$ & $13.08 \pm 0.14$ \\
    N$\;$\textsc{v} & $13.83 \pm 0.93$ & $> 13.89 $ & $> 13.86 $ & $> 13.93 $ & $> 13.95 $ & $> 13.90 $ & $> 13.83 $ & $< 14.06 $ \\
    O$\;$\textsc{iv} & $> 14.74 $ & $15.02 \pm 0.04$ & $14.54 \pm 0.22$ & $14.93 \pm 0.14$ & $14.59 \pm 0.11$ & $14.51 \pm 0.09$ & $14.12 \pm 0.15$ & $13.68 \pm 0.12$ \\
    O$\;$\textsc{iv}* & $14.45 \pm 0.06$ & $14.10 \pm 0.12$ & $< 13.46 $ & $14.22 \pm 0.07$ & $< 13.46 $ & $< 13.46 $ & $< 13.46 $ & $< 13.46 $ \\
    O$\;$\textsc{v} & $> 14.24 $ & $> 14.34 $ & $> 14.39 $ & $> 14.24 $ & $> 14.14 $ & $> 14.22 $ & $> 14.22 $ & $13.62 \pm 0.09$ \\
    O$\;$\textsc{vi} & $> 13.90 $ & $> 14.21 $ & $> 14.22 $ & $> 14.25 $ & $> 14.34 $ & $> 14.35 $ & $> 14.28 $ & $< 14.12 $ \\
    Ne$\;$\textsc{viii} & $14.94 \pm 0.07$ & $14.91 \pm 0.19$ & $15.06 \pm 0.29$ & $15.10 \pm 0.15$ & $> 14.83 $ & $> 15.01 $ & $> 14.91 $ & $< 13.48 $ \\
    Mg$\;$\textsc{x} & $< 13.46 $ & $14.50 \pm 0.12$ & $14.80 \pm 0.28$ & $14.68 \pm 0.39$ & $14.78 \pm 0.19$ & $14.54 \pm 0.54$ & $15.29 \pm 0.09$ & $< 13.46 $ \\
    S$\;$\textsc{v} & $13.14 \pm 0.07$ & $12.97 \pm 0.08$ & $< 12.40 $ & $12.90 \pm 0.09$ & $< 12.40 $ & $< 12.40 $ & $< 12.40 $ & $< 12.40 $ \\
    Ar$\;$\textsc{viii} & $13.19 \pm 0.13$ & $13.28 \pm 0.13$ & $< 12.82 $ & $13.34 \pm 0.09$ & $< 12.82 $ & $13.26 \pm 0.22$ & $13.04 \pm 0.29$ & $< 12.82 $ \\
    \hline
    & \multicolumn{8}{c}{$b \left(\textrm{km~s}^{-1}\right)$} \\
    \\
    \cline{2-9}
    ion & $v_1$ & $v_2$ & $v_3$ & $v_4$ & $v_5$ & $v_6$ & $v_7$ & $v_8$ \\
    \hline
    H$\;$\textsc{i} & $24$ & $34$ & $35$ & $29$ & $26$ & $28$ & $33$ & ... \\
    N$\;$\textsc{iii} & ... & ... & ... & $34 \pm 12$ & ... & ... & ... & ... \\
    N$\;$\textsc{iv} & ... & $27 \pm 3$ & $27 \pm 13$ & $20 \pm 3$ & $15 \pm 4$ & $18 \pm 6$ & $25 \pm 6$ & $10 \pm 7$ \\
    N$\;$\textsc{v} & $18 \pm 106$ & ... & ... & ... & ... & ... & ... & ... \\
    O$\;$\textsc{iv} & ... & $26$ & $27$ & $20$ & $15$ & $18$ & $25$ & $10$ \\
    O$\;$\textsc{iv}* & $20 \pm 4$ & $33 \pm 14$ & ... & $17 \pm 6$ & ... & ... & ... & ... \\
    O$\;$\textsc{v} & ... & ... & ... & ... & ... & ... & ... & $13 \pm 5$ \\
    O$\;$\textsc{vi} & ... & ... & ... & ... & ... & ... & ... & ... \\
    Ne$\;$\textsc{viii} & $19 \pm 2$ & $26 \pm 7$ & $77 \pm 76$ & $29 \pm 12$ & ... & ... & ... & ... \\
    Mg$\;$\textsc{x} & ... & $14 \pm 6$ & $171 \pm 78$ & $54 \pm 33$ & $31 \pm 12$ & $109 \pm 96$ & $25 \pm 5$ & ... \\
    S$\;$\textsc{v} & $18 \pm 5$ & $17 \pm 6$ & ... & $23 \pm 8$ & ... & ... & ... & ... \\
    Ar$\;$\textsc{viii} & $16 \pm 9$ & $41 \pm 19$ & ... & $25 \pm 9$ & ... & $49 \pm 22$ & $22 \pm 19$ & ... \\
    \hline
\end{tabular}
\end{table*}

\begin{figure*}
	\centering
	\includegraphics[width=18cm, natwidth=25.4cm, natheight=32.14cm]{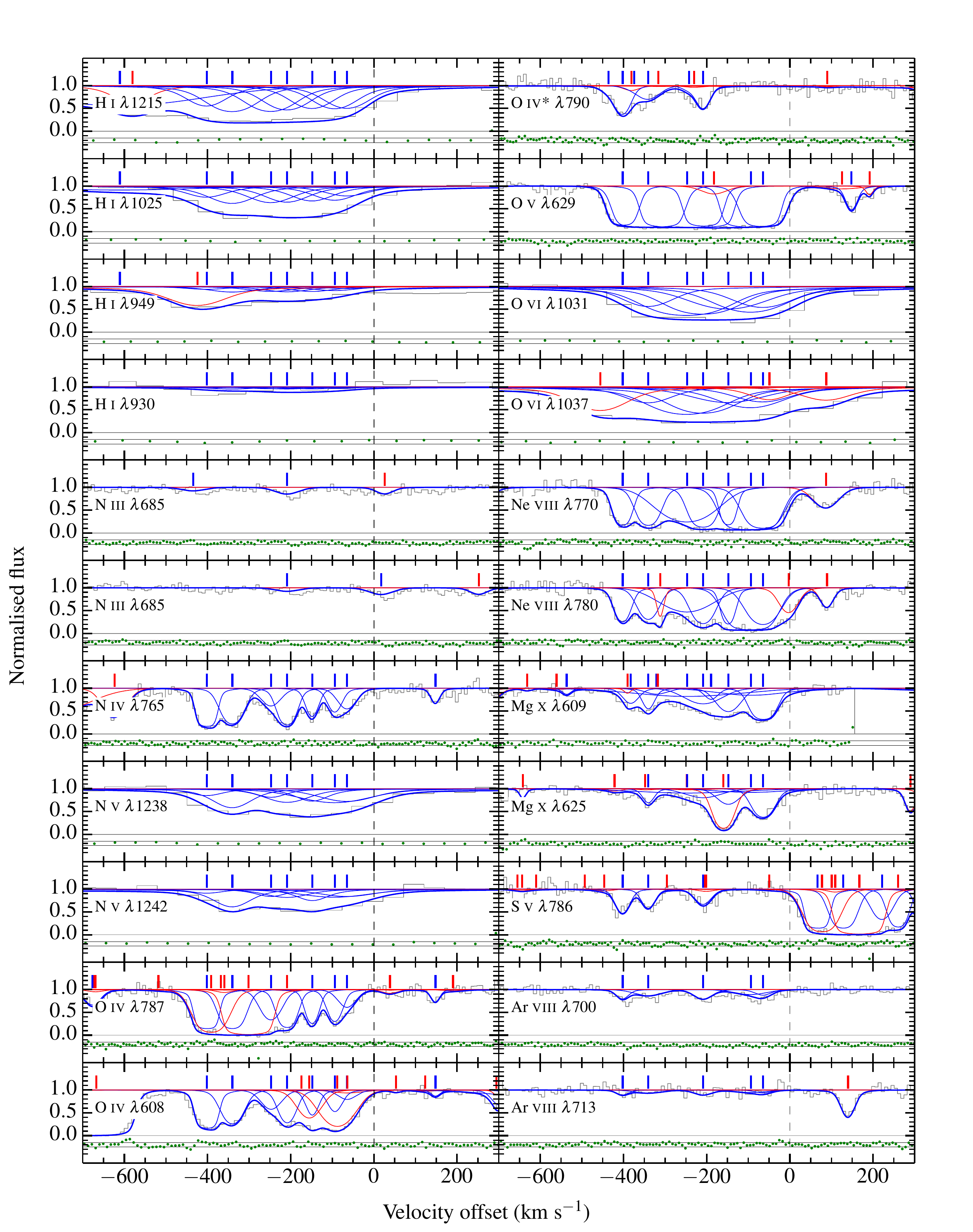}
	\caption{Voigt profiles fitted to the \ac{aals} in the \ac{hst}/\ac{cos} spectrum of Q0209. Thin blue lines are the individual velocity components for each labelled ion. Thin red lines are the blended components. The thick blue line in each panel is the total summed profile. The dashed line represents the rest-frame velocity of the \ac{qso}. Green points are the residuals between the model and the data, with the solid black lines representing the $\pm 1 \sigma$ standard deviation.}
\label{aals}
\end{figure*}

We can try to determine a plausible range in $T$ by choosing two ions that are assumed to trace the same gas, and whose individual components would then likely possess the same $b_{\textrm{turb}}$. We note that \ion{O}{4} and \ion{N}{4} have almost identical ionization destruction potentials, and very similar ionization creation potentials (54.9 eV and 47.4 eV respectively), making it likely that these ions will satisfy this criterion. We can then require that $b_{\textrm{turb}}$ be the same for each, such that
\begin{equation}
    b(\textrm{N}\;\textsc{iv}) = \sqrt{b^2(\textrm{O}\;\textsc{iv}) + 2kT\left(\frac{1}{m(\textrm{N})} - \frac{1}{m(\textrm{O})}\right)}.
\label{same_bturb}
\end{equation}
We proceed by fitting Voigt profiles to \ion{O}{4} and \ion{N}{4}, requiring that these ions share the same 8 velocity components as before, but additionally force the $b$ values to scale as in \cref{same_bturb}, for a range of temperatures between $\sim 10^4$ and $\sim 10^6~\textrm{K}$. The resulting normalised $\chi^2$ values on the fit as a function of gas temperature are shown as the black curve in \Cref{temp}. Above temperatures of $\sim 10^{5.5}$ K, normalised $\chi^2$ values on the fit start to increase dramatically, and we conclude that the data favours gas temperatures less than this value. Turbulence dominates the line broadening at these temperatures, and so scaling $b$ values between pairs of ions based on their masses alone will not be sufficient. Green and red dotted curves in \Cref{temp} show, in the case of a collisionally ionized medium, the expected ion fractions of \ion{N}{4} and \ion{O}{4} respectively as a function of temperature \citep{Mazzotta:1998hi}. Both ion fractions peak at $\sim 10^5~\textrm{K}$, and we are therefore not able to rule out the possibility that the gas is in \ac{cie} based on line widths alone.

We proceed by assuming the gas traced by \ion{N}{4} and \ion{O}{4} has a characteristic temperature of $3 \times 10^4~\textrm{K}$ (a reasonable value based on \Cref{temp}) and use this information to constrain the $b$ values in \ion{H}{1}. This procedure minimises the uncertainty in the \ion{H}{1} column densities, crucial for later photoionization modelling. We first assume that $b_{\textrm{turb}}$ for \ion{N}{4} be roughly the same as that for \ion{O}{4} and \ion{H}{1}, then apply a $b$ scaling between these ions like that in \cref{same_bturb}. The assumption that \ion{N}{4} and \ion{O}{4} trace the same gas is already well motivated based on the similarity in their ionization potentials. To extend this argument to \ion{H}{1}, with ionization potential $> 5$ times smaller than these ions, we assume that most of the \ion{H}{1} is locked up in the same gas as traced by \ion{N}{4} and \ion{O}{4}. We note that this assumption is justified later in \Cref{properties} upon consideration of the measured column densities and the ionization fractions derived from photoionization models. It is reassuring to note that the column density measurements resulting from this approach are insensitive to the assumed temperature over the range $10^4 \lesssim T \lesssim 10^5$~K, where the normalised $\chi^2$ values in \Cref{temp} are minimised. Therefore, the gas temperature we assume in the \ion{O}{4}/\ion{N}{4} gas is (nearly) independent of any constraints later obtained in \Cref{properties}.

All ions are now fitted simultaneously, tying the velocity structure to the 8 components identified in \ion{N}{4} (\Cref{vstructure}) as before, but with an additional $b$ scaling between \ion{N}{4}, \ion{O}{4} and \ion{H}{1} as described. All other $b$ values are allowed to float. The resulting fit has a normalised $\chi^2$ value of 1.37 and is shown in \Cref{aals}. Transitions are ordered first by atomic mass, and second by oscillator strength. The dashed vertical line indicates the rest-frame velocity of the \ac{qso}. Individual blue Voigt profiles represent those components attributable to the labelled ion and transition, whereas red profiles represent components from unrelated blended transitions. Most of these blends are constrained by accompanying transitions of the same ion. The two most prominent blends with \ion{O}{4} $\lambda$608 and the blend with \ion{Mg}{10} $\lambda$625 at $\sim 400$~\kms, are assumed to be \ion{H}{1} \lya. The thick blue line represents the overall model absorption trough, and the green points show the residuals on the fit, with the solid black lines marking the $\pm 1 \sigma$ standard deviation. Line saturation is evident from the \lya\ absorption trough, although we note that the column densities are well measured due to a large number of observed transitions in the \ion{H}{1} Lyman series.

Column densities and Doppler broadening parameters are listed on a component by component basis ($v_1$ -- $v_8$, ascending in velocity offset), together with their $1 \sigma$ error bars in \Cref{voigt_parameters}. For components where no absorption line is detected in any given ion, we calculate the upper bound on the equivalent width at the $3\sigma$ significance level, derived using equations (4) -- (5), (7) and (9) -- (10) in \citet{Keeney:2012kg}, then perform the conversion to column density assuming a linear curve of growth. For column densities with $1 \sigma$ uncertainties greater than 1 (typically when $\tau \gg 1$, i.e. where the lines are saturated), we quote lower limits on the column densities, based on the apparent optical depth at the line centres (calculated using \cref{tau}) and assuming a $b$ value of 25~\kms, which is approximately typical of the well-measured lines. We do not list $b$ values for components that have upper or lower limits on the column densities. Error bars are not presented for inferred $b$ values.

\section{Properties of the associated absorbers}
\label{properties}
In the following section, we first present constraints on the electron number density in the clouds based on an analysis of the fine-structure transition \ion{O}{4}*. We then consider both photoionization and \ac{cie} models, and use these to put constraints on the properties of the associated gas clouds. Models are generated using version c13.00 of \cloudy, described in \citep{Ferland:2013wla}. We note that, in general, absorbers may not be in ionization equilibrium \cite[e.g.][]{Oppenheimer:2013cr,Oppenheimer:2013dx}. This possibility is discussed later in \Cref{equilibrium}.

\subsection{Electron number density in the absorbing clouds}
\label{density}
\begin{figure}
    \centering
    \includegraphics[width=8.4cm, natwidth=20.32cm, natheight=16.27cm]{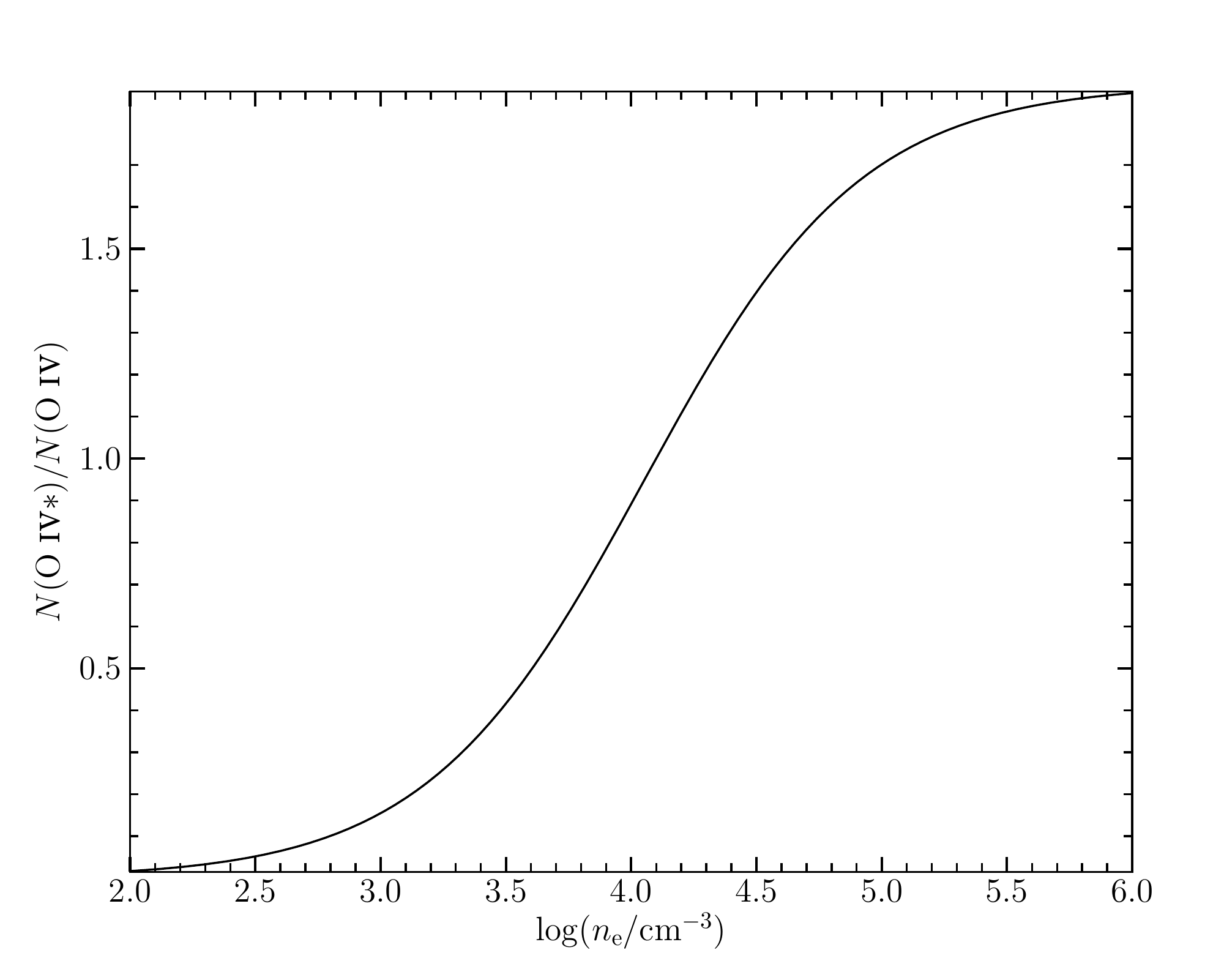}
    \caption{The electron number density $n_\textrm{e}$ as a function of the column density ratio between $\textrm{O}\;\textsc{iv}^*$ and $\textrm{O}\;\textsc{iv}$ based on predictions from a theoretical level population assuming a temperature of $10^4~\textrm{K}$.}
\label{e_density_fig}
\end{figure}

In this section we present analysis on the absorption due to fine structure, metastable transitions in \ion{O}{4}, which enables us to estimate the electron number density in the clouds giving rise to the absorption by \ion{O}{4}. First, we note that \ion{O}{4}* arises from doublet fine-structure ($J = 1/2$ and $3/2$) in the ground state, which should behave approximately as a two-level atom, where the level populations are controlled by collisional processes and forbidden radiative decays \citep{Bahcall:1968ig}. If we neglect stimulated emission, the only acting processes are collisional excitation, collisional de-excitation and radiative decay. The energy level spacing corresponds to $25.91~\mu\textrm{m}$, or $1.2 \times 10^{13}~\textrm{Hz}$, so we find this to be a fair approximation on the basis of \Cref{sed}, together with the fact that stimulated emission is extremely forbidden. We denote the ground state as level 0, the excited state as level 1, and let $n_j$ be the number density ($\textrm{cm}^{-3}$) of \ion{O}{4} in level $j$. If we assume that collisional excitation is dominated by electrons, then the population of the excited state must satisfy
\begin{equation}
    \frac{\mathrm{d}n_1}{\mathrm{d}t} = n_e n_0 k_{01} - n_e n_1 k_{10} - n_1 A_{10}
\end{equation}
\citep{Draine:2011tr}, where $n_e$ is the electron number density ($\textrm{cm}^{-3}$), $k_{01}$ and $k_{10}$ are the upward and downward rate coefficients ($\textrm{cm}^{3}~\textrm{s}^{-1}$) respectively, and $A_{10}$ is the spontaneous decay rate ($\textrm{s}^{-1}$). For a steady state ($\mathrm{d}n_1 / \mathrm{d}t = 0$), we then require
\begin{equation}
    \frac{N(\textrm{O}\;\textsc{iv}^*)}{N(\textrm{O}\;\textsc{iv})} = \frac{n_e k_{01}}{n_e k_{10} + A_{10}},
\end{equation}
where we have now replaced the number densities by the observed column densities. We can write $k_{01}$ in terms of $k_{10}$ as follows:
\begin{equation}
    k_{01} = \frac{g_1}{g_0} k_{10} e^{-E_{10} / kT},
\end{equation}
where $g_0$ and $g_1$ are the level degeneracies, $E_{10}$ is the energy level difference and $T$ is the kinetic temperature of the gas. We can additionally define the critical electron density, at which the collisional de-excitation rate equals the radiative de-excitation rate:
\begin{equation}
    n_{\textrm{crit}} \equiv \frac{A_{10}}{k_{10}}.
\end{equation}
The electron number density can then be written as
\begin{equation}
    n_e = n_{\textrm{crit}}\left(\frac{N(\textrm{O}\;\textsc{iv})}{N(\textrm{O}\;\textsc{iv}^*)}\frac{k_{01}}{k_{10}} - 1\right)^{-1}.
\label{e_density}
\end{equation}
Assuming a temperature of $10^4~\textrm{K}$, we can write
\begin{equation}
    k_{10} = 8.629 \times 10^{-8} \frac{\Omega_{10}}{g_1}~\textrm{cm}^3~\textrm{s}^{-1},
\end{equation}
where $\Omega_{10}$ is the collision strength connecting levels 1 and 0. For electrons at this temperature, the collision strength is $\Omega_{10} = 2.144$ \citep{Tayal:2006jo}. We take $A_{10} = 5.19 \times 10^{-4}~\textrm{s}^{-1}$ from the NIST atomic spectra database\footnote{\url{http://www.nist.gov/pml/data/asd.cfm}}. The resulting dependence of $n_e$ on the ratio $N(\textrm{O}\;\textsc{iv}^*) / N(\textrm{O}\;\textsc{iv})$ is shown in \Cref{e_density_fig}. Comparing to fig. 12 in \cite{Arav:2013be}, we find an excellent agreement. For a highly ionized plasma, assuming solar metallicity and abundances, we can relate $n_e$ to the hydrogen number density $n_{\textrm{H}}$ as 
\begin{equation}
    n_e = n_{\textrm{H}}\left(1 + 2\frac{n_{\textrm{He}}}{n_{\textrm{H}}} + \sum_{i \geq 3} Z_i \frac{n_{Z_i}}{n_{\textrm{H}}}\right) \approx \frac{1 + X}{2X} n_{\textrm{H}},
\end{equation}
where $Z_i$ is the atomic number of element $i$ and $X$ is the mass fraction in hydrogen. For a value $X = 0.71$ we have $n_e \simeq 1.2 n_{\textrm{H}}$. We calculate $n_{\textrm{H}}$ in the gas traced by \ion{O}{4} for each velocity component using \cref{e_density}, and summarise the results in \Cref{density_table}. Where there is a lower limit on the \ion{O}{4} column density, or an upper limit on the \ion{O}{4}* density, this results in an upper limit on $n_{\textrm{H}}$. We caution that these upper limits are only approximate, as they are sensitive to the adopted significance level used to calculate limiting equivalent widths, and the assumed $b$ value used to calculate the lower limit on the \ion{O}{4} column density. We find that a value $\log (n_{\textrm{H}} / \textrm{cm}^{-3}) \sim 3$ is representative for the AAL region as a whole.

\begin{table*}
\caption{Total hydrogen number density in each velocity component.}
\label{density_table}
\begin{tabular}{| c c c c c c c c |}
    \hline
    \multicolumn{8}{c}{$\log \left(n_{\textrm{H}} / \textrm{cm}^{-3}\right)$} \\
    \\
    \cline{1-8}
    $v_1$ & $v_2$ & $v_3$ & $v_4$ & $v_5$ & $v_6$ & $v_7$ & $v_8$ \\
    \hline
    $\lesssim 3.54$ & $2.80 \pm 0.13$ & $\lesssim 2.63$ & $3.03 \pm 0.17$ & $\lesssim 2.70$ & $\lesssim 2.66$ & $\lesssim 3.09$ & $\lesssim 3.64$ \\
    \hline
\end{tabular}
\end{table*}

\subsection{Photoionization analysis}
In the following we assume that the gas clouds are in photoionization equilibrium with either of the \ac{qso} \ac{sed}s determined in \Cref{sed_section}. The two main parameters that best describe the photoionization structure are the total hydrogen column density ($N_{\textrm{H}}$) and the ionization parameter ($U$). The latter is defined as the dimensionless ratio between the number density of hydrogen atoms and the number density of photons that ionize hydrogen at the illuminated face of the absorbing gas clouds. We express this as
\begin{equation}
    U \equiv \frac{1}{4 \pi c R^2 n_{\textrm{H}}} \int^{\infty}_{\nu_{\textrm{LL}}} \frac{L_{\nu}}{h \nu} d \nu,
\label{ionization_parameter}
\end{equation}
where $\nu$ is the frequency, $c$ is the speed of light, $L_{\nu}$ is the luminosity density of the \ac{qso} (erg s$^{-1}$ cm$^{-2}$ Hz$^{-1}$), $\nu_{\textrm{LL}}$ is the frequency corresponding to the Lyman limit (912~\AA), $R$ is the radial distance between the absorber and the \ac{qso}, and $n_{\textrm{H}}$ is the total hydrogen density (i.e. \ion{H}{1} $+$ \ion{H}{2}). The equations of ionization and thermal balance are then solved using \cloudy. We run the code multiple times to form three-dimensional grids of predicted quantities over $(U, N_{\textrm{H}}, Z)$ parameter space, where $Z$ is the metallicity of the gas, normalised to solar metallicity. We vary the parameters $U$ and $N_{\textrm{H}}$ in steps of 0.1 dex, and $Z$ in steps of 1 dex, assuming a constant total hydrogen density. The results from these grids are the subject of this section and \Cref{metallicity_section}.

\begin{figure}
    \centering
    \includegraphics[width=8.4cm, natwidth=20.32cm, natheight=20.32cm]{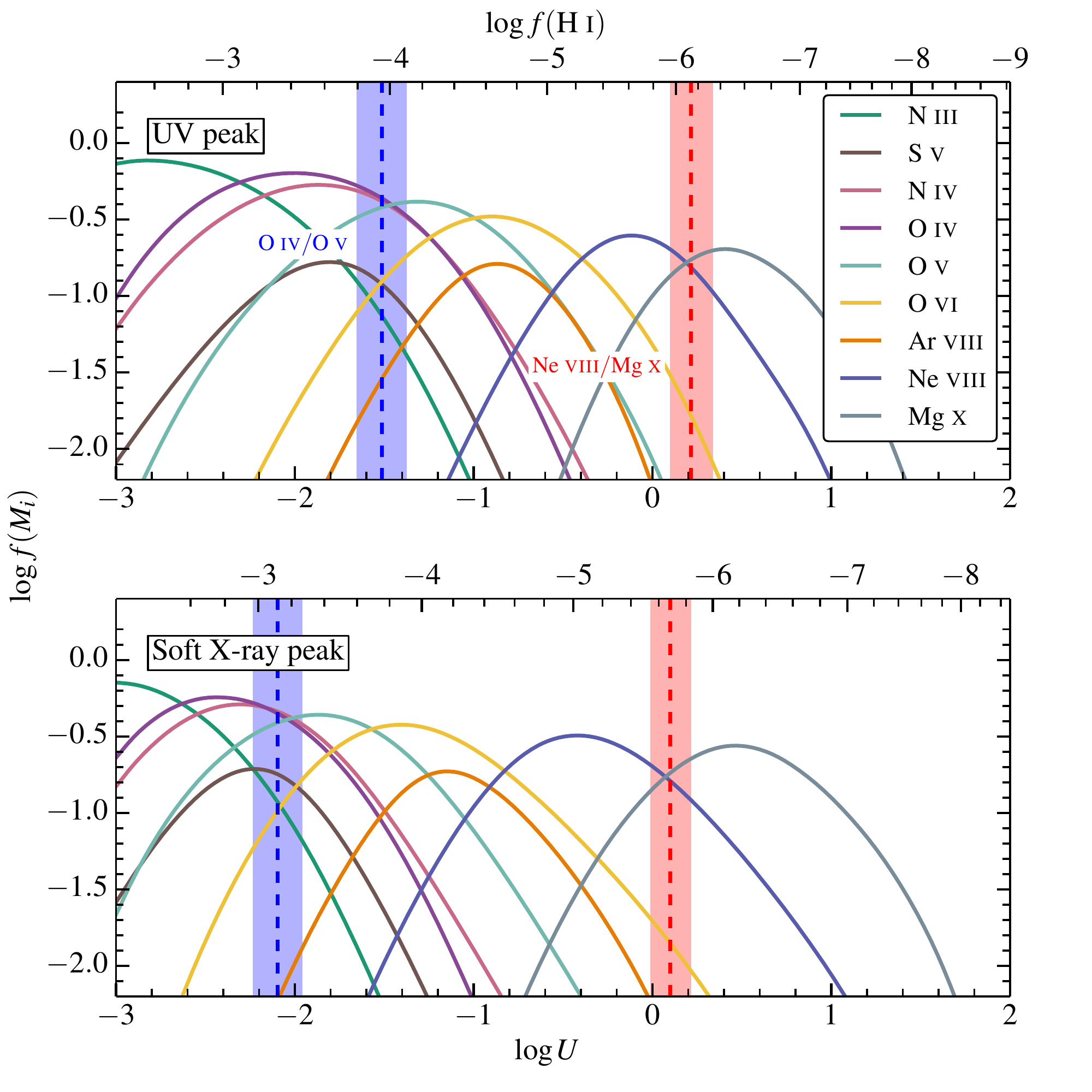}
    \caption{Theoretical ionization fractions for a range of metal ions $f(M_i)$ in optically thin clouds in photoionization equilibrium with the `UV peak' \ac{sed} (top panel) and the `Soft X-ray peak' \ac{sed} (bottom panel), plotted as a function of ionization parameter. Curves are colour coded by ion. The \ion{H}{1} fraction is shown on the top axis. Vertical bars indicate the range in ionization parameter allowed for by measured column density ratios in components where those ratios are best measured. For $N(\textrm{Ne}\;\textsc{viii}) / N(\textrm{Mg}\;\textsc{x})$ we scale by relative solar abundances. These ratios indicate there is a range in ionization parameter covering nearly two orders of magnitude under photoionization equilibrium.}
\label{ionization_plot}
\end{figure}

Shown in \Cref{ionization_plot} are theoretical ionization fractions of \ion{H}{1} and various metal ions for which we have reliable column density estimates as a function of $U$. We show model curves for clouds in photoionization equilibrium with both the `UV peak' and `Soft X-ray peak' \ac{sed}s (see \Cref{sed_section}). These results are not sensitive to the abundances used in the calculations (in this case we have used solar abundances), and neither are they sensitive to the total column densities. This is because the model clouds are optically thin in the Lyman continuum, which means that there are no steep gradients in ionization (e.g. due to shielding from an \ion{H}{2} -- \ion{H}{1} recombination front). We note that this situation is a good approximation to the real one, given the absence of absorption at the Lyman limit. We calculate the column density ratios $N(\textrm{O}\;\textsc{iv}) / N(\textrm{O}\;\textsc{v})$ and $N(\textrm{Ne}\;\textsc{viii}) / N(\textrm{Mg}\;\textsc{x})$ in the components where those ratios are best measured, then find the range in ionization parameter over which each is predicted within their $1 \sigma$ uncertainties. For $N(\textrm{Ne}\;\textsc{viii}) / N(\textrm{Mg}\;\textsc{x})$ we scale by relative solar abundances. The resulting constraints on $U$ are plotted as a series of vertical bars.

From \Cref{ionization_plot} it is clear that the gas has a range in ionization parameter that covers around two orders of magnitude. At a fixed distance $R$ from the \ac{qso}, this corresponds to a range in gas density that covers around two orders of magnitude. Measuring the same column density ratios in different velocity components to those plotted (including column density lower limits) gives constraints on the ionization parameter that are fully consistent with those obtained above. This range ionization parameter is therefore representative for the AAL region as a whole.

\subsection{Collisional ionization equilibrium}
\begin{figure}
    \centering
    \includegraphics[width=8.4cm, natwidth=20.32cm, natheight=12.2cm]{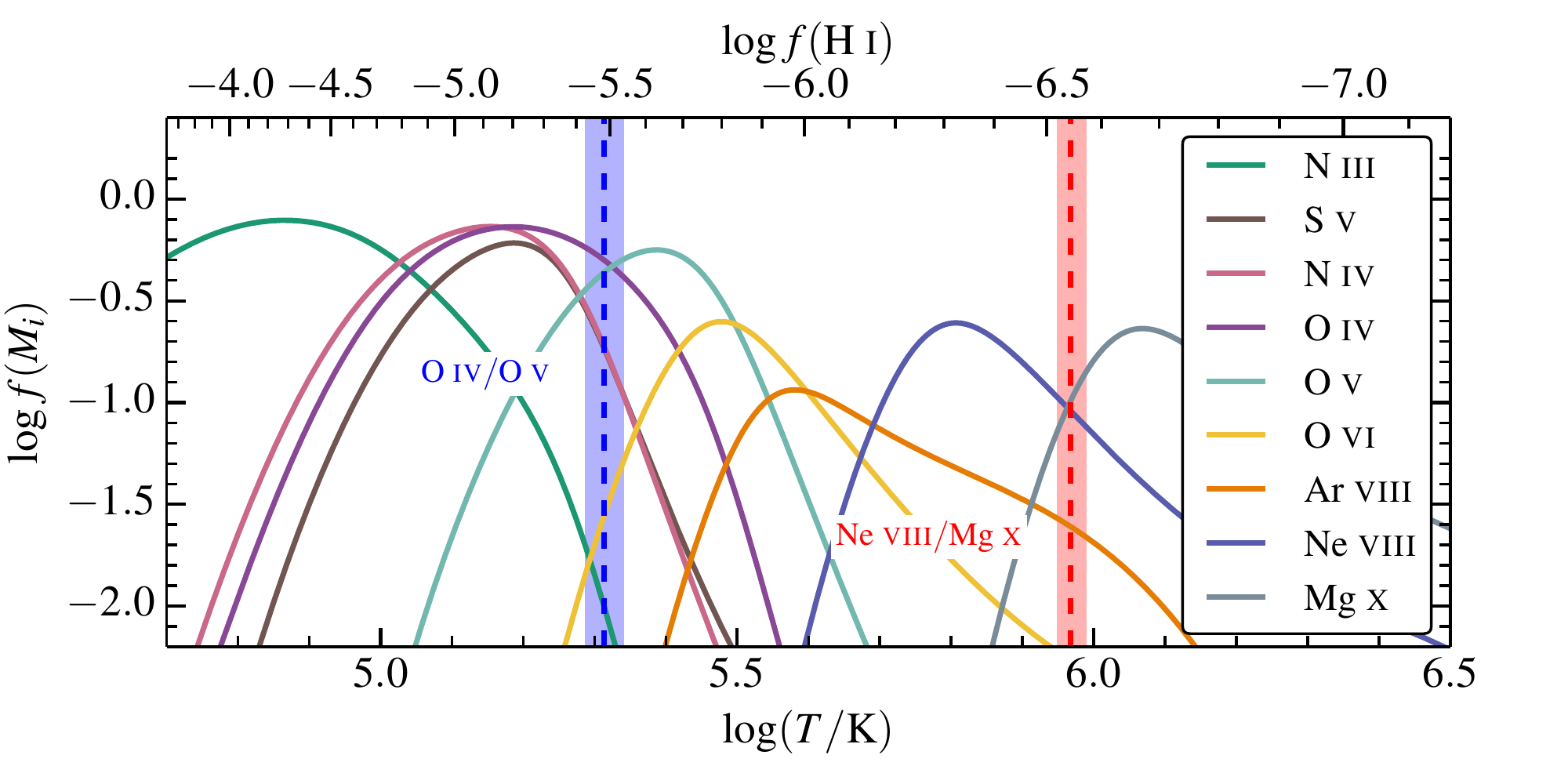}
    \caption{Theoretical ionization fractions for a range of metal ions $f(M_i)$ under \ac{cie} as a function of temperature. Curves are colour coded by ion. Vertical bars indicate the range in temperature allowed for by measured column density ratios in components where those ratios are best measured. For $N(\textrm{Ne}\;\textsc{viii}) / N(\textrm{Mg}\;\textsc{x})$ we scale by relative solar abundances. These ratios indicate a range covering more than an order of magnitude in gas temperature under \ac{cie}.}
\label{cie}
\end{figure}

We next consider a situation whereby the AAL clouds are in \ac{cie}. To simulate \ac{cie}, we run \cloudy\ multiple times with $\log U = -5$ (so that the effect of the radiative field is negligible), holding temperature constant for values in the range $4.5 \leq T \leq 6.5$. Shown in \Cref{cie} are theoretical ionization fractions of various metal ions for which we have reliable column density estimates under \ac{cie} as a function of temperature. Like \Cref{ionization_plot}, the results are not sensitive to the the specific abundances adopted, or the total column density. Vertical bars indicate the range in temperature permitted according to the best-measured column density ratios $N(\textrm{O}\;\textsc{iv}) / N(\textrm{O}\;\textsc{v})$ and $N(\textrm{Ne}\;\textsc{viii}) / N(\textrm{Mg}\;\textsc{x})$ within their $1 \sigma$ uncertainties. Again, for $N(\textrm{Ne}\;\textsc{viii}) / N(\textrm{Mg}\;\textsc{x})$ we scale by relative solar abundances. It is clear that temperatures cover more than an order of magnitude in the AAL region if the gas is in \ac{cie}. Similarly to the case of photoionization, constraints on the temperature are consistent between different velocity components.

\begin{figure}
    \centering
    \includegraphics[width=8.4cm, natwidth=20.32, natheight=16.26cm]{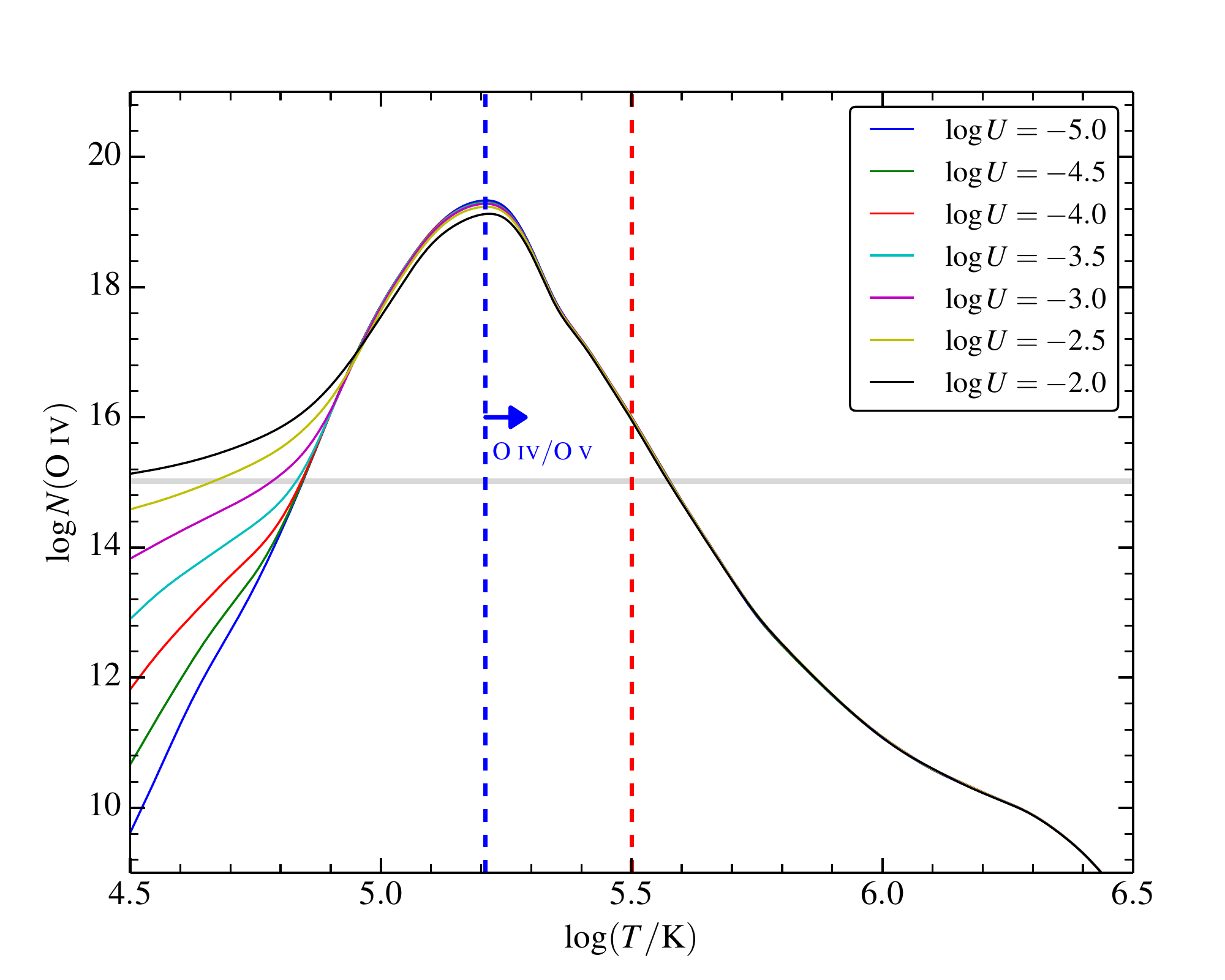}
    \caption{The predicted $\textrm{O}\;\textsc{iv}$ column density as a function of gas temperature for a range of ionization parameters and a fixed $\textrm{H}\;\textsc{i}$ column density of $10^{15}~\textrm{cm}^{-2}$. The dashed blue vertical line and accompanying arrow denotes a lower limit on the temperature of the gas clouds giving rise to the majority of the $\textrm{O}\;\textsc{iv}$ absorption under \ac{cie}, inferred from the ratio $N(\textrm{O}\;\textsc{iv}) / N(\textrm{O}\;\textsc{v})$. The vertical red dashed line marks the maximum temperature allowed from an analysis of the absorption line widths. The horizontal grey shaded bar represents the $1 \sigma$ constraints on the \ion{O}{4} column density from velocity component $v_2$. Theoretical curves match this column density at temperatures nearly an order of magnitude lower than those inferred from \ac{cie} models, favouring a scenario in which the AAL gas traced by $\textrm{O}\;\textsc{iv}$ is predominantly photoionized.}
\label{oiv_cie}
\end{figure}

To determine whether \ac{cie} is allowed by the data, we model clouds illuminated by the `UV peak' incident continuum, assuming an \ion{H}{1} column density of $10^{15}$ cm$^{-2}$ (matching the observed value in velocity component $v_2$), a total hydrogen density of $\log (n_{\textrm{H}} / \textrm{cm}^{-3}) = 2.8$, and solar metallicity, for a range of ionization parameters $U$, holding $T$ constant, and repeating to create a grid of values ($U$, $T$). We then plot the predicted \ion{O}{4} column density as a function of $T$ for a range of values of $U$ in \Cref{oiv_cie}. \ac{cie} is achieved in the limit of high $T$ and/or low $U$. The horizontal grey bar represents the $1 \sigma$ constraint on the \ion{O}{4} column density in velocity component $v_2$. The dashed blue vertical line and accompanying arrow represents the lower limit on the temperature based on the column density ratio $\textrm{O}\;\textsc{iv} / \textrm{O}\;\textsc{v}$. This is then an indication of the characteristic temperature of the gas hosting \ion{O}{4} in CIE. Temperatures to the right of the red dashed line at $10^{5.5}$~K have already been ruled out based on the observed line-widths (see \Cref{temp}).

Given the lower and upper bounds on the gas temperature, it is clear that the predicted \ion{O}{4} column density in \Cref{cie} is at least an order of magnitude larger than that observed. All theoretical curves match the observed \ion{O}{4} column density for $T \lesssim 10^{4.85}~\textrm{K}$. We note that decreasing $U$ further has no effect on this result, since by $\log U = -5$, the solution has converged to a value $T \approx 10^{4.85}~\textrm{K}$. The assumed \ion{H}{1} column density is close to the maximum allowed by the data, although this value would have to be reduced by at least an order of magnitude to give agreement between the predicted temperatures. Similarly, the metallicity would have to be lowered by at least an order of magnitude. In photoionization equilibrium, the ionization parameter for the gas traced by \ion{O}{4} is $\log U \sim -2$, which upon inspection of \Cref{oiv_cie} gives a temperature of $T \approx 10^{4.5}~\textrm{K}$, fully consistent with that predicted by the photoionization models. Changing the incident continuum to the `Soft X-ray peak' model gives similar results. We therefore conclude that the AAL gas traced by \ion{O}{4} (and ions with similar ionization potential) is predominantly photoionized. We are not able to rule out \ac{cie} in the more highly ionized species like \ion{Ne}{8} and \ion{Mg}{10}, since the required temperatures would produce negligible accompanying \ion{O}{4} absorption, below the detection limits of the present data.

\subsection{Gas metallicity and total column density}
\label{metallicity_section}
The metallicity $[M / \textrm{H}]$ of the gas giving rise to the \ac{aals} can be expressed as
\begin{equation}
    \left[\frac{M}{\textrm{H}}\right] = \log \left(\frac{N\left(M_i\right)}{N\left(\textrm{H}\;\textsc{i}\right)}\right) + \log \left(\frac{f\left(\textrm{H}\;\textsc{i}\right)}{f\left(M_i\right)}\right) + \log\left(\frac{\textrm{H}}{M}\right)_{\odot}
\label{metallicity}
\end{equation}
\citep{Hamann:1999ky}, where $(\textrm{H} / M)_{\odot}$ is the solar abundance ratio of hydrogen to some metal species $M$, $N(\textrm{H}\;\textsc{i})$ and $f(\textrm{H}\;\textsc{i})$ respectively are the column density and ionization fraction in \ion{H}{1}, and $N(M_i)$ and $f(M_i)$ respectively are the column density and ionization fraction in some ion $M_i$ of metal species $M$. If the gas is well characterised by a single ionization parameter, \cref{metallicity} can be implemented using the measured column densities in hydrogen and some arbitrary metal ion, together with the inferred ionization fractions in each, and assuming solar abundance ratios. When there is a range in ionization parameter, the situation is more complicated, since the measured column density in each ion will be the sum of the column densities arising in each region (each characterised by a different value of $U$). In this case, the measured column density in some ion $M_i$ will be expressed as
\begin{equation}
    N(M_i) = \sum_k N(M)_k f(M_i)_k,
\end{equation}
where $N(M)_k$ and $f(M_i)_k$ are the $k$ total column densities of element $M$, and ionization fractions of ion $M_i$ respectively, for a set of $k$ ionization parameters. In the limit where there is a continuous distribution over ionization parameter, this becomes
\begin{equation}
    N(M_i)_{\textrm{obs}} = \int_{\xi_{\textrm{min}}}^{\xi_{\textrm{max}}} \frac{\textrm{d}N(M)}{\textrm{d}\xi} f(M_i)~\textrm{d}\xi,
\label{continuous_U}
\end{equation}
where $\xi = \log U$.

\begin{figure*}
    \centering
    \includegraphics[width=18cm, natwidth=25.4cm, natheight=31.5cm]{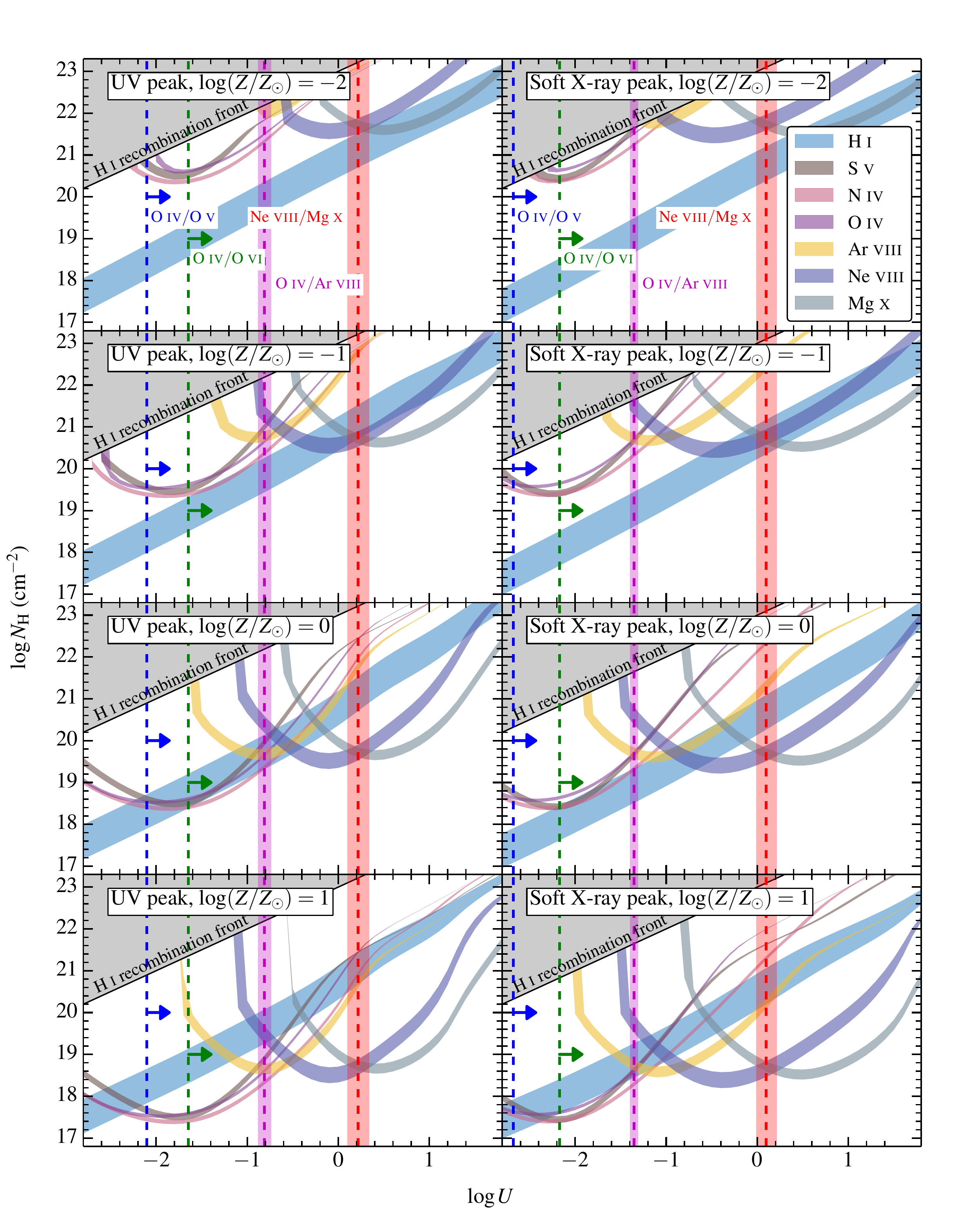}
    \caption{Ionization solutions for component $v_2$ in the case of the `UV peak' incident \ac{sed} (left panels) and the `Soft X-ray peak' \ac{sed} (right panels) for a range of metallicities. Coloured regions represent the values of $\log U$ and $\log N_{\textrm{H}}$ that predict the $\pm 1 \sigma$ bounds on the column densities for each labelled ion. Vertical bars indicate the plausible range in $U$ for each of the ionization components identified by the column density ratios labelled. The crossing points between contours, defining the pair ($U$, $N_{\textrm{H}}$) physically characterise these components. Dashed vertical lines and arrows represent lower limits on $U$.}
\label{phase_plot_v2}
\end{figure*}

\begin{figure*}
    \centering
    \includegraphics[width=18cm, natwidth=25.4cm, natheight=31.5cm]{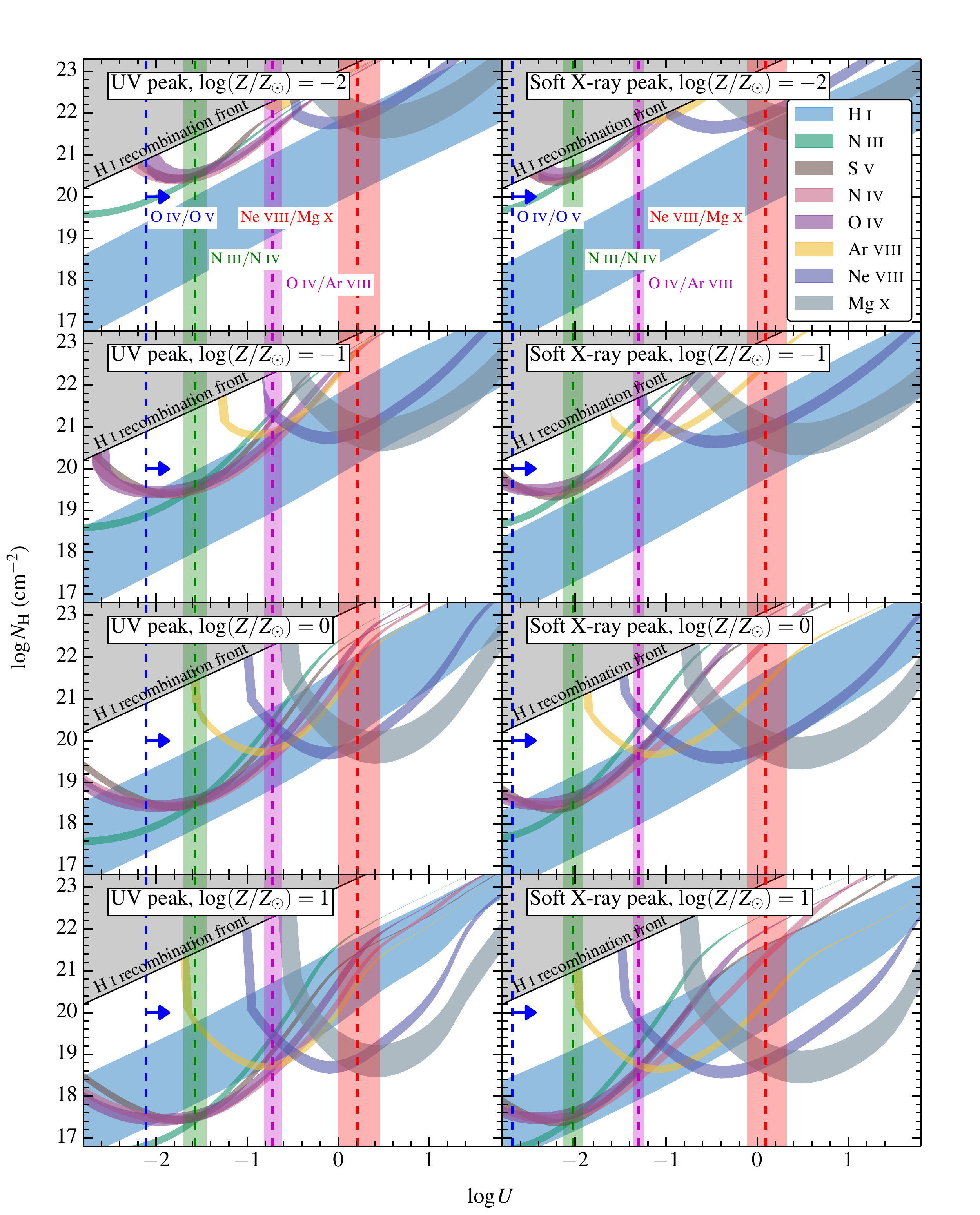}
    \caption{Ionization solutions for component $v_4$ in the case of the `UV peak' incident \ac{sed} (left panels) and the `Soft X-ray peak' \ac{sed} (right panels) for a range of metallicities. This figure has the same format as \Cref{phase_plot_v2}.}
\label{phase_plot_v4}
\end{figure*}

To estimate total column densities in velocity components $v_2$ and $v_4$ (where we have the largest number of well-measured ions), we assume that all of the gas is photoionized, and plot the locus of points that mark the observed column densities in $\log U$ -- $\log N_{\textrm{H}}$ space as shown in \Cref{phase_plot_v2,phase_plot_v4}, where the width of each contour indicates the $1\sigma$ uncertainty. The left-hand panels represent the case where the gas is photoionized by the `UV peak' \ac{sed}, with assumed metallicities $[M / \textrm{H}] = -2, -1, 0$, and $1$. The right-hand panels represent the equivalent scenario, but with the `Soft X-ray peak' \ac{sed}. Various zones within the absorbing region are physically characterised by the pair $(U, N_{\textrm{H}})$ that best predict the observed column densities. Constraints on the ionization parameter are identified on the basis of column density ratios that are labelled similarly to those in \Cref{ionization_plot}. Ratios that involve one ion having a lower limit on the column density give corresponding lower limits on the ionization parameter in \Cref{phase_plot_v2,phase_plot_v4}, plotted with vertical dashed lines and corresponding arrows. In velocity component $v_2$, we identify a minimum of two ionization components $(U, N_{\textrm{H}})$ that can account for all of the observed column densities. Ionization parameters in these components are inferred from the ratios $N(\textrm{O}\;\textsc{iv}) / N(\textrm{Ar}\;\textsc{viii})$ and $N(\textrm{Ne}\;\textsc{viii}) / N(\textrm{Mg}\;\textsc{x})$. This conclusion comes with the caveat that we require there to be an under-abundance of nitrogen (and an over-abundance of sulphur, in the case of the UV peak SED) by factors of a few with respect to the solar values to explain why the \ion{O}{4}, \ion{N}{4}, \ion{S}{5} and \ion{Ar}{8} contours do not all cross in the $\log U$ range defined by $N(\textrm{O}\;\textsc{iv}) / N(\textrm{Ar}\;\textsc{viii})$. In velocity component $v_4$, we require there to be at least three ionization components, with a similar caveat on the assumed overabundance of sulphur. This result contrasts with that from some previous, similar studies, that find a maximum of two discrete ionization components are required to adequately fit the data \cite[e.g.][]{Moe:2009do,Edmonds:2011fz,Borguet:2012ca,Arav:2013be}. In this case, the ionization components inferred are no longer robust, since there are non-negligible fractional abundances of e.g. \ion{O}{4} and \ion{N}{4} in both of the lower ionization components. The limiting scenario, is one where there exists a continuous distribution of ionization parameters through the absorbing region, which at a fixed distance $R$ corresponds to a smooth variation in gas density. This scenario is described by \cref{continuous_U}. To simplify the problem, we assume that a (close-to) continuous distribution in $U$ exists in the AAL region, and find an approximate solution to \cref{continuous_U} assuming that each ion forms largely at the peak in its fractional abundance, $f(M_i)_{\textrm{max}}$. In this case, \cref{continuous_U} becomes
\begin{equation}
    N(M) \simeq \frac{N(M_i)}{f(M_i)_{\textrm{max}}},
\end{equation}
and we can read off a corresponding ionization parameter from \Cref{ionization_plot}.

Assuming that most of the gas hosting \ion{H}{1} also hosts most of the low ionization species in our sample (e.g. \ion{N}{3}, \ion{N}{4}, \ion{O}{4} and \ion{S}{5}), we can identify peak fractional abundances of these ions with fractional abundances of \ion{H}{1} in \Cref{ionization_plot} and find the metallicity of the gas using \cref{metallicity}. Incorporating uncertainty in the SED, we conservatively conclude that $0 \lesssim [\textrm{O} / \textrm{H}] \lesssim 1$. We find this metallicity to be representative across all velocity components, and find the same result for $[\textrm{N} / \textrm{H}]$ and $[\textrm{S} / \textrm{H}]$. Peak fractional abundances correspond to column density loci minima in \Cref{phase_plot_v2,phase_plot_v4}. Splitting the $\log U$ range in half at $\log U = -1$, it is clear that the high ionization gas traced by e.g. \ion{Ne}{8} and \ion{Mg}{10} has a factor of $\sim 10$ higher contribution to the total column density compared to the low ionization gas traced by e.g. \ion{N}{3} and \ion{O}{4}. Constructing plots like these for all other velocity components (not shown), we conservatively conclude that, for each of the velocity components $v_1$ -- $v_7$, the total hydrogen column density is $10^{17} \lesssim N_{\textrm{H}} \lesssim 10^{18.5}~\textrm{cm}^{-2}$ in the low ionization gas, and $10^{18.5} \lesssim N_{\textrm{H}} \lesssim 10^{20}~\textrm{cm}^{-2}$ in the high ionization gas. In velocity component $v_8$ we detect only \ion{O}{4}, \ion{O}{5} and \ion{N}{4} absorption, and the total hydrogen column density through this region is, conservatively, $10^{16.5} \lesssim N_{\textrm{H}} \lesssim 10^{17.5}~\textrm{cm}^{-2}$.

\subsection{Distance and size constraints}
Given the estimate on $n_{\textrm{H}}$, combined with information on $U$ and $N_{\textrm{H}}$, we can put a constraint on both the distance to the absorbing clouds from the \ac{qso} and their geometry. To determine the distance, $R$, we simply rearrange \Cref{ionization_parameter} to obtain
\begin{equation}
    R = \sqrt{\frac{Q_{\textrm{H}}}{4 \pi c n_{\textrm{H}} U}},
\end{equation}
where
\begin{equation}
    Q_{\textrm{H}} \equiv \int^{\infty}_{\nu_{\textrm{LL}}} \frac{L_{\nu}}{h \nu} d \nu
\end{equation}
expresses the rate of emission of photons having energies sufficient to ionize hydrogen. Incorporating the uncertainty in $Q_{\textrm{H}}$ (due to uncertainties in the \ac{sed}), and conservatively estimating from \Cref{ionization_plot} that absorption due to \ion{O}{4} and \ion{O}{4}* is expected to arise in gas with $-2.4 \lesssim \log U \lesssim -2.0$ (from peak fractional abundances), we find that $2.3 \lesssim R \lesssim 6.0$ kpc. Since the velocity structure is consistent with being the same across all ions, this distance is likely to be representative for the AAL region as a whole. Therefore, if all of the gas were photoionized, given the expression for $U$ in \cref{ionization_parameter}, we should expect clouds with differing densities. The ionization parameter is around two orders of magnitude different between the gas hosting the majority of \ion{O}{4} and the gas hosting the most highly ionized species (\ion{Ne}{8}, \ion{Mg}{10}), so this gas should have a characteristic density around two orders of magnitude smaller, i.e. $n_{\textrm{H}} \sim 10~\textrm{cm}^{-3}$. We note that the most highly ionized species may also form at temperatures higher than those typical of photoionization equilibrium. In such a case, densities may be higher than implied by the reasoning presented above.

We can now estimate the absorption path length through the AAL region as $l_{\textrm{abs}} = N_{\textrm{H}} / n_{\textrm{H}}$. Adopting $n_{\textrm{H}} \sim 10^3~\textrm{cm}^{-3}$ through the least ionized gas ($10^{17} \lesssim N_{\textrm{H}} \lesssim 10^{18.5}~\textrm{cm}^{-2}$) and $n_{\textrm{H}} \sim 10~\textrm{cm}^{-3}$ through the most highly ionization gas we detect ($10^{18.5} \lesssim N_{\textrm{H}} \lesssim 10^{20}~\textrm{cm}^{-2}$), we derive characteristic absorption path lengths of $10^{-4.5} \lesssim l_{\textrm{abs}} \lesssim 10^{-3}~\textrm{pc}$ and $0.1 \lesssim l_{\textrm{abs}} \lesssim 1~\textrm{pc}$ respectively in each velocity component. Note that the absorption path length is even smaller for velocity component $v_8$. If we assume that the absorbing gas completely fills the volume it encompasses, then these path lengths are also representative of the cloud sizes along the line-of-sight. If instead there are many clouds contributing to the total column density in each component, then the cloud sizes will be \emph{smaller}, and the volume they encompass \emph{larger} than the absorption path length. Additionally, based on indications in \Cref{partial_covering_section} that the \ac{qso} continuum is only partially covered by the absorbing clouds, the transverse sizes of these gas clouds are likely to be $l_{\textrm{trans}} \lesssim 10^{-2.5}~\textrm{pc}$ \citep{JimenezVicente:2012es}.

\section{Discussion and conclusions}
\label{discussion}
The main results from the data analysis and photoionization/collisional ionization equilibrium models are as follows:
\begin{enumerate}
    \item{The gas in the least ionized AAL region is predominantly photoionized.}
    \item{Under photoionization equilibrium, multiple ionization parameters are required to reproduce the column density ratios seen in the data.}
    \item{Based on the observed column densities and ionization fractions implied from characteristic ionization parameters, incorporating uncertainties in the shape of the \ac{qso} \ac{sed}, the gas metallicity is conservatively $0 \lesssim [\textrm{O} / \textrm{H}] \lesssim 1$.}
    \item{Given the range in possible gas metallicity, the total hydrogen column density in each velocity component is $10^{17} \lesssim N_{\textrm{H}} \lesssim 10^{18.5}~\textrm{cm}^{-2}$ in the least ionized gas (with slightly smaller values for velocity component $v_8$) and $10^{18.5} \lesssim N_{\textrm{H}} \lesssim 10^{20}~\textrm{cm}^{-2}$ through the most highly ionized gas we detect.}
    \item{Taking the column density ratio between \ion{O}{4}* and \ion{O}{4}, assuming the fine structure excited states are populated mostly due to collisions with electrons, the total hydrogen density in the gas traced by these ions is found to be $\log (n_{\textrm{H}} / \textrm{cm}^{-3}) \sim 3$ for solar metallicity.}
    \item{Given the total hydrogen density, and the plausible range in ionization parameter for the gas traced by \ion{O}{4} and \ion{O}{4}*, the distance to the absorbing clouds from the centre of the \ac{qso} is found to be $2.3 \lesssim R \lesssim 6.0~\textrm{kpc}$. An empirically identified, shared velocity structure amongst all ions, suggests this distance determination is likely to be representative for the AAL region as a whole.}
    \item{Under photoionization equilibrium, the total hydrogen density in the most highly ionized AAL gas is found to be two orders of magnitude lower than that implied for the low ionization gas. Alternatively, models in which this gas is under \ac{cie} allow for densities to be similar across the AAL region probed by these data.}
    \item{The ratio $N_{\textrm{H}} / n_{\textrm{H}}$ sets limits on the absorption pathlength through the least and mostly highly ionized regions of $10^{-4.5} \lesssim l_{\textrm{abs}} \lesssim 10^{-3}~\textrm{pc}$ and $0.1 \lesssim l_{\textrm{abs}} \lesssim 1~\textrm{pc}$ respectively in each velocity component.}
    \item{Covering fractions less than unity (in all cases where they can be reliably measured), suggest that the continuum region is only partially covered, requiring clouds with transverse sizes $l_{\textrm{trans}} \lesssim 10^{-2.5}~\textrm{pc}$.}
\end{enumerate}
In summary, the analysis of the previous sections has revealed the presence of metal enriched (to at least solar), highly ionized gas clouds a few kpc from the centre of Q0209 that are likely to be very small (sub-pc scale). In the following sections we place these results in a wider context, and speculate on the origins and fate of the absorbing gas. For simplicity, we shall speak of two, co-spatial regions in ionization equilibrium: a low ionization, photoionized region with $\log U \lesssim -1$, and a high ionization region with $\log U \gtrsim -1$ if photoionized, or temperatures $T \gtrsim 10^{5.5}~\textrm{K}$ if collisionally ionized.

\subsection{Gas structure and dynamics}
\label{structure_dynamics}
A redshift measurement for Q0209 of $z_{\textrm{QSO}} = 1.13194 \pm 0.00001$ implies that the AAL gas is mostly outflowing from the \ac{qso} with velocities up to $\sim 400$~\kms\ (see \Cref{aals}). This is unusually small, compared to the majority of the \ac{aals} and \ac{bals} in the literature with high ionization species such as \ion{Ne}{8} and \ion{Mg}{10}, which are typically outflowing with velocities closer to a few thousand or few tens of thousand \kms\ \cite[e.g.][]{Hamann:1995ff,Telfer:1998gx,Petitjean:1999vh,Arav:1999ka,Muzahid:2012kl,Muzahid:2013dm}, although see \cite{Hamann:2000bi} for a more similar example. If we assume that the gas is moving with a constant radial velocity $v$ and originates close to the \ac{smbh}, then the time-scale for reaching its current radius $R$ is at least
\begin{equation}
    t \approx 10^7 \left(\frac{R}{2.3~\textrm{kpc}}\right)\left(\frac{200~\textrm{km~s}^{-1}}{v}\right)~\textrm{yrs}.
\label{flow_time}
\end{equation}
Different velocity components in the AAL gas are moving at different speeds, so the overall region should possess an appreciable radial thickness after a time $t$, even though we derive densities and ionization parameters that are consistent with one another across the different velocity components.

Given the small cloud sizes in the low ionization gas, a key question is how long they are expected to survive. The free-fall time-scale for these clouds can approximately be expressed as
\begin{equation}
    t_{\textrm{ff}} \equiv \frac{1}{\sqrt{G \rho}} \sim 1.0 \times 10^{15}~\textrm{s} \left(\frac{n_{\textrm{H}}}{\textrm{cm}^{-3}}\right)^{-1/2},
\end{equation}
where $G$ is the gravitational constant, $\rho$ is the gas density, assuming that all of the mass is baryonic, and setting the mass fraction in helium to 0.28 (assuming solar abundances). In addition, for a characteristic cloud size $l$, the sound crossing time in a highly ionized plasma can be approximated by
\begin{equation}
    t_{\textrm{sc}} \equiv \frac{l}{c_{\textrm{s}}} \sim 2.1 \times 10^{15}~\textrm{s} \left(\frac{l}{\textrm{kpc}}\right) T_4^{-1/2},
\label{sc}
\end{equation}
where $c_{\textrm{s}}$ is the sound speed in an ideal monatomic gas and $T = T_4 \times 10^4~\textrm{K}$ \cite[e.g.][]{Schaye:2001dv}. For a stable cloud in hydrostatic equilibrium, $t_{\textrm{sc}} \sim t_{\textrm{ff}}$. We take a value of $n_{\textrm{H}} = 10^3~\textrm{cm}^{-2}$, and a value of $l = 10^{-6}~\textrm{kpc}$ (assuming $l_\textrm{abs} \approx l$). The photoionization models indicate that $T_4 \approx 2$ in this gas and so we find $t_{\textrm{ff}} \sim 3.2 \times 10^{13}~\textrm{s} \gg t_{\textrm{sc}} \sim 1.5 \times 10^9~\textrm{s}$. This implies that the clouds will expand on the sound crossing time-scale, so they should have lifetimes of $\lesssim 100$~years. This is considerably less than the characteristic flow time in \cref{flow_time}, and so the probability of observing these clouds at their implied distance from the \ac{qso} is extremely small in this case.

The analysis presented above poses a problem, which may be overcome if the clouds are being held in pressure equilibrium. This may be a thermal pressure equilibrium with higher temperature, lower density, more highly ionized gas, equivalent to the statement $n_{\textrm{H}1} T_1 = n_{\textrm{H}2} T_2$, where $n_{\textrm{H}1}$, $T_1$ and $n_{\textrm{H}2}$, $T_2$ are the total hydrogen densities and temperatures of the low and high ionization regions respectively. From \Cref{properties}, under photoionization equilibrium we found that $n_{\textrm{H}2} \sim 10~\textrm{cm}^{-2}$, and these models also indicate that $T_2 \approx 6 \times 10^4~\textrm{K}$. In this case, $n_{\textrm{H}1} T_1 > n_{\textrm{H}2} T_2$, and the high ionization gas cannot pressure support the low ionization gas. If the former is collisionally ionized, we now have temperatures that differ by more than an order of magnitude. Densities in the high ionization region may be low enough to allow for pressure support. Nevertheless, the high ionization gas itself, accounting for the possibility that it is photoionized, should have a lifetime $\lesssim 10^5$ years, which is still short enough to suggest that this gas may also require pressure support from even more highly ionized gas that we do not detect in the UV, and which would require larger total column density, higher temperature and lower density.

Massive galaxies are expected to host hot gas coronae, well within the implied location of the AAL region, with $T \sim 10^6~\textrm{K}$ and $n_{\textrm{H}} \sim 10^{-2}~\textrm{cm}^{-2}$ \cite[e.g.][]{White:1991in,Fukugita:2006dg}. Pressure from this external medium, together with additional pressure support from magnetic confinement \citep{deKool:1995gu} may help to alleviate the problems outlined above, although pressure supporting gas with varying internal pressure is clearly a complex issue. We note that the analysis above does not incorporate the effects of turbulence, which is almost certainly present given the observed line widths (see \Cref{column_densities}). In addition, gas outflowing from a \ac{qso} will likely encounter the \ac{ism} of the host galaxy on its journey out into the halo. At supersonic velocities, shocks will likely occur at the interface between the outflowing gas and the \ac{ism}, heating the gas close to this interface. The resulting mix of hot and cool gas creates instabilities that can destroy the clouds before they reach the halo \cite[see for example arguments in][]{FaucherGiguere:2012jk}. These authors suggest an alternative scenario, in which small clouds may be formed in situ from moderately dense \ac{ism} clouds within hot, recently shocked gas. These clouds become shredded by a passing blast wave and gain momentum from an accompanying shock. The resulting `cloudlets' in this model have sizes and densities comparable to those derived here, and can possess a range of velocities that may explain the multi-component velocity structure in the absorption profiles.

Models such as these may offer a more promising route to explain the structure and dynamics of \ac{aals} with properties (density, cloud size, velocity structure, distance from the \ac{qso}) similar to those found in Q0209 \cite[e.g.][]{Petitjean:1999vh,Hamann:2000bi,Edmonds:2011fz,Borguet:2012ca,Arav:2013be,Muzahid:2013dm}. Any viable model must additionally reproduce the covering fractions seen in the present data. Covering fractions less than unity, and with little variation, are seen in ions spanning a range in ionization potential from a few tens to a few hundreds of eV, tracing gas with more than one possible ionization mechanism. Vastly differing absorption path lengths through the AAL region, as hinted at in the analysis of the previous sections, make it very difficult to account for the near constancy in covering fraction across all ions using simple geometrical models. We also note that these results differ from e.g. \citet{Hamann:2000bi} and \citet{Borguet:2012ca}, who find more complete coverage in high-ionization \ac{uv} transitions compared to those at lower ionization potentials. Covering fractions less than unity across our sample also go against general trends for more complete coverage with lower outflow velocities, as identified in the \ac{cos} sample presented by \citet{Muzahid:2013dm}. It is therefore clear that simple trends such as these may not produce robust predictions for individual systems, which further highlights the apparent complexity in these absorbers.

\subsection{Are the AAL clouds out of equilibrium?}
\label{equilibrium}
Up to this point, our analysis and discussion has assumed that the AAL clouds are in ionization equilibrium. However, in general, absorbers may be out of equilibrium when close to an AGN, due to recombination time-scales that can be long compared to typical AGN lifetimes and duty cycles \cite[e.g.][]{1995ApJ...447..512K,Nicastro:1999fj,Arav:2012gv,Oppenheimer:2013cr,Oppenheimer:2013dx}. The resulting recombination lag can lead to situations where high ionization stages like \ion{O}{6}, \ion{Ne}{8} and \ion{Mg}{10} are enhanced relative to the expectation from equilibrium models. We examine these issues here.

We define the the photoionization rate $\Gamma_{M_i}$ (s$^{-1}$) for a given ion $M_i$ as
\begin{equation}
    \Gamma_{M_i} \equiv \int_{\nu_{0, M_i}}^{\infty} \frac{4 \pi J_{\nu}}{h \nu} \sigma_{M_i} (\nu)~\textrm{d}\nu.
\label{photoionization_rate}
\end{equation}
Here $\nu$ is the frequency, $\nu_{0, M_i}$ is the ionization frequency, $J_{\nu}$ is the intensity of the \ac{qso} radiation field (erg s$^{-1}$ cm$^{-2}$ Hz$^{-1}$ sr$^{-1}$), $\sigma_{M_i}$ is the photoionization cross-section, and $h$ is Planck's constant. The recombination rate (s$^{-1}$) into an ion $M_i$ is given by
\begin{equation}
    R_{M_i} \equiv \alpha_{M_i} n_e,
\end{equation}
where $n_e$ is the electron number density, and $\alpha_{M_i}$ is the temperature dependent recombination rate coefficient (cm$^3$ s$^{-1}$) for that ion. Finally, the collisional ionization rate for an ion $M_i$ is
\begin{equation}
    C_{M_i} \equiv \beta_{M_i} n_e,
\end{equation}
where $\beta_{M_i}$ is the temperature dependent collisional ionization rate coefficient (cm$^3$~s$^{-1}$). Neglecting Auger ionization and charge transfer, the population in an ion $M_i$ is then
\begin{align}
    \frac{\textrm{d}n_{M_i}}{\textrm{d}t} &= -n_{M_i} (\Gamma_{M_i} + R_{M_{i - 1}} + C_{M_i}) + n_{M_{i + 1}} R_{M_i} \nonumber \\ &\qquad {} + n_{M_{i - 1}} (\Gamma_{M_{i - 1}} + C_{M_{i - 1}}).
\label{time_dependent_ionization}
\end{align}

Now suppose that an absorber is in photoionization equilibrium, i.e. $\textrm{d}n_{M_i} / \textrm{d}t = 0$, at time $t = 0$, but there is a sudden change in the ionizing flux, such that $\Gamma_{M_i}(t > 0) = (1 + \delta) \Gamma_{M_i}(t = 0)$, where $-1 \leq \delta \leq \infty$. Taking the collisional ionisation rate to be negligible (a reasonable approximation for a photoionized plasma), it can then be shown that the $e$-folding time-scale for change in the ionic fraction is given by
\begin{equation}
    t_{\textrm{change}} = \left[ -\delta \alpha_{M_i} n_e \left( \frac{n_{M_{i + 1}}}{n_{M_i}} - \frac{\alpha_{M_{i - 1}}}{\alpha_{M_i}} \right) \right]^{-1},
\label{time-scale}
\end{equation}
\citep{Arav:2012gv}, where negative time-scales indicate a decrease in the ionic fraction, and positive time-scales indicate an increase. For changes in the ionizing flux within an order of magnitude ($0.1 < 1 + \delta < 10$), these time-scales are typically  $\sim 10$ years for the densities implied by the \ion{O}{4}* analysis of \Cref{density}, assuming $T \sim 10^4$~K. Since the gas densities in the AAL region are much higher than those typical of the diffuse \ac{igm} and \ac{cgm}, these time-scales are much shorter than the typical AGN lifetime \cite[$\sim~\textrm{Myr}$ time-scales; e.g.][]{Novak:2011gc}. Therefore, photoionization equilibrium might be a valid assumption in our case, so long as these time-scales are also short compared to the time-scale over which the \ac{qso} luminosity changes.

However, we must also consider the dynamical evolution of the AAL clouds. In the previous section, we found that, unless the clouds traced by \ion{O}{4} are pressure supported by an external medium, they will expand on time-scales $\lesssim 100$ years. If this process is occurring, then the clouds may be entering a non-equilibrium state due to a recombination lag. To determine whether or not this scenario is likely, we numerically solve the coupled, time-dependent ionization equations (\cref{time_dependent_ionization}) for a set of elements using a 4th order Runge-Kutta method. We assume that the AAL region possesses a gas density of $n_{\textrm{H}} = 10^3~\textrm{cm}^{-3}$ and is illuminated by the `UV peak' SED at a distance of 2.3~kpc. Assuming a gas temperature $T = 10^4$~K, we then calculate recombination rate coefficients using the \citet{2006ApJS..167..334B} fits (assuming case A recombination), and collisional ionization rate coefficients using the data in \cite{1997ADNDT..65....1V}. We perform the integral in \cref{photoionization_rate} using the same photoionization cross-sections as used in \cloudy\ c13.00. Equilibrium values of $n_{M_i}$ are then calculated, assuming solar abundances. We assume these values to hold at a time $t = 0$. We next perturb the gas density such that $n(t > 0) = (1 + \delta) n(t = 0)$, where $-1 \leq \delta \leq \infty$, identical to the case of flux changes considered above. We do this for four different values of $(1 + \delta) = 0.1$, 0.01, 0.001 and 10. The latter is included for the interest of comparing both increasing and decreasing density. For example, if the AAL clouds are the result of shocked ISM clouds, we might expect them to be crushed prior to their subsequent expansion \cite[e.g.][]{FaucherGiguere:2012jk}. The resulting time-dependent evolution in the number density of \ion{H}{1}, \ion{O}{4}, \ion{O}{5}, \ion{O}{6}, \ion{Ne}{8} and \ion{Mg}{10} is shown in \Cref{non_equilibrium_figure} in \Cref{time_dependent_ionization_section}.

Simple inspection of \Cref{non_equilibrium_figure} indicates that, for a decrease of more than an order of magnitude in gas density, with the exception of \ion{H}{1}, time-scales for reaching a new equilibrium are $> 100$~years. Comparing this to the expansion time-scale, we conclude that non-equilibrium effects are important if the density in the AAL region is dropping by orders of magnitude. Time-scales are orders of magnitude shorter for an increase in density, and so we expect that a decrease in density should form the dominant contribution to any non-equilibrium behaviour. We note that the situation described is not physical -- we expect a smooth change in density with time, not a step-function change. Nevertheless, lacking a physical motivation for the exact functional form, we present these results as an approximation. The non-equilibrium behaviour shown in \Cref{non_equilibrium_figure} indicates considerable variation in the rate of change of ionic number density, with the more highly ionized species changing more rapidly. Under photoionization equilibrium, the AAL region was found to possess a range in density and absorption path-length covering orders of magnitude. This situation was required due to the co-spatial existence of ionic species tracing gas with multiple ionization parameters. If ionic number densities are changing at highly variable rates, then is possible to envisage scenarios whereby the fractional abundances of ions spanning a large range in ionization potential can all be high. We therefore speculate that it may be possible to find non-equilibrium models that reproduce all of the observed column densities in a single phase, with a single density and absorption path length. Such a scenario may be desirable, as it is in better concordance with the near constant covering fraction seen across these ions.

It is important to note that the calculations leading to inferred cloud lifetimes of $\lesssim 100$~years assume that the cloud sizes $l$ are approximately equal to the absorption path-length $l_{\textrm{abs}}$. However, in general, $l < l_{\textrm{abs}}$, which would make the cloud lifetimes even shorter. Scenarios where $l < l_{\textrm{abs}}$ have been put forward multiple times in the literature, typically to explain situations where the derived covering fractions in the data depend on velocity across the absorption profiles, and the ionization and/or true optical depth in the lines. This situation is referred to as inhomogenous partial coverage \cite[e.g.][]{deKool:2002by,Hamann:2004tu,Arav:2008fa}, where there are many small clouds having a power-law dependence in optical depth across their transverse extent. Covering fractions that vary in the way just described are not found in our data. Nevertheless, we cannot rule out the possibility that there are multiple small clouds along the line-of-sight. In such a scenario, the case for non-equilibrium evolution in the ionic number densities becomes more compelling.

In summary, if the AAL region is not pressure supported, then explicit numerical calculations suggest that non-equilibrium effects may be important in these clouds. A cloud expansion time-scale of $\lesssim 100$ years is sufficiently short, that these effects might be confirmed with repeat observations. These observations will be crucial in determining appropriate non-equilibrium models for the data.

\subsection{The connection to associated X-ray absorption}
We next consider the link between associated \ac{uv} absorption and so-called `warm absorbers', often characterised by both bound-bound and bound-free absorption in X-rays. A key question is whether or not this absorption is predicted by the \ac{uv} absorption lines characterised here. In the most highly ionized gas, under photoionization equilibrium, the maximum predicted total hydrogen column density is $\log (N_{\textrm{H}} / \textrm{cm}^{-2}) \approx 20$, and the ionization parameter $\log U \approx 0.5$. We assume an upper limit on the gas metallicity of $[\textrm{O} / \textrm{H}] \approx 1$. Explicit photoionization calculations using these parameters predict the presence of bound-bound transitions, but no significant bound-free absorption in X-rays. The same result is obtained in \ac{cie} calculations. In this respect, and also in terms of their relatively small velocity shift from the \ac{qso}, the AALs in Q0209 are similar to those reported in e.g. \citet{Hamann:2000bi}. If there is continuous X-ray absorption, it should be in much more highly ionized gas with larger total column density. This gas may trace the bulk of an outflow that produces the gas condensations described in the previous section. However, in general, the gas giving rise to warm absorbers need not be co-spatial with \ac{uv} and/or optical \ac{aals}, especially since these absorption systems arise in gas with a wide range of physical parameters \cite[outflow velocities, ionization, covering fractions, distance from the \ac{qso} etc.; see for example][]{Ganguly:1999ia,Misawa:2007dd,Nestor:2008gb,Ganguly:2013hd,Muzahid:2013dm,Sharma:2013hk}. It is intriguing nonetheless, that the \ac{aals} in Q0209 show nearly identical velocity component structure over $\sim 300$~eV in ionization potential, suggesting that absorption lines from many ionization stages can indeed arise co-spatially. Simultaneous observations in the UV and X-rays will likely be required to gain deeper insights into the connection between warm absorbers and associated absorption lines in general \cite[e.g.][]{DiGesu:2013cf,Lee:2013jj}.

\subsection{Outflow models}
Before considering potential origins for the outflowing gas, we first perform a rough estimate of the mass and kinetic energy in the \ac{aals}. We assume the geometry of the outflowing gas traced by these data to be that of a thin, partially filled shell of material moving radially outwards from the centre of the \ac{qso}, the flux from which is modelled by the `UV peak' SED. Under this geometry, the mass depends on the distance from the \ac{qso} ($R \sim 2.3$~kpc), the total column density through the AAL region (we find a value $N_{\textrm{H}} \approx 2 \times 10^{20}~\textrm{cm}^{-2}$), and crucially, the global covering fraction, $\Omega$, of the AAL gas, as opposed to the line-of-sight covering fraction that we measure. A rough estimate of this quantity comes from the incidence rate of associated absorption systems like that seen in Q0209. \citet{Muzahid:2013dm} presented a sample of 20 quasar spectra observed with \ac{cos}, from which associated absorbers were selected based on the presence of \ion{Ne}{8} absorption. The incidence rate of these absorption systems was found to be 40\%, and we consider this to be the closest representative sample in the literature at present, although the general conclusions below are not sensitive to this value. We therefore express the total gas mass in the \ac{uv} AAL region as
\begin{equation}
    M \approx 6 \times 10^7 \left(\frac{\Omega}{0.4}\right) \left(\frac{N_{\textrm{H}}}{2 \times 10^{20}~\textrm{cm}^{-2}}\right) \left(\frac{R}{2.3~\textrm{kpc}}\right)^2 M_{\odot},
\end{equation}
where we have assumed a mean molecular weight per proton of $\mu_{\textrm{H}} = 1.4$. The total kinetic energy in this gas is then
\begin{equation}
    K \approx 2 \times 10^{55} \left(\frac{M}{6 \times 10^7M_{\odot}}\right) \left(\frac{v}{200~\textrm{km s}^{-1}}\right)^2~\textrm{erg}.
\end{equation}
We can derive the average mass outflow rate $\dot{M}$, by dividing $M$ by the dynamical time-scale $R / v$, and subsequently derive the kinetic luminosity of the gas as $\dot{K} = 0.5\dot{M}v^2$. This gives values of $\dot{M} \approx 5 M_{\odot}~\textrm{yr}^{-1}$ and $\dot{K} \approx 7 \times 10^{40}~\textrm{erg s}^{-1}$. It is instructive to bear in mind that, while these quantities are useful, there are good reasons to believe that the gas clouds traced by the \ac{aals} may not travel a distance $R$, and are instead accelerated close to their observed location (see \Cref{structure_dynamics}).

We consider two primary sources for the gas flow generating the \ac{aals}: (i) line-driven winds, and (ii) supernova-driven winds. Line-driven winds, initially accelerated through radiation pressure from the \ac{smbh} accretion disc, are commonly invoked to explain the winds traced by \ac{bals} and \ac{aals} with velocities of a few 1000 \kms, and are a major source of energy injection into the \ac{ism} in models of AGN feedback \cite[e.g.][]{Scannapieco:2004es,DiMatteo:2005hl,Hopkins:2010cf}. Specifically, these models require kinetic luminosities to be $\dot{K} \gtrsim 0.1$\% of the Eddington luminosity, $L_{\textrm{Edd}}$. For Q0209, $\log(L / L_{\textrm{Edd}}) \gtrsim 0$ (Done et al., in prep), which implies the kinetic luminosity in the AALs is at least two orders of magnitude below this level. In addition, models involving line-driven winds suggest they must be launched close to the \ac{smbh} at velocities of a few 100 \kms\ \cite[e.g.][]{Risaliti:2010jh}, which is already the velocity of the \ac{aals} seen here at much larger distances. If the \ac{aals} are pressure confined in a line driven wind such as this, they must encounter significant drag from a surrounding medium to slow them down, or keep them from accelerating to much larger velocities. In the more likely case that the \ac{aals} are formed in situ, a variety of velocities could in principle be observed. For example, in the radiative shock model of \cite{FaucherGiguere:2012jk}, AAL clouds will take a finite time to accelerate up to the velocity of the passing blast wave (see their equation (12)). Although the \ac{uv} AAL clouds in Q0209 contribute only a small percentage of the kinetic luminosity required for significant \ac{agn} feedback into the surrounding \ac{ism} (and \ac{igm}), a much larger percentage may be carried by an associated, much more highly ionized warm absorber, with higher total column density, detectable as bound-free absorption in X-rays \cite[e.g.][]{Crenshaw:2003hz,Gabel:2005jz,Arav:2007dx}. We note that bound-free X-ray absorption is by no means ubiquitous, with some warm absorbers now being detected via absorption \emph{lines} such as \ion{O}{7}. These can have $N_{\textrm{H}}$ consistent with that seen in associated UV absorption lines \cite[e.g.][]{DiGesu:2013cf}.

Supernova driven winds are thought to drive fountains of gas a few kpc into the halos of massive galaxies, some of which is then expected to fall back on ballistic trajectories \cite[e.g.][]{Bregman:1980gn,Fraternali:2006eo,Fraternali:2008id,Marinacci:2010ho}. We find that the distance, velocity and density of the AAL gas in Q0209 is typical of galactic winds \cite[e.g.][]{Veilleux:2005ec,Creasey:2013gu}. The infalling velocity component $v_8$ also indicates that some of the AAL gas may be on a return trajectory back towards the disc of the \ac{qso} host. The time-scale derived in \cref{flow_time} is consistent with expected \ac{qso} lifetimes \cite[e.g.][]{Novak:2011gc}, so if the clouds are a result of supernova driven winds, it is possible these winds were launched during a starburst phase. If we assume that the starburst proceeded at $\sim 10 M_{\odot}~\textrm{yr}^{-1}$ and that this star formation results in one supernova per $100 M_{\odot}$ with energy $E_{\textrm{SN}} \sim 10^{51}~\textrm{erg}$, then $\dot{E} \sim 10^{42}~\textrm{erg s}^{-1}$. Only a fraction of this energy will be converted into the kinetic energy powering an outflow, and so the kinetic luminosity we derive for the \ac{aals} in Q0209 may be consistent with this simple estimation.

At present, the data are consistent with both an outflow driven by AGN and starburst activity in Q0209. Future X-ray observations of this \ac{qso} may help to distinguish between these two possibilities. In particular, associated bound-free X-ray absorption tracing gas with high total column density and kinetic luminosity would favour an origin closely tied with the \ac{agn}.

\section*{Acknowledgments}
We thank Joop Schaye, Gabriel Altay and Ewan Hemingway for helpful discussions and comments, and the anonymous referee, whose suggestions improved this paper.

C.W.F would like to acknowledge the support of an STFC studentship (ST/J201013/1). F.H. acknowledges support from the USA National Science Foundation grant AST-1009628. T.T. acknowledges support from the Interuniversity Attraction Poles Programme initiated by the Belgian Science Policy Office ([AP P7/08 CHARM]). Support for M.F. was provided by NASA through Hubble Fellowship grant HF-51305.01-A.

We thank the contributors to SciPy\footnote{\url{http://www.scipy.org}}, Matplotlib\footnote{\url{http://matplotlib.org}} and the Python programming language\footnote{\url{http://www.python.org}}, the free and open-source community, and the NASA Astrophysics Data system\footnote{\url{http://adswww.harvard.edu}} for software and services. This work also made use of Astropy; a community-developed core Python package for Astronomy \citep{2013A&A...558A..33A}.

The work in this paper was mainly based on observations made with the NASA/ESA Hubble Space Telescope under program GO 12264, obtained at the Space Telescope Science Institute, which is operated by the Association of Universities for Research in Astronomy Inc., under NASA contract NAS 5-26555; and with the 6.5m Magellan Telescopes located at Las Campanas Observatory, Chile.

This work also made use of the DiRAC Data Centric system at Durham University, operated by the Institute for Computational Cosmology on behalf of the STFC DiRAC HPC Facility\footnote{\url{http://www.dirac.ac.uk}}. This equipment was funded by BIS National E-infrastructure capital grant ST/K00042X/1, STFC capital grant ST/H008519/1, and STFC DiRAC Operations grant ST/K003267/1 and Durham University. DiRAC is part of the National E-Infrastructure.

The raw data from \ac{hst}/\ac{cos} may be accessed from the MAST archive\footnote{\url{http://archive.stsci.edu}}. Data from Magellan/FIRE are available from the lead author upon request.

\bibliographystyle{mn2e}
\bibliography{q0209_AAL_paper}

\begin{thebibliography}{}

\bibitem[\protect\citeauthoryear{Aguirre, Schaye \& Theuns}{Aguirre
  et~al.}{2002}]{Aguirre:2002is}
Aguirre A.,  Schaye J.,    Theuns T.,  2002, ApJ, 576, 1

\bibitem[\protect\citeauthoryear{Ahn et~al.,}{Ahn et~al.}{2013}]{Ahn:2013wx}
Ahn C.~P.  et~al., 2013, arXiv

\bibitem[\protect\citeauthoryear{Altay \& Theuns}{Altay \&
  Theuns}{2013}]{Altay:2013dua}
Altay G.,  Theuns T.,  2013, Monthly Notices of the Royal Astronomical Society,
  Volume 434, Issue 1, p.748-764, 434, 748

\bibitem[\protect\citeauthoryear{Arav, Borguet, Chamberlain, Edmonds \&
  Danforth}{Arav et~al.}{2013}]{Arav:2013be}
Arav N.,  Borguet B.,  Chamberlain C.,  Edmonds D.,    Danforth C.~W.,  2013,
  MNRAS, p.~2561

\bibitem[\protect\citeauthoryear{Arav et~al.,}{Arav et~al.}{2012}]{Arav:2012gv}
Arav N.  et~al., 2012, A{\&}A, 544, 33

\bibitem[\protect\citeauthoryear{Arav et~al.,}{Arav et~al.}{2007}]{Arav:2007dx}
Arav N.  et~al., 2007, ApJ, 658, 829

\bibitem[\protect\citeauthoryear{Arav, Korista, de Kool, Junkkarinen \&
  Begelman}{Arav et~al.}{1999}]{Arav:1999ka}
Arav N.,  Korista K.~T.,  de Kool M.,  Junkkarinen V.~T.,    Begelman M.~C.,
  1999, ApJ, 516, 27

\bibitem[\protect\citeauthoryear{Arav, Moe, Costantini, Korista, Benn \&
  Ellison}{Arav et~al.}{2008}]{Arav:2008fa}
Arav N.,  Moe M.,  Costantini E.,  Korista K.~T.,  Benn C.,    Ellison S.,
  2008, ApJ, 681, 954

\bibitem[\protect\citeauthoryear{{Astropy Collaboration} et~al.,}{{Astropy
  Collaboration} et~al.}{2013}]{2013A&A...558A..33A}
{Astropy Collaboration} et~al., 2013, A{\&}A, 558, 33

\bibitem[\protect\citeauthoryear{Badnell}{Badnell}{2006}]{2006ApJS..167..334B}
Badnell N.~R.,  2006, ApJS, 167, 334

\bibitem[\protect\citeauthoryear{Bahcall \& Wolf}{Bahcall \&
  Wolf}{1968}]{Bahcall:1968ig}
Bahcall J.~N.,  Wolf R.~A.,  1968, ApJ, 152, 701

\bibitem[\protect\citeauthoryear{Balashev, Petitjean, Ivanchik, Ledoux,
  Srianand, Noterdaeme \& Varshalovich}{Balashev
  et~al.}{2011}]{Balashev:2011cc}
Balashev S.~A.,  Petitjean P.,  Ivanchik A.~V.,  Ledoux C.,  Srianand R.,
  Noterdaeme P.,    Varshalovich D.~A.,  2011, MNRAS, 418, 357

\bibitem[\protect\citeauthoryear{Barlow \& Sargent}{Barlow \&
  Sargent}{1997}]{Barlow:1997eb}
Barlow T.~A.,  Sargent W. L.~W.,  1997, AJ, 113, 136

\bibitem[\protect\citeauthoryear{Benson, Bower, Frenk, Lacey, Baugh \&
  Cole}{Benson et~al.}{2003}]{Benson:2003ch}
Benson A.~J.,  Bower R.~G.,  Frenk C.~S.,  Lacey C.~G.,  Baugh C.~M.,    Cole
  S.,  2003, ApJ, 599, 38

\bibitem[\protect\citeauthoryear{Borguet, Edmonds, Arav, Dunn \& Kriss}{Borguet
  et~al.}{2012}]{Borguet:2012ca}
Borguet B. C.~J.,  Edmonds D.,  Arav N.,  Dunn J.,    Kriss G.~A.,  2012, ApJ,
  751, 107

\bibitem[\protect\citeauthoryear{Bower, Benson, Malbon, Helly, Frenk, Baugh,
  Cole \& Lacey}{Bower et~al.}{2006}]{Bower:2006fj}
Bower R.~G.,  Benson A.~J.,  Malbon R.,  Helly J.~C.,  Frenk C.~S.,  Baugh
  C.~M.,  Cole S.,    Lacey C.~G.,  2006, MNRAS, 370, 645

\bibitem[\protect\citeauthoryear{Brandt, Laor \& Wills}{Brandt
  et~al.}{2000}]{Brandt:2000de}
Brandt W.~N.,  Laor A.,    Wills B.~J.,  2000, ApJ, 528, 637

\bibitem[\protect\citeauthoryear{Bregman}{Bregman}{1980}]{Bregman:1980gn}
Bregman J.~N.,  1980, ApJ, 236, 577

\bibitem[\protect\citeauthoryear{Cardelli, Clayton \& Mathis}{Cardelli
  et~al.}{1989}]{Cardelli:1989dp}
Cardelli J.~A.,  Clayton G.~C.,    Mathis J.~S.,  1989, ApJ, 345, 245

\bibitem[\protect\citeauthoryear{Carswell, Whelan, Smith, Boksenberg \&
  Tytler}{Carswell et~al.}{1982}]{Carswell:1982uz}
Carswell R.~F.,  Whelan J. A.~J.,  Smith M.~G.,  Boksenberg A.,    Tytler D.,
  1982, MNRAS, 198, 91

\bibitem[\protect\citeauthoryear{Creasey, Theuns \& Bower}{Creasey
  et~al.}{2013}]{Creasey:2013gu}
Creasey P.,  Theuns T.,    Bower R.~G.,  2013, MNRAS, 429, 1922

\bibitem[\protect\citeauthoryear{Crenshaw, Kraemer \& George}{Crenshaw
  et~al.}{2003}]{Crenshaw:2003hz}
Crenshaw D.~M.,  Kraemer S.~B.,    George I.~M.,  2003, ARA{\&}A, 41, 117

\bibitem[\protect\citeauthoryear{Cuillandre et~al.,}{Cuillandre
  et~al.}{2012}]{Cuillandre:2012jv}
Cuillandre J.-C.~J.  et~al., 2012, in Observatory Operations: Strategies,
  Processes and Systems IV.

\bibitem[\protect\citeauthoryear{Cushing, Vacca \& Rayner}{Cushing
  et~al.}{2004}]{Cushing:2004bq}
Cushing M.~C.,  Vacca W.~D.,    Rayner J.~T.,  2004, PASP, 116, 362

\bibitem[\protect\citeauthoryear{de Kool \& Begelman}{de~Kool \&
  Begelman}{1995}]{deKool:1995gu}
de Kool M.,  Begelman M.~C.,  1995, ApJ, 455, 448

\bibitem[\protect\citeauthoryear{de Kool, Korista \& Arav}{de~Kool
  et~al.}{2002}]{deKool:2002by}
de Kool M.,  Korista K.~T.,    Arav N.,  2002, ApJ, 580, 54

\bibitem[\protect\citeauthoryear{Di~Gesu et~al.,}{Di~Gesu
  et~al.}{2013}]{DiGesu:2013cf}
Di~Gesu L.  et~al., 2013, A{\&}A, 556, 94

\bibitem[\protect\citeauthoryear{Di~Matteo, Springel \& Hernquist}{Di~Matteo
  et~al.}{2005}]{DiMatteo:2005hl}
Di~Matteo T.,  Springel V.,    Hernquist L.,  2005, Nature, 433, 604

\bibitem[\protect\citeauthoryear{Dickey \& Lockman}{Dickey \&
  Lockman}{1990}]{Dickey:1990df}
Dickey J.~M.,  Lockman F.~J.,  1990, ARA{\&}A, 28, 215

\bibitem[\protect\citeauthoryear{D'Odorico, Cristiani, Romano, Granato \&
  Danese}{D'Odorico et~al.}{2004}]{DOdorico:2004jd}
D'Odorico V.,  Cristiani S.,  Romano D.,  Granato G.~L.,    Danese L.,  2004,
  MNRAS, 351, 976

\bibitem[\protect\citeauthoryear{Done, Davis, Jin, Blaes \& Ward}{Done
  et~al.}{2012}]{Done:2012eq}
Done C.,  Davis S.~W.,  Jin C.,  Blaes O.,    Ward M.,  2012, MNRAS, 420, 1848

\bibitem[\protect\citeauthoryear{Draine}{Draine}{2011}]{Draine:2011tr}
Draine B.~T.,  2011, {Physics of the Interstellar and Intergalactic Medium}.
Princeton Series in Astrophysics, Princeton University Press

\bibitem[\protect\citeauthoryear{Edmonds et~al.,}{Edmonds
  et~al.}{2011}]{Edmonds:2011fz}
Edmonds D.  et~al., 2011, ApJ, 739, 7

\bibitem[\protect\citeauthoryear{Faucher-Gigu{\`e}re, Quataert \&
  Murray}{Faucher-Gigu{\`e}re et~al.}{2012}]{FaucherGiguere:2012jk}
Faucher-Gigu{\`e}re C.-A.,  Quataert E.,    Murray N.,  2012, MNRAS, 420, 1347

\bibitem[\protect\citeauthoryear{Ferland et~al.,}{Ferland
  et~al.}{2013}]{Ferland:2013wla}
Ferland G.~J.  et~al., 2013, Rev. Mex. Astron. Astrofis., 49, 137

\bibitem[\protect\citeauthoryear{Foltz, Weymann, Peterson, Sun, Malkan \&
  Chaffee}{Foltz et~al.}{1986}]{Foltz:1986fg}
Foltz C.~B.,  Weymann R.~J.,  Peterson B.~M.,  Sun L.,  Malkan M.~A.,
  Chaffee F. H.~J.,  1986, ApJ, 307, 504

\bibitem[\protect\citeauthoryear{Fraternali \& Binney}{Fraternali \&
  Binney}{2006}]{Fraternali:2006eo}
Fraternali F.,  Binney J.~J.,  2006, MNRAS, 366, 449

\bibitem[\protect\citeauthoryear{Fraternali \& Binney}{Fraternali \&
  Binney}{2008}]{Fraternali:2008id}
Fraternali F.,  Binney J.~J.,  2008, MNRAS, 386, 935

\bibitem[\protect\citeauthoryear{Fraternali, Marasco, Marinacci \&
  Binney}{Fraternali et~al.}{2013}]{Fraternali:2013dd}
Fraternali F.,  Marasco A.,  Marinacci F.,    Binney J.~J.,  2013, ApJ, 764,
  L21

\bibitem[\protect\citeauthoryear{Fukugita \& Peebles}{Fukugita \&
  Peebles}{2006}]{Fukugita:2006dg}
Fukugita M.,  Peebles P. J.~E.,  2006, ApJ, 639, 590

\bibitem[\protect\citeauthoryear{Fumagalli, O'Meara, Prochaska \&
  Worseck}{Fumagalli et~al.}{2013}]{Fumagalli:2013ee}
Fumagalli M.,  O'Meara J.~M.,  Prochaska J.~X.,    Worseck G.,  2013, ApJ, 775,
  78

\bibitem[\protect\citeauthoryear{Gabel et~al.,}{Gabel
  et~al.}{2005}]{Gabel:2005jz}
Gabel J.~R.  et~al., 2005, ApJ, 623, 85

\bibitem[\protect\citeauthoryear{Gabel, Arav \& Kim}{Gabel
  et~al.}{2006}]{Gabel:2006fw}
Gabel J.~R.,  Arav N.,    Kim T.~S.,  2006, ApJ, 646, 742

\bibitem[\protect\citeauthoryear{Ganguly, Eracleous, Charlton \&
  Churchill}{Ganguly et~al.}{1999}]{Ganguly:1999ia}
Ganguly R.,  Eracleous M.,  Charlton J.~C.,    Churchill C.~W.,  1999, AJ, 117,
  2594

\bibitem[\protect\citeauthoryear{Ganguly et~al.,}{Ganguly
  et~al.}{2013}]{Ganguly:2013hd}
Ganguly R.  et~al., 2013, MNRAS, p.~2039

\bibitem[\protect\citeauthoryear{Gehrels}{Gehrels}{1986}]{Gehrels:1986cx}
Gehrels N.,  1986, ApJ, 303, 336

\bibitem[\protect\citeauthoryear{Ghavamian et~al.,}{Ghavamian
  et~al.}{2009}]{Ghavamian:2009tr}
Ghavamian P.  et~al., 2009, STScI ISR COS 2009-01

\bibitem[\protect\citeauthoryear{Gorenstein}{Gorenstein}{1975}]{Gorenstein:1975jp}
Gorenstein P.,  1975, ApJ, 198, 95

\bibitem[\protect\citeauthoryear{Green et~al.,}{Green
  et~al.}{2012}]{Green:2012dj}
Green J.~C.  et~al., 2012, ApJ, 744, 60

\bibitem[\protect\citeauthoryear{Hall, Anosov, White, Brandt, Gregg, Gibson,
  Becker \& Schneider}{Hall et~al.}{2011}]{Hall:2011ej}
Hall P.~B.,  Anosov K.,  White R.~L.,  Brandt W.~N.,  Gregg M.~D.,  Gibson
  R.~R.,  Becker R.~H.,    Schneider D.~P.,  2011, MNRAS, 411, 2653

\bibitem[\protect\citeauthoryear{Hamann, Chartas, McGraw, Rodriguez~Hidalgo,
  Shields, Capellupo, Charlton \& Eracleous}{Hamann
  et~al.}{2013}]{Hamann:2013cq}
Hamann F.,  Chartas G.,  McGraw S.,  Rodriguez~Hidalgo P.,  Shields J.,
  Capellupo D.,  Charlton J.,    Eracleous M.,  2013, MNRAS, 435, 133

\bibitem[\protect\citeauthoryear{Hamann \& Sabra}{Hamann \&
  Sabra}{2004}]{Hamann:2004tu}
Hamann F.,  Sabra B.,  2004, in AGN Physics with the Sloan Digital Sky Survey.
  p.~203

\bibitem[\protect\citeauthoryear{Hamann}{Hamann}{1997}]{Hamann:1997iu}
Hamann F.~W.,  1997, ApJS, 109, 279

\bibitem[\protect\citeauthoryear{Hamann, Barlow, Beaver, Burbidge, Cohen,
  Junkkarinen \& Lyons}{Hamann et~al.}{1995}]{Hamann:1995ff}
Hamann F.~W.,  Barlow T.~A.,  Beaver E.~A.,  Burbidge E.~M.,  Cohen R.~D.,
  Junkkarinen V.,    Lyons R.,  1995, ApJ, 443, 606

\bibitem[\protect\citeauthoryear{Hamann, Barlow, Chaffee, Foltz \&
  Weymann}{Hamann et~al.}{2001}]{Hamann:2001eg}
Hamann F.~W.,  Barlow T.~A.,  Chaffee F.~C.,  Foltz C.~B.,    Weymann R.~J.,
  2001, ApJ, 550, 142

\bibitem[\protect\citeauthoryear{Hamann \& Ferland}{Hamann \&
  Ferland}{1993}]{Hamann:1993jb}
Hamann F.~W.,  Ferland G.,  1993, ApJ, 418, 11

\bibitem[\protect\citeauthoryear{Hamann \& Ferland}{Hamann \&
  Ferland}{1999}]{Hamann:1999ky}
Hamann F.~W.,  Ferland G.,  1999, ARA{\&}A, 37, 487

\bibitem[\protect\citeauthoryear{Hamann, Netzer \& Shields}{Hamann
  et~al.}{2000}]{Hamann:2000bi}
Hamann F.~W.,  Netzer H.,    Shields J.~C.,  2000, ApJ, 536, 101

\bibitem[\protect\citeauthoryear{Hopkins \& Elvis}{Hopkins \&
  Elvis}{2010}]{Hopkins:2010cf}
Hopkins P.~F.,  Elvis M.,  2010, MNRAS, 401, 7

\bibitem[\protect\citeauthoryear{Jimenez-Vicente, Mediavilla, Mu{\~n}oz \&
  Kochanek}{Jimenez-Vicente et~al.}{2012}]{JimenezVicente:2012es}
Jimenez-Vicente J.,  Mediavilla E.,  Mu{\~n}oz J.~A.,    Kochanek C.~S.,  2012,
  ApJ, 751, 106

\bibitem[\protect\citeauthoryear{Jin, Ward, Done \& Gelbord}{Jin
  et~al.}{2012}]{Jin:2012eu}
Jin C.,  Ward M.,  Done C.,    Gelbord J.,  2012, MNRAS, 420, 1825

\bibitem[\protect\citeauthoryear{Kalberla, Burton, Hartmann, Arnal, Bajaja,
  Morras \& P{\"o}ppel}{Kalberla et~al.}{2005}]{Kalberla:2005de}
Kalberla P. M.~W.,  Burton W.~B.,  Hartmann D.,  Arnal E.~M.,  Bajaja E.,
  Morras R.,    P{\"o}ppel W. G.~L.,  2005, A{\&}A, 440, 775

\bibitem[\protect\citeauthoryear{Kaspi et~al.,}{Kaspi
  et~al.}{2002}]{Kaspi:2002cr}
Kaspi S.  et~al., 2002, ApJ, 574, 643

\bibitem[\protect\citeauthoryear{Keeney, Danforth, Stocke, France \&
  Green}{Keeney et~al.}{2012}]{Keeney:2012kg}
Keeney B.~A.,  Danforth C.~W.,  Stocke J.~T.,  France K.,    Green J.~C.,
  2012, PASP, 124, 830

\bibitem[\protect\citeauthoryear{King}{King}{2003}]{King:2003gt}
King A.,  2003, ApJ, 596, L27

\bibitem[\protect\citeauthoryear{Kriss}{Kriss}{2011}]{Kriss:2011um}
Kriss G.~A.,  2011, STScI ISR COS 2011-01

\bibitem[\protect\citeauthoryear{Krolik \& Kriss}{Krolik \&
  Kriss}{1995}]{1995ApJ...447..512K}
Krolik J.~H.,  Kriss G.~A.,  1995, ApJ, 447, 512

\bibitem[\protect\citeauthoryear{Lee et~al.,}{Lee et~al.}{2013}]{Lee:2013jj}
Lee J.~C.  et~al., 2013, MNRAS, 430, 2650

\bibitem[\protect\citeauthoryear{Marinacci, Binney, Fraternali, Nipoti, Ciotti
  \& Londrillo}{Marinacci et~al.}{2010}]{Marinacci:2010ho}
Marinacci F.,  Binney J.~J.,  Fraternali F.,  Nipoti C.,  Ciotti L.,
  Londrillo P.,  2010, MNRAS, 404, 1464

\bibitem[\protect\citeauthoryear{Mathur, Wilkes \& Elvis}{Mathur
  et~al.}{1998}]{Mathur:1998fo}
Mathur S.,  Wilkes B.,    Elvis M.,  1998, ApJ, 503, L23

\bibitem[\protect\citeauthoryear{Mathur, Wilkes, Elvis \& Fiore}{Mathur
  et~al.}{1994}]{Mathur:1994ds}
Mathur S.,  Wilkes B.,  Elvis M.,    Fiore F.,  1994, ApJ, 434, 493

\bibitem[\protect\citeauthoryear{Matsuoka et~al.,}{Matsuoka
  et~al.}{2013}]{Matsuoka:2013ey}
Matsuoka K.  et~al., 2013, ApJ, 771, 64

\bibitem[\protect\citeauthoryear{Mazzotta, Mazzitelli, Colafrancesco \&
  Vittorio}{Mazzotta et~al.}{1998}]{Mazzotta:1998hi}
Mazzotta P.,  Mazzitelli G.,  Colafrancesco S.,    Vittorio N.,  1998, A{\&}AS,
  133, 403

\bibitem[\protect\citeauthoryear{Misawa, Charlton, Eracleous, Ganguly, Tytler,
  Kirkman, Suzuki \& Lubin}{Misawa et~al.}{2007}]{Misawa:2007dd}
Misawa T.,  Charlton J.~C.,  Eracleous M.,  Ganguly R.,  Tytler D.,  Kirkman
  D.,  Suzuki N.,    Lubin D.,  2007, ApJS, 171, 1

\bibitem[\protect\citeauthoryear{Moe, Arav, Bautista \& Korista}{Moe
  et~al.}{2009}]{Moe:2009do}
Moe M.,  Arav N.,  Bautista M.~A.,    Korista K.~T.,  2009, ApJ, 706, 525

\bibitem[\protect\citeauthoryear{Morris, Weymann, Foltz, Turnshek, Shectman,
  Price \& Boroson}{Morris et~al.}{1986}]{Morris:1986fa}
Morris S.~L.,  Weymann R.~J.,  Foltz C.~B.,  Turnshek D.~A.,  Shectman S.,
  Price C.,    Boroson T.~A.,  1986, ApJ, 310, 40

\bibitem[\protect\citeauthoryear{Muzahid, Srianand, Arav, Savage \&
  Narayanan}{Muzahid et~al.}{2013}]{Muzahid:2013dm}
Muzahid S.,  Srianand R.,  Arav N.,  Savage B.~D.,    Narayanan A.,  2013,
  MNRAS, 431, 2885

\bibitem[\protect\citeauthoryear{Muzahid, Srianand, Savage, Narayanan, Mohan \&
  Dewangan}{Muzahid et~al.}{2012}]{Muzahid:2012kl}
Muzahid S.,  Srianand R.,  Savage B.~D.,  Narayanan A.,  Mohan V.,    Dewangan
  G.~C.,  2012, MNRAS, 424, L59

\bibitem[\protect\citeauthoryear{Nestor, Hamann \& Rodriguez~Hidalgo}{Nestor
  et~al.}{2008}]{Nestor:2008gb}
Nestor D.,  Hamann F.,    Rodriguez~Hidalgo P.,  2008, MNRAS, 386, 2055

\bibitem[\protect\citeauthoryear{Nicastro, Fiore, Perola \& Elvis}{Nicastro
  et~al.}{1999}]{Nicastro:1999fj}
Nicastro F.,  Fiore F.,  Perola G.~C.,    Elvis M.,  1999, ApJ, 512, 184

\bibitem[\protect\citeauthoryear{Novak, Ostriker \& Ciotti}{Novak
  et~al.}{2011}]{Novak:2011gc}
Novak G.~S.,  Ostriker J.~P.,    Ciotti L.,  2011, ApJ, 737, 26

\bibitem[\protect\citeauthoryear{Oppenheimer \& Schaye}{Oppenheimer \&
  Schaye}{2013a}]{Oppenheimer:2013cr}
Oppenheimer B.~D.,  Schaye J.,  2013a, MNRAS, 434, 1063

\bibitem[\protect\citeauthoryear{Oppenheimer \& Schaye}{Oppenheimer \&
  Schaye}{2013b}]{Oppenheimer:2013dx}
Oppenheimer B.~D.,  Schaye J.,  2013b, MNRAS, 434, 1043

\bibitem[\protect\citeauthoryear{Osterman et~al.,}{Osterman
  et~al.}{2011}]{Osterman:2011gj}
Osterman S.  et~al., 2011, Ap{\&}SS, 335, 257

\bibitem[\protect\citeauthoryear{Ostriker, Choi, Ciotti, Novak \&
  Proga}{Ostriker et~al.}{2010}]{Ostriker:2010ik}
Ostriker J.~P.,  Choi E.,  Ciotti L.,  Novak G.~S.,    Proga D.,  2010, ApJ,
  722, 642

\bibitem[\protect\citeauthoryear{Petitjean, Rauch \& Carswell}{Petitjean
  et~al.}{1994}]{Petitjean:1994ti}
Petitjean P.,  Rauch M.,    Carswell R.~F.,  1994, A{\&}A, 291, 29

\bibitem[\protect\citeauthoryear{Petitjean \& Srianand}{Petitjean \&
  Srianand}{1999}]{Petitjean:1999vh}
Petitjean P.,  Srianand R.,  1999, A{\&}A, 345, 73

\bibitem[\protect\citeauthoryear{Risaliti \& Elvis}{Risaliti \&
  Elvis}{2010}]{Risaliti:2010jh}
Risaliti G.,  Elvis M.,  2010, A{\&}A, 516, 89

\bibitem[\protect\citeauthoryear{Sargent, Boksenberg \& Young}{Sargent
  et~al.}{1982}]{Sargent:1982kl}
Sargent W. L.~W.,  Boksenberg A.,    Young P.,  1982, ApJ, 252, 54

\bibitem[\protect\citeauthoryear{Scannapieco \& Oh}{Scannapieco \&
  Oh}{2004}]{Scannapieco:2004es}
Scannapieco E.,  Oh S.~P.,  2004, ApJ, 608, 62

\bibitem[\protect\citeauthoryear{Schaye}{Schaye}{2001}]{Schaye:2001dv}
Schaye J.,  2001, ApJ, 559, 507

\bibitem[\protect\citeauthoryear{Sharma, Nath \& Chand}{Sharma
  et~al.}{2013}]{Sharma:2013hk}
Sharma M.,  Nath B.~B.,    Chand H.,  2013, MNRAS, p.~L67

\bibitem[\protect\citeauthoryear{Shull, Stevans \& Danforth}{Shull
  et~al.}{2012}]{Shull:2012ki}
Shull J.~M.,  Stevans M.,    Danforth C.~W.,  2012, ApJ, 752, 162

\bibitem[\protect\citeauthoryear{Silk \& Rees}{Silk \&
  Rees}{1998}]{Silk:1998up}
Silk J.,  Rees M.~J.,  1998, A{\&}A, 331, L1

\bibitem[\protect\citeauthoryear{Simcoe et~al.,}{Simcoe
  et~al.}{2013}]{Simcoe:2013kh}
Simcoe R.~A.  et~al., 2013, PASP, 125, 270

\bibitem[\protect\citeauthoryear{Skrutskie et~al.,}{Skrutskie
  et~al.}{2006}]{Skrutskie:2006hl}
Skrutskie M.~F.  et~al., 2006, AJ, 131, 1163

\bibitem[\protect\citeauthoryear{Srianand}{Srianand}{2000}]{Srianand:2000jd}
Srianand R.,  2000, ApJ, 528, 617

\bibitem[\protect\citeauthoryear{Srianand \& Petitjean}{Srianand \&
  Petitjean}{2000}]{Srianand:2000tq}
Srianand R.,  Petitjean P.,  2000, A{\&}A, 357, 414

\bibitem[\protect\citeauthoryear{Srianand \& Petitjean}{Srianand \&
  Petitjean}{2001}]{Srianand:2001hh}
Srianand R.,  Petitjean P.,  2001, A{\&}A, 373, 816

\bibitem[\protect\citeauthoryear{Srianand \& Shankaranarayanan}{Srianand \&
  Shankaranarayanan}{1999}]{Srianand:1999hj}
Srianand R.,  Shankaranarayanan S.,  1999, ApJ, 518, 672

\bibitem[\protect\citeauthoryear{Tayal}{Tayal}{2006}]{Tayal:2006jo}
Tayal S.~S.,  2006, ApJS, 166, 634

\bibitem[\protect\citeauthoryear{Tejos et~al.,}{Tejos
  et~al.}{2014}]{2014MNRAS.437.2017T}
Tejos N.  et~al., 2014, MNRAS, 437, 2017

\bibitem[\protect\citeauthoryear{Telfer, Kriss, Zheng, Davidsen \&
  Green}{Telfer et~al.}{1998}]{Telfer:1998gx}
Telfer R.~C.,  Kriss G.~A.,  Zheng W.,  Davidsen A.~F.,    Green R.~F.,  1998,
  ApJ, 509, 132

\bibitem[\protect\citeauthoryear{Tripp, Lu \& Savage}{Tripp
  et~al.}{1996}]{Tripp:1996ge}
Tripp T.~M.,  Lu L.,    Savage B.~D.,  1996, ApJS, 102, 239

\bibitem[\protect\citeauthoryear{Trump et~al.,}{Trump
  et~al.}{2006}]{Trump:2006ht}
Trump J.~R.  et~al., 2006, ApJS, 165, 1

\bibitem[\protect\citeauthoryear{Veilleux, Cecil \& Bland-Hawthorn}{Veilleux
  et~al.}{2005}]{Veilleux:2005ec}
Veilleux S.,  Cecil G.,    Bland-Hawthorn J.,  2005, ARA{\&}A, 43, 769

\bibitem[\protect\citeauthoryear{V{\'e}ron-Cetty, Joly \&
  V{\'e}ron}{V{\'e}ron-Cetty et~al.}{2004}]{VeronCetty:2004ix}
V{\'e}ron-Cetty M.~P.,  Joly M.,    V{\'e}ron P.,  2004, A{\&}A, 417, 515

\bibitem[\protect\citeauthoryear{Vivek, Srianand, Petitjean, Noterdaeme, Mohan,
  Mahabal \& Kuriakose}{Vivek et~al.}{2012}]{Vivek:2012hh}
Vivek M.,  Srianand R.,  Petitjean P.,  Noterdaeme P.,  Mohan V.,  Mahabal A.,
    Kuriakose V.~C.,  2012, MNRAS, 423, 2879

\bibitem[\protect\citeauthoryear{Voges et~al.,}{Voges
  et~al.}{1999}]{Voges:1999ws}
Voges W.  et~al., 1999, A{\&}A, 349, 389

\bibitem[\protect\citeauthoryear{Voronov}{Voronov}{1997}]{1997ADNDT..65....1V}
Voronov G.~S.,  1997, Atomic Data and Nuclear Data Tables, 65, 1

\bibitem[\protect\citeauthoryear{Weymann, Morris, Foltz \& Hewett}{Weymann
  et~al.}{1991}]{Weymann:1991cn}
Weymann R.~J.,  Morris S.~L.,  Foltz C.~B.,    Hewett P.~C.,  1991, ApJ, 373,
  23

\bibitem[\protect\citeauthoryear{Weymann, Williams, Peterson \&
  Turnshek}{Weymann et~al.}{1979}]{Weymann:1979bt}
Weymann R.~J.,  Williams R.~E.,  Peterson B.~M.,    Turnshek D.~A.,  1979, ApJ,
  234, 33

\bibitem[\protect\citeauthoryear{White \& Frenk}{White \&
  Frenk}{1991}]{White:1991in}
White S. D.~M.,  Frenk C.~S.,  1991, ApJ, 379, 52

\bibitem[\protect\citeauthoryear{Wild et~al.,}{Wild et~al.}{2008}]{Wild:2008hn}
Wild V.  et~al., 2008, MNRAS, 388, 227

\bibitem[\protect\citeauthoryear{Williams, Strittmatter, Carswell \&
  Craine}{Williams et~al.}{1975}]{Williams:1975ge}
Williams R.~E.,  Strittmatter P.~A.,  Carswell R.~F.,    Craine E.~R.,  1975,
  ApJ, 202, 296

\bibitem[\protect\citeauthoryear{Wilms, Allen \& McCray}{Wilms
  et~al.}{2000}]{Wilms:2000en}
Wilms J.,  Allen A.,    McCray R.,  2000, ApJ, 542, 914

\bibitem[\protect\citeauthoryear{Woo \& Urry}{Woo \& Urry}{2002}]{Woo:2002kw}
Woo J.~H.,  Urry C.~M.,  2002, ApJ, 579, 530

\bibitem[\protect\citeauthoryear{Young, Sargent, Boksenberg, Carswell \&
  Whelan}{Young et~al.}{1979}]{Young:1979if}
Young P.~J.,  Sargent W. L.~W.,  Boksenberg A.,  Carswell R.~F.,    Whelan J.
  A.~J.,  1979, ApJ, 229, 891

\end{thebibliography}

\appendix

\section[]{Time-dependent ionization calculations}
\label{time_dependent_ionization_section}
The set of time-dependent ionization equations (\cref{time_dependent_ionization}) may be compactly written as
\begin{equation}
    \partial_t \mathbf{n} = \mathbf{An},
\label{pde}
\end{equation}
where $\mathbf{n}$ is a length $N + 1$ vector that specifies the ionic number densities for an element with $N$ electrons, and $\mathbf{A}$ is the $(N + 1) \times (N + 1)$ matrix containing the photoionization, collisional ionization and recombination rate coefficients, derived assuming some temperature, incident radiation field and electron number density, $n_e$. These equations are closed by the condition that
\begin{equation}
    \sum_{i = 0}^N n_{M_i} = n_{\textrm{total}},
\end{equation}
where $n_{\textrm{total}}$ can be related to $n_e$ in a highly ionized plasma, assuming solar metallicity and abundances, by
\begin{equation}
    n_{\textrm{total}} \approx \frac{2X}{1 + X} A_Z n_e.
\end{equation}
Here $A_Z$ is the absolute elemental abundance relative to hydrogen, and $X$ is the mass fraction in hydrogen. In equilibrium, $\partial_t \mathbf{n} = 0$, and we find that
\begin{equation}
    \frac{n_{M_{i + 1}}}{n_{M_i}} = \frac{\beta_{M_i} n_e + \Gamma_{M_i}}{n_e \alpha_{M_i}} \equiv a_{M_i}.
\end{equation}
It is then straightforward to show that we can write any $n_{M_i}$ for $i > 0$ in terms of $n_{M_0}$ via
\begin{equation}
    n_{M_i} = \left(\prod_{j = 0}^{i - 1} a_{M_j}\right) n_{M_0},
\end{equation}
where the scaling factors $a_{M_j}$ are defined above. Therefore
\begin{align}
    &n_{\textrm{total}} = n_{M_0} + \sum_{i = 1}^N \left(\prod_{j = 0}^{i - 1} a_{M_j}\right) n_{M_0} = \frac{2X}{1 + X} A_Z n_e \nonumber \\
&\implies n_{M_0} = \frac{2 X A_Z n_e}{(1 + X)\left(1 + \sum_{i = 1}^N\left(\prod_{j = 0}^{i - 1} a_j\right)\right)}.
\end{align}
Eliminating $n_{M_0}$, we finally arrive at an expression that defines the equilibrium number density of some ion $n_{M_i}$ as
\begin{equation}
    n_{M_i,\textrm{eq}} = \prod_{j = 0}^{i - 1} a_{M_j} \frac{2 X A_Z n_e}{(1 + X)\left(1 + \sum_{k = 1}^N\left(\prod_{j = 0}^{k - 1} a_{M_j}\right)\right)},
\label{equilibrium_value}
\end{equation}
for $i > 0$.

We first calculate the equilibrium set $n_{M_i,\textrm{eq}}$ using \cref{equilibrium_value}, assuming $n_e = 10^3~\textrm{cm}^{-3}$ (approximately that inferred from the analysis of \Cref{density}) and $X = 0.28$. To examine the effect of changing gas density, we then perturb these values and $n_e$ by a factor $(1 + \delta)$. For changes in the incident ionizing flux, we perturb the photoionization rates $\Gamma_{M_i}$ in an identical fashion. The time-dependent evolution in the number densities, $n$, of \ion{H}{1}, \ion{O}{4}, \ion{O}{5}, \ion{O}{6}, \ion{Ne}{8} and \ion{Mg}{10} are then calculated by numerically solving \cref{pde}, set to initially contain the perturbed $\mathbf{n}$ and/or $\mathbf{A}$ (note that both $\mathbf{n}$ and $\mathbf{A}$ change if $n_e$ changes). For $\mathbf{A}$, we additionally assume a temperature $T = 10^4$~K and illumination by the `UV peak' SED at a distance of 2.3~kpc.

\begin{figure}
    \centering
    \includegraphics[width=8.4cm]{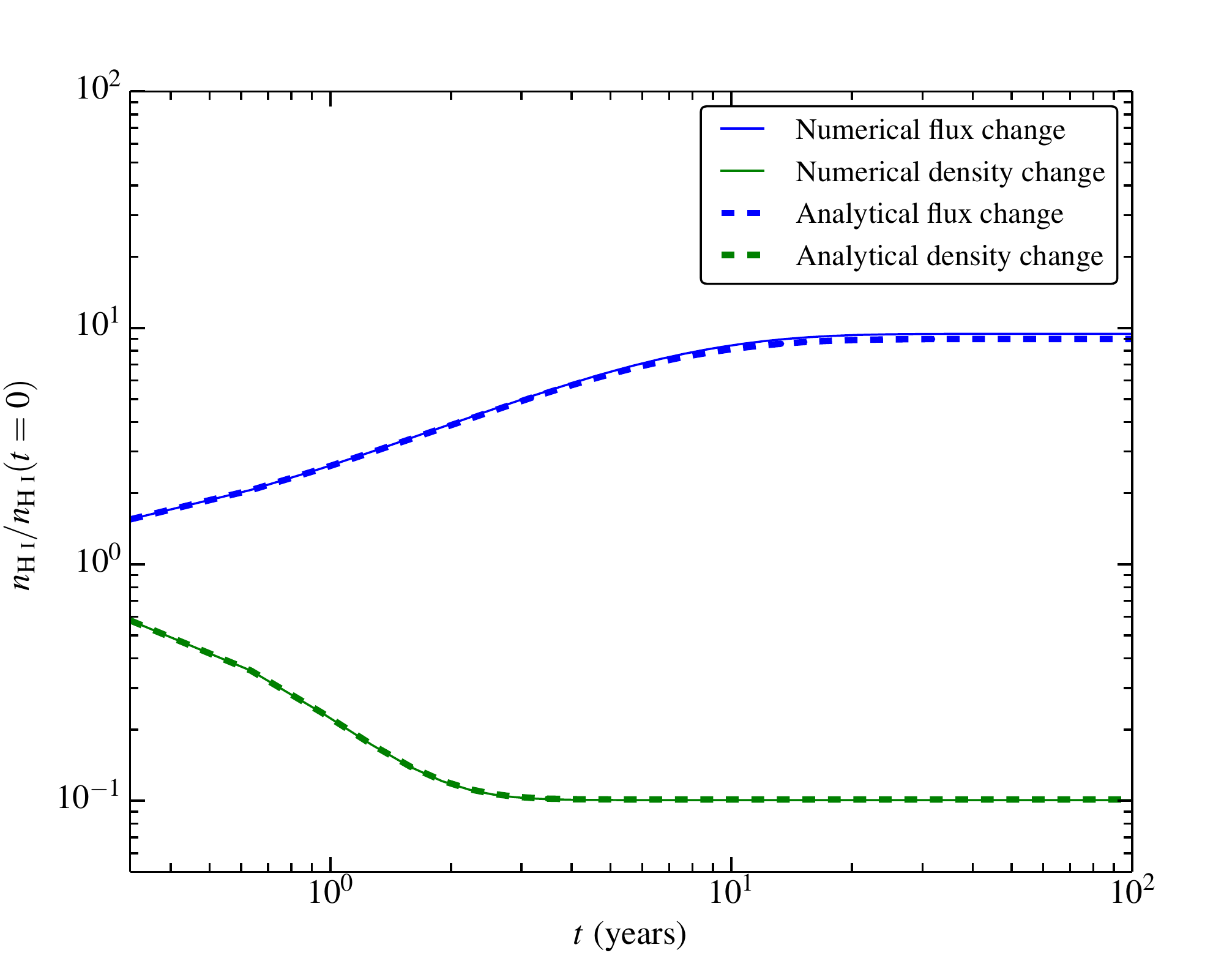}
    \caption{Non-equilibrium evolution in $n_{\textrm{H}\;\textsc{i}}$ following a step-function, order of magnitude change in flux (blue) and density (green) in a pure hydrogen gas. Numerical results are shown with solid lines, and analytical results are shown with dashed lines. The results are normalised with respect to the starting value at $t = 0$. We find an excellent agreement between the numerical and analytical calculations.}
\label{check}
\end{figure}

\begin{figure*}
    \centering
    \includegraphics[width=18cm]{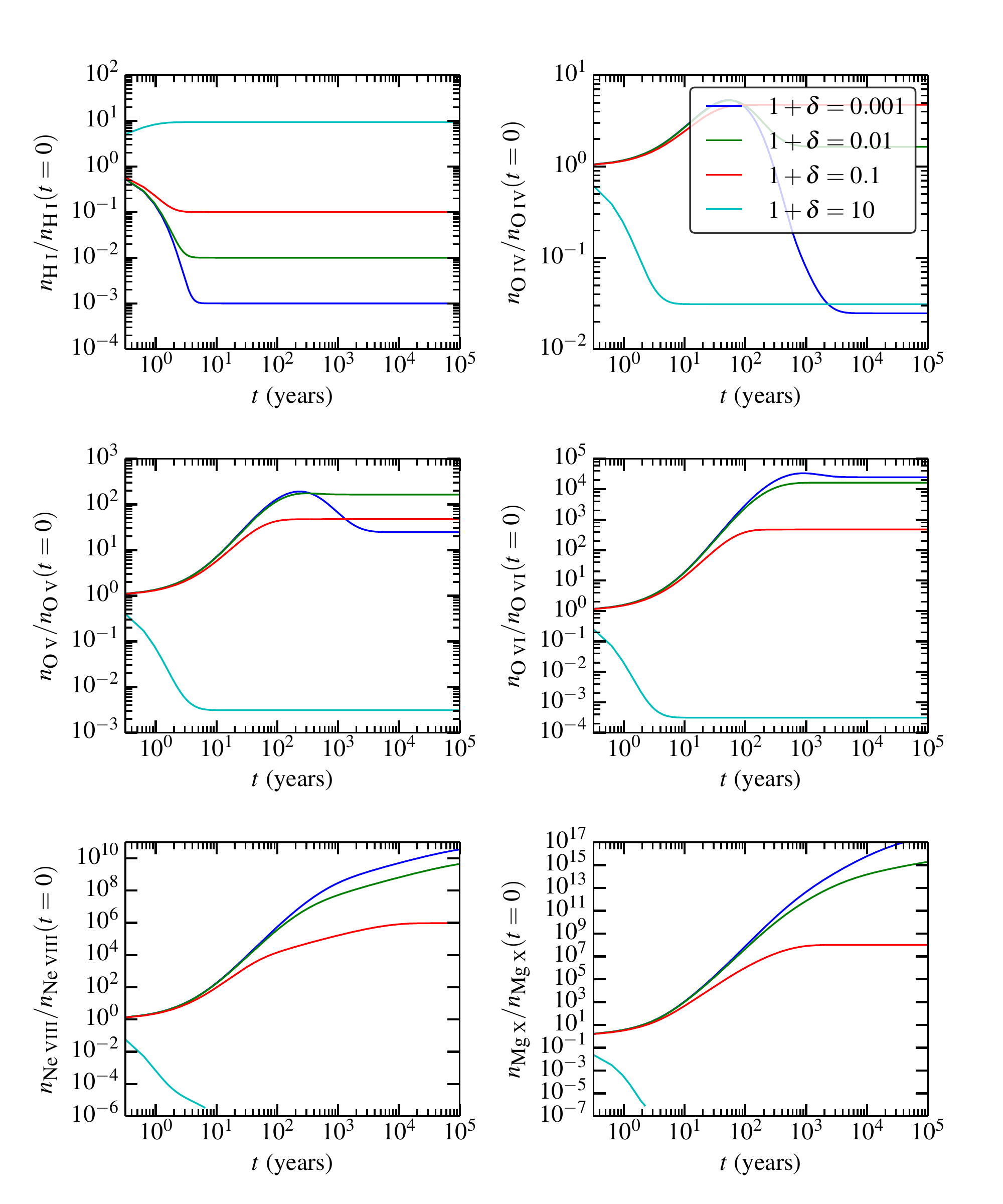}
    \caption{Non-equilibrium evolution in the number density, $n$, of ions $\textrm{H}\;\textsc{i}$, $\textrm{O}\;\textsc{iv}$, $\textrm{O}\;\textsc{v}$, $\textrm{O}\;\textsc{vi}$, $\textrm{Ne}\;\textsc{viii}$ and $\textrm{Mg}\;\textsc{x}$ relative to their starting values, following a step function change in density given by $n(t > 0) = (1 + \delta) n(t = 0)$. We set $n$ at $t = 0$ to the equilibrium values for a temperature $T = 10^4$~K, a distance of 2.3~kpc from the QSO modelled by the `UV peak' SED, and with $n_e = 10^3~\textrm{cm}^{-3}$. Numerical results for a range of $(1 + \delta)$ are presented with different coloured lines. Time-scales for a restored equilibrium are $> 100$~years for a drop in gas density of more than an order of magnitude, with the exception of \ion{H}{1}. Time-scales are orders of magnitude shorter for an increase in gas density.}
\label{non_equilibrium_figure}
\end{figure*}

\begin{figure*}
    \centering
    \includegraphics[width=18cm]{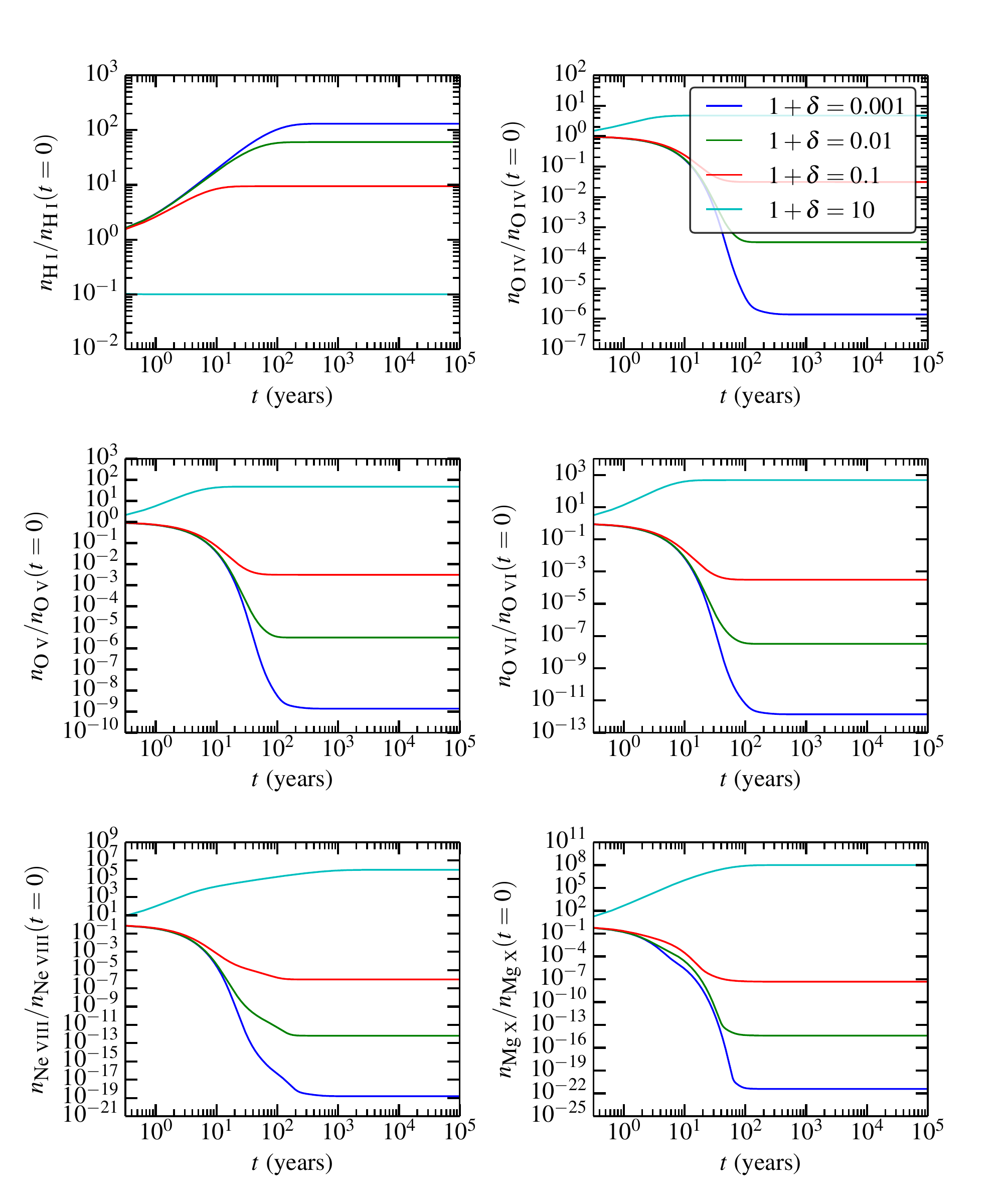}
    \caption{Non-equilibrium evolution in the number density, $n$, of ions $\textrm{H}\;\textsc{i}$, $\textrm{O}\;\textsc{iv}$, $\textrm{O}\;\textsc{v}$, $\textrm{O}\;\textsc{vi}$, $\textrm{Ne}\;\textsc{viii}$ and $\textrm{Mg}\;\textsc{x}$ relative to their starting values, following a step function change in incident flux given by $\Gamma(t > 0) = (1 + \delta) \Gamma(t = 0)$. We set $\Gamma$ at $t = 0$ to the photoionization rate at a distance of 2.3~kpc from the QSO modelled by the `UV peak' SED. Numerical results for a range of $(1 + \delta)$ are presented with different coloured lines. Time-scales for a restored equilibrium are, typically, $< 100$~years. These results agree well with the typical $e$-folding time-scales derived in \cref{time-scale}.}
\label{non_equilibrium_figure2}
\end{figure*}

In the case of \ion{H}{1}, we can obtain an analytical solution to the rate of change in the neutral fraction $x \equiv n_{\textrm{H}\;\textsc{i}} / n_{\textrm{H}}$, determined by the hydrogen photoionization rate $\Gamma$, collisional ionization rate coefficient $\beta$, and recombination rate coefficient $\alpha$, as well as the electron number density $n_e$, according to
\begin{equation}
    \frac{\textrm{d}x}{\textrm{d}t} = -\left(\Gamma + \beta n_e\right) + \alpha n_e(1 - x).
\end{equation}
For a pure hydrogen gas ($X = 1$), we can express the electron number density as $n_e = (1 - x)n_{\textrm{H}}$ and write $\textrm{d}x/\textrm{d}t$ in the form of a Riccati equation:
\begin{align}
    &\frac{\textrm{d}x}{\textrm{d}t} = Rx^2 + Qx + P \\
    &R \equiv (\beta + \alpha)n_{\textrm{H}} \\ \nonumber
    &Q \equiv -\left(\Gamma + \alpha n_{\textrm{H}} + R\right) \\ \nonumber
    &P \equiv \alpha n_{\textrm{H}}
\end{align}
\citep{Altay:2013dua}. Assuming that $P$, $Q$ and $R$ are all constant, the time-dependent solution can then be found by separation of variables:
\begin{equation}
    x(t) = x_- + \left(x_0 - x_-\right) \frac{\left(x_+ - x_-\right)F}{\left(x_+ - x_0\right) + \left(x_0 - x_-\right)F},
\label{analytical_calculation}
\end{equation}
where $x_+$ and $x_-$ are the roots of the quadratic term in equation (A9), $x_0 \equiv x(t = 0)$ is the initial value, and
\begin{align}
    &F(t) \equiv \exp \left(\frac{-\left(x_+ - x_-\right)t}{t_{\textrm{rec}}}\right) \\ \nonumber
    &t_{\textrm{rec}} \equiv \frac{1}{(\alpha + \beta)n_{\textrm{H}}}.
\end{align}
We identify $t_{\textrm{rec}}$ as the recombination time-scale. It can be shown that $x_-$ represents the physical equilibrium solution \citep{Altay:2013dua}, so for the density changes described, both $x (t = \infty)$ and $x (t = 0)$ take the form of $x_-$, but with a different value of $n_{\textrm{H}}$. The same applies for incident ionizing flux changes, but with different values of $\Gamma$ instead.

First, to check the validity of our numerical calculations, we compare the numerically calculated non-equilibrium evolution in the number density of \ion{H}{1} for a pure hydrogen gas ($X = 1$) with that computed from \cref{analytical_calculation}. We do so for an order of magnitude step-function decrease in both the incident ionizing flux and the gas density separately. The results of this comparison are shown in \Cref{check}. We find an excellent agreement between these two calculations, which confirms that our numerical results are robust. Full numerical results for the evolution in $n$ following a step-function change in density are plotted in \Cref{non_equilibrium_figure} relative to the starting values $n(t = 0)$, for a range of values of $(1 + \delta)$. A similar calculation for changes in the incident ionizing flux (described in \Cref{equilibrium}) is presented in \Cref{non_equilibrium_figure2}. We find that the numerical results in the latter agree well with the typical $e$-folding time-scales derived in \cref{time-scale}.

\label{lastpage}
\end{document}